\documentclass[11pt]{article}
\usepackage{a4wide,amsmath,epsfig,longtable}
\usepackage{amssymb}

\allowdisplaybreaks[4]

\begin{document}

\rightline{LU TP 11-20}

\vspace*{1cm}

\begin{center}
{\LARGE \textbf{Resonance saturation in the odd-intrinsic\\
parity sector of low-energy QCD\\[2cm]
}}

{\large \textbf{Karol Kampf$^{\,1,2}$ and Ji\v{r}\'{\i} Novotn\'y$^{\,2}$}\\[%
1.2 cm]
} $\ ^{1}$\textit{Department of Astronomy and Theoretical Physics. Lund
University, S\"olvegatan 14A, SE 223-62 Lund, Sweden}\\[0.4cm]
$\ ^{2}$\textit{Institute of Particle and Nuclear Physics, Faculty of
Mathematics and Physics,\\[0pt]
Charles University in Prague, 18000 Prague, Czech Republic.}
\end{center}

\vspace*{2.0cm}
\begin{abstract}
Using the large $N_C$ approximation we have constructed the most
general chiral resonance Lagrangian in the odd-intrinsic parity
sector that can generate low energy chiral constants up to
$\mathcal{O}(p^6)$. Integrating out the resonance fields these
$\mathcal{O}(p^6)$ constants are expressed in terms of resonance
couplings and masses. The role of $\eta^{\prime }$ is discussed
and its contribution is explicitly factorized. Using the resonance
basis we have also calculated two QCD Green functions of currents:
$\langle VVP\rangle$ and $\langle VAS \rangle$ and found, imposing
high energy constraints, additional relations for resonance
couplings. We have studied several phenomenological implications
based on these correlators from which let us mention here our
prediction for the $\pi^0$-pole contribution to the muon $g-2$
factor: $a_\mu^{\pi^0} = 65.8(1.2)\times 10^{-11}$.
\end{abstract}
\newpage
\tableofcontents

\setcounter{footnote}{0}

%%%%%%%%%%%%%%%%%%%%%%%%%%%%%%%%%%%%%%

\section{Introduction}

As is well known, there are two regimes where the QCD dynamics of the
current correlators is well understood. The first one corresponds to the
high energies where the asymptotic freedom allows to use the perturbative
approach in terms of the strong coupling constant $\alpha _{s}$ and where
the asymptotics of the correlators for large euclidean momenta is governed
by operator product expansion (OPE). The second well understood region is
that of low external momenta where the dynamics is constrained by the
spontaneously broken chiral symmetry. As a consequence, the dominant
contributions to the correlators and related amplitudes of the processes
under interest come from the octet of the lightest pseudoscalar mesons ($\pi
$, $K$, $\eta $) \ which are the corresponding (pseudo)Goldstone bosons
(GB). The correlators can be studied here by means of Chiral Perturbation
Theory (ChPT)\cite{Weinberg:1978kz,Gasser:1983yg,Gasser:1984gg}, which is
the effective Lagrangian field theory for this region, in terms of
systematic simultaneous expansion in powers (and logs) of the momenta and
quark masses. The applicability of ChPT extends up to the hadronic scale $%
\Lambda _{H}\sim 1\mathrm{GeV}$ which corresponds to the onset of
non-Goldstone resonances and where the ChPT expansion fails to converge.

OPE and ChPT provides us with asymptotic behaviour of the correlators in
different regimes, however, both these approaches need further
non-perturbative long-distance piece of information which is not known from
the first principles, namely the values of the vacuum condensates for OPE
and the values of the effective low-energy constants (LECs) for ChPT. In the
latter case the LECs parameterize our lack of detailed information on the
non-perturbative dynamics of the degrees of freedom above the hadronic scale
$\Lambda _{H}$ and are connected with the order parameters of the
spontaneously broken chiral symmetry. The predictivity of ChPT heavily
relies on their determination. At the order $O(p^6)$, which corresponds to
the recent accuracy of the NNLO ChPT calculation (for a comprehensive review
and further references see \cite{Bijnens:2006zp}), 90+4 LECs in the even
intrinsic parity sector \cite{Bijnens:1999sh,Bijnens:1999hw} and 23 LECs in
the odd sector\footnote{%
These numbers of LECs are relevant for $SU(3)$ variant of ChPT. In the $%
SU(2) $ case we get 53+4 LECs in the intrinsic parity even sector and 5(13)
LECs in the odd sector.} \cite{Ebertshauser:2001nj,Bijnens:2001bb} appear in
the effective Lagrangian. Though only special linear combinations of them
are relevant for particular physical amplitudes, the uncertainty in their
estimation is usually the weakest point of the interconnection between the
theory and experiment.

Dispersion representation of those correlators which are order parameters of
the chiral symmetry breaking (and therefore do not get any genuine
perturbative contribution) enables to make use of information on the
asymptotics both in the low and high energy regions and to relate the
unknown LECs to the properties of the corresponding spectral functions in
terms of the chiral sum rules \cite%
{Gasser:1983yg,Donoghue:1993xb,Davier:1998dz,Moussallam:1994xp,
Knecht:2001xc}. These are usually assumed to be saturated by the low-lying
resonant states; such an assumption (known as resonance saturation
hypothesis) connects the LECs to the phenomenology of resonances in the
intermediate energy region $1\mathrm{GeV}\leq E< 2\mathrm{GeV}$. Though the
inclusion of only finite number of resonances has been questioned in the
literature \cite{Masjuan:2007ay,Golterman:2006gv}, it proved to be
consistent in the $O(p^4)$ case with other phenomenological determinations
of LECs.

The necessary ingredient of the resonance saturation approach to the
determination / estimation of LECs is the phenomenological information on
the physics of the lowest resonances. It can be conveniently parameterized
by means of suitable phenomenological Lagrangian. Along with the chiral
symmetry the guiding theoretical principles for its construction are those
based on the large $N_{C}$ expansion of QCD \cite{'tHooft:1973jz}. Within
the leading order in $1/N_{C}$ the correlators of the quark bilinears are
given by an infinite sum of contributions of narrow meson resonance states
the mass of which scales as $O(N_{C}^{0})$ \ and the interaction of which is
suppressed by an appropriate power of $1/\sqrt{N_{C}}$. Such a large $N_{C}$
\ representation of the correlators can be reconstructed using effective
Lagrangian $\mathcal{L}_{\infty }$ including GB and infinite tower of
resonance fields with couplings of the order $O(N_{C}^{1-n/2})$ according to
the number $n$ of the resonance fields in the interaction vertices. The $%
1/N_{C}$ expansion is equivalent to the quasi-classical expansion, thus at
the leading order only the tree graphs contribute and each additional loop
is suppressed by one power of $1/N_{C}$.

Though $\mathcal{L}_{\infty }$ is not known from the first principles, the
information on the large $N_{C}$ hierarchy of the individual operators
together with general symmetry assumptions allows one to construct all the
relevant terms necessary to determine the LECs in the leading order of the
large $N_{C}$ expansion up to given chiral order. The large $N_{C}$
approximation of LECs can be then formally achieved by means of the
integrating out the resonance fields from the Lagrangian $\mathcal{L}%
_{\infty }$. Formally one gets LECs expressed in terms of the (from the
first principles unknown) masses and couplings of the infinite tower of
resonances.

The large $N_{C}$ \ inspired phenomenological Lagrangian suitable for the
resonance saturation program for LECs can be then obtained as an
approximation to $\mathcal{L}_{\infty }$ where only finite number of
resonances is kept. \ Such a truncation of $\mathcal{L}_{\infty }$ seems to
be legitimate at low energies where the contribution of the higher
resonances is expected to be suppressed. However, the lack of effective
cut-off scale which could play here a role analogous to $\Lambda _{H}$ \ for
ChPT prevents us to interpret the resonance phenomenological Lagrangian as a
well defined effective theory in the usual sense. It is rather a QCD
inspired phenomenological model which should share as much common features
with QCD as possible. The latter principle generally puts various
constraints on its effective couplings. For instance, the finite number of
resonances involved generally corrupts the asymptotic behaviour of the
correlators required by perturbative QCD and OPE. However, it is natural to
expect that for the correlators which are order parameters of the
spontaneous chiral symmetry breaking that the latter behaviour extends down
to the region of applicability of the phenomenological Lagrangian and thus
it is desirable to ensure the correct asymptotics by means of adjusting its
couplings. This is however not enough to fix all of them (often it is even
not possible to satisfy all the OPE requirements at once by a finite set of
resonances) therefore further phenomenological input is needed.

At the leading order in $1/N_{C}$, the above strategy for determination of
LECs is essentially equivalent to the similar approach known as Minimal
Hadronic Ansatz (MHA)\cite{Peris:2000tw}. Within this approach the
correlators are approximated by meromorphic function with correct pole
structure corresponding to the resonance poles and the free parameters are
fixed both by OPE constraints and experimental inputs. Only minimal number
of resonances is taken into account, just those necessary to satisfy all the
relevant OPE (when only the lowest resonances in each channel are included,
the method is called Lowest Meson Dominance (LMD) ansatz \cite%
{Peris:2000tw,Knecht:2001xc}, but in this case not all OPE constraints are
guaranteed to be met \cite{Knecht:2001xc,Bijnens:2003rc}). Matching this
ansatz to the low energy ChPT expansion enables to determine relevant linear
combinations of LECs.

The method based on the resonance Lagrangian is however little bit more
general than MHA or LMD. On one hand it enables to determine (at least in
principle) the individual LECs, not only their linear combinations connected
with particular correlators, on the other hand it provides a natural
framework for going beyond the leading order in $1/N_{C}$ by means of
integrating out the resonances at one loop level \cite%
{Rosell:2005ai,Rosell:2006dt,Pich:2008jm,Rosell:2009yb,Pich:2010sm} which
also takes correctly into account the renormalization scale dependence of
the LECs.

The above principles of construction of phenomenological Lagrangian with
resonances are known since 1989 when the seminal paper \cite{Ecker:1988te}
on what is now known as Resonance Chiral Theory (R$\chi$T) was published. In
this paper the resonance saturation of the $O(p^{4})$ LECs has been studied
systematically while the $O(p^{6})$ LECs of the even intrinsic parity sector
of ChPT has been systematically analyzed 17 years later in \cite%
{Cirigliano:2006hb}. For a recent review and further references see \cite%
{Portoles:2010yt}.

The study of the odd intrinsic parity sector of R$\chi$T with vector
resonances and corresponding saturation of the LECs for the $O(p^{6})$
anomaly sector of ChPT started in \cite{Pallante:1992qe} and \cite%
{Prades:1993ys,Knecht:2001xc,RuizFemenia:2003hm}, where also axial vector
resonances has been included and where the particular operator basis of the R%
$\chi$T Lagrangian contributing to the correlators under interest has been
constructed. The influence of pseudoscalar resonances on the odd intrinsic
parity LECs has been studied in \cite{Moussallam:1994xp,Knecht:2001xc} and
corresponding part of R$\chi$T Lagrangian has been constructed in \cite%
{Ananthanarayan:2002kj} (see also \cite{Kampf:2009tk}). In this paper we
resume this effort and construct the most general odd intrinsic parity
sector of the R$\chi$T Lagrangian including the lowest multiplets of the
vector $V(1^{--})$, axial-vector $A(1^{++})$, scalar $S(0^{++})$ and
pseudoscalar $P(0^{-+})$ resonances. In the $0^{-+}$ channel we introduce
thus beside the GB also the lowest non-GB resonance multiplet and therefore
we go beyond the LMD approximation (our correlators then correspond to what
is called in \cite{Knecht:2001xc} as LMD+P ansatz). The resulting Lagrangian
is then used for the lowest resonance saturation of the $O(p^{6})$ anomaly
sector of ChPT. We also illustrate the general strategy of matching the
correlators with OPE on the concrete example of $\langle VVP\rangle $ and $%
\langle VAS\rangle $ three point functions and discuss related
phenomenological applications.

The paper is organized as follows. In Sect. 2 we fix our notation
and remind briefly the principles of the construction of the
Lagrangian of the $R\chi T$. Sec. 3 is devoted to the presentation
of the complete basis of the odd intrinsic parity sector of $R\chi
T$. In Sect. 4 we discuss related phenomenological applications
and in Sect. 5 we give the result of the resonance saturation of
the odd intrinsic parity $O(p^{6})$ LECs. A brief summary is given
in Sec 6. The large $N_C$ counting of the relevant operators is
discussed in Appendix A and the operator redefinitions and
reduction of the Lagrangian is studied in Appendix B.

\section{The Resonance Chiral Theory}

In what follows we will work in the chiral limit. The standard basic
building block which includes the octet of GB (here we assume that $\eta ^{{%
\prime }}$ has been already integrated out from our effective Lagrangian,
for details see Appendix~\ref{apA}) is
\begin{equation}
u(\phi )=\exp \left( i\frac{\phi }{\sqrt{2}F}\right)\,,
\end{equation}%
where $\phi =\frac{1}{\sqrt{2}}\lambda ^{a}\phi ^{a}$, $\lambda^i$ being a
standard Gell-Mann matrix and
\begin{equation}
\phi (x)\,\,=\,\,\left(
\begin{array}{ccc}
\frac{1}{\sqrt{2}}\pi ^{0}+\frac{1}{\sqrt{6}}\eta _{8} & \pi ^{+} & K^{+} \\
\pi ^{-} & -\frac{1}{\sqrt{2}}\pi ^{0}+\frac{1}{\sqrt{6}}\eta _{8} & K^{0}
\\
K^{-} & \overline{K}^{0} & -\frac{2}{\sqrt{6}}\eta _{8}%
\end{array}%
\right) \,.
\end{equation}%
One can form the basic covariant tensors \cite{ColemanCallan}, \cite%
{Bijnens:1999sh}
\begin{align}
& u_{\mu }=u_{\mu }^{\dagger }=i\,\{u^{\dagger }(\partial _{\mu }-ir_{\mu
})u\,-\,u\,(\partial _{\mu }-i\ell _{\mu })u^{\dagger }\}\,,  \notag \\
& \chi _{\pm }=u^{\dagger }\,\chi \,u^{\dagger }\,\pm \,u\,\chi ^{\dagger
}\,u\,,  \notag \\
& f_{\pm }^{\mu \nu }=u\,F_{L}^{\mu \nu }\,u^{\dagger }\,\pm \,u^{\dagger
}\,F_{R}^{\mu \nu }\,u\,,  \notag \\
& h_{\mu \nu }=\nabla _{\mu }u_{\nu }+\nabla _{\nu }u_{\mu }\,,
\end{align}%
with $\chi =2B_{0}(s+ip)$, where $s$ and $p$ stand for the scalar and
pseudo-scalar external sources. Vector source $v^{\mu }$ and axial-vector
source $a^{\mu }$ are then related to the right and left sources $r^{\mu }$
and $\ell ^{\mu }$ by relations $v^{\mu }=\frac{1}{2}(r^{\mu }+\ell ^{\mu })$
and $a^{\mu }=\frac{1}{2}(r^{\mu }-\ell ^{\mu })$ respectively, and $%
F_{L,R}^{\mu \nu }$ the corresponding left and right field-strength tensors:
\begin{equation}
F_{R}^{\mu \nu }=\partial ^{\mu }r^{\nu }-\partial ^{\nu }r^{\mu }-i[r^{\mu
},r^{\nu }]\,,\qquad F_{L}^{\mu \nu }=\partial ^{\mu }l^{\nu }-\partial
^{\nu }l^{\mu }-i[l^{\mu },l^{\nu }]\,.
\end{equation}%
The covariant derivative is defined by
\begin{equation}
\nabla _{\mu }X=\partial _{\mu }+[\Gamma _{\mu },X]\,,
\end{equation}%
where the chiral connection is
\begin{equation}
\Gamma _{\mu }=\frac{1}{2}\{u^{\dagger }(\partial _{\mu }-ir_{\mu
})u+u(\partial _{\mu }-il_{\mu })u^{\dagger }\}\,.
\end{equation}

Inspired by the large $N_{C}$ limit the GB couple to massive $U(3)$
multiplets of the type $V(1^{--})$, $A(1^{++})$, $S(0^{++})$ and $P(0^{-+})$%
, denoted generically as a nonet field $R$. This field can be decomposed
into octet $R_{8}$ and singlet $R_{0}$ via
\begin{equation}
R=\frac{1}{\sqrt{3}}R_{0}+\sum_{i}\frac{\lambda _{i}}{\sqrt{2}}R_{i}\,.
\end{equation}%
The explicit form of the vector multiplet $V(1^{--})$ is
\begin{equation}
V_{\mu \nu }\,\,=\,\,\left(
\begin{array}{ccc}
\frac{1}{\sqrt{2}}\rho ^{0}+\frac{1}{\sqrt{6}}\omega _{8}+\frac{1}{\sqrt{3}}%
\omega _{1} & \rho ^{+} & K^{\ast \,+} \\
\rho ^{-} & -\frac{1}{\sqrt{2}}\rho ^{0}+\frac{1}{\sqrt{6}}\omega _{8}+\frac{%
1}{\sqrt{3}}\omega _{1} & K^{\ast \,0} \\
K^{\ast \,-} & \overline{K}^{\,\ast \,0} & -\frac{2}{\sqrt{6}}\omega _{8}+%
\frac{1}{\sqrt{3}}\omega _{1}%
\end{array}%
\right) _{\mu \nu }\,,
\end{equation}%
(and similarly for other types). We use here the antisymmetric tensor field
for description of the spin-1 resonances. The reason is that though it is in
principle equivalent to the Proca field formalism (see \cite{Ecker:1989yg}
and \cite{Kampf:2006yf} for the general discussion of the equivalence at the
order $O(p^{4})$ and $O(p^{6})$ respectively and \cite{Kampf:2009jh} for
particular discussion of the one-loop equivalence), the antisymmetric tensor
field naturally couples to the lowest order $O(p^{2})$ chiral building
blocks without derivatives and therefore it does not require additional
contact terms necessary to compensate the wrong high energy behaviour of
amplitudes and form factors under interest. Moreover, when using the Proca
field without such contact terms it is not possible to saturate the $O(p^{4})
$ LECs in the even intrinsic parity sector and for the similar reason also
the LECs in the $O(p^{6})$ odd intrinsic parity sector.

According to the large $N_{C}$ counting of interaction vertices with
resonances we can organize the Lagrangian $\mathcal{L}_{R\chi T}$ of R$\chi$%
T as an expansion in the number of resonance fields,
\begin{equation}
\mathcal{L}_{R\chi T}=\mathcal{L}_{GB}+\mathcal{L}_{RR,\mathrm{kin}}+%
\mathcal{L}_{R}+\mathcal{L}_{RR^{\prime }}+\mathcal{L}_{RR^{\prime
}R^{\prime \prime }}+\dots  \label{NC}
\end{equation}%
Here $\mathcal{L}_{GB}$ contains only Goldstone bosons and external sources
and includes terms with the same structure as the usual $SU(3)_{L}\times
SU(3)_{R}$ ChPT Lagrangian, but the coupling constants are generally
different. The resonance kinetic terms $\mathcal{L}_{RR,\mathrm{kin}}$,
which are of the order $O(N_{C}^{0})$, have the form
\begin{equation}
\mathcal{L}_{RR,\mathrm{kin}}\,\,=\,\,-\frac{1}{2}\langle \nabla ^{\mu
}R_{\mu \nu }\nabla _{\alpha }R^{\alpha \nu }\rangle +\frac{1}{4}%
M_{R}^{2}\langle R_{\mu \nu }R^{\mu \nu }\rangle +\frac{1}{2}\langle \nabla
^{\alpha }R^{\prime }\nabla _{\alpha }R^{\prime }\rangle -\frac{1}{2}%
M_{R^{\prime }}^{2}\langle R^{\prime }R^{\prime }\rangle\,,
\end{equation}%
where $R$ stands for $V^{\mu \nu }$ and $A^{\mu \nu }$ while $R^{\prime }$
stands for $S$ and $P$. The terms $\mathcal{L}_{R}$, $\mathcal{L}%
_{RR^{\prime }}$ and $\mathcal{L}_{RR^{\prime }R^{\prime \prime }}$ collect
the interaction vertices linear, quadratic and cubic in the resonance
fields, respectively.

There is also another type of expansion for $\mathcal{L}_{R\chi T}$. It is
based on the ordering according to the contribution to chiral coupling
constants. Within this counting, the resonance fields are effectively of the
order
\begin{equation}
R=O(p^{2})\,,  \label{R_effective_counting}
\end{equation}%
while the chiral building blocks with GB only are counted in a usual way.
For $\mathcal{L}_{GB}$ it is therefore just the usual chiral power counting.
Combining this with the large $N_{C}$ expansion (\ref{NC}) we can write
\begin{equation}
\mathcal{L}_{R\chi T}=\mathcal{L}_{GB}^{(2)}+\mathcal{L}_{GB}^{(4)}+\mathcal{%
L}_{RR,kin}^{(4)}+\mathcal{L}_{RR,kin}^{(6)}+\mathcal{L}_{R}^{(4)}+\mathcal{L%
}_{GB}^{(6)}+\mathcal{L}_{R}^{(6)}+\mathcal{L}_{RR^{\prime }}^{(6)}+\mathcal{%
L}_{RR^{\prime }R^{\prime \prime }}^{(6)}\dots\,,
\end{equation}%
where the subscript $\,^{(n)}$ stands for the contribution to $\mathcal{O}%
(p^{n})$ chiral constant. For our further discussion we will explicitly need
\begin{equation}
\mathcal{L}_{GB}^{(2)}=\frac{F^{2}}{4}\langle u_{\mu }u^{\mu }+\chi
_{+}\rangle \,.
\end{equation}%
The leading order of the odd intrinsic parity sector of $\mathcal{L}%
_{GB}^{(4)}$ coincides with the Wess-Zumino-Witten Lagrangian \cite%
{Witten:1983tw} $\mathcal{L}_{WZW}^{(4)}$. For the explicit form of even
parity part $\mathcal{L}_{GB}^{(4)}$ and complete $\mathcal{L}_{GB}^{(6)}$
see \cite{Gasser:1984gg, Bijnens:1999sh, Bijnens:2001bb}, (see also \cite%
{Ebertshauser:2001nj}).

The most general interaction Lagrangian $\mathcal{L}_{R}^{(4)}$ which is
relevant for the saturation of the $O(p^{4})$ LECs \cite{Ecker:1988te} is
linear in resonance fields, namely
\begin{align}
\mathcal{L}_{R}^{(4)}=& \;c_{d}\langle Su^{\mu }u_{\mu }\rangle
+c_{m}\langle S\chi _{+}\rangle +id_{m}\langle P\chi _{-}\rangle +i\frac{%
d_{m0}}{N_F}\langle P\rangle \langle \chi _{-}\rangle +  \notag \\
& +\frac{F_{V}}{2\sqrt{2}}\langle V_{\mu \nu }f_{+}^{\mu \nu }\rangle +\frac{%
iG_{V}}{2\sqrt{2}}\langle V_{\mu \nu }[u^{\mu },u^{\nu }]\rangle +\frac{F_{A}%
}{2\sqrt{2}}\langle A_{\mu \nu }f_{-}^{\mu \nu }\rangle  \label{LR}
\end{align}%
and all the couplings are of the order $O(N_{C}^{1/2})$. This is true also
for the last term of the first line with two traces which is enhanced due to
the $\eta ^{{\prime }}$ exchange (see Appendix~\ref{apA}, esp.~(\ref{dm0con}%
)). This term with $d_{m0}$ (depending solely on the singlet component of $P$%
) has not yet been studied in the phenomenology as it always contributes to
the saturation of LECs together with the large $N_C$ enhanced $\eta^{\prime }
$ exchange. The complete operator basis of the $O(p^{6})$ even intrinsic
parity of R$\chi$T has been constructed in \cite{Cirigliano:2006hb}.

Integrating out the resonance fields at the tree level we reconstruct the
Lagrangian $\mathcal{L}_{\chi PT}$ of ChPT, schematically
\begin{equation}
\exp \left( i\int d^{4}x\mathcal{L}_{\chi PT}\right) =\int \mathcal{D}R\exp
\left( i\int d^{4}x\mathcal{L}_{R\chi T}\right) .  \label{PI}
\end{equation}%
Effectively up to the order $O(p^{6})$ the integration over $R$ is
equivalent to the insertion of the solution $R^{(2)}$ of the lowest order
equation of motion (i.e. those derived from $\mathcal{L}_{RR,kin}^{(4)}+%
\mathcal{L}_{R}^{(4)}$) for resonance field $R$ \ into the Lagrangian $%
\mathcal{L}_{R\chi T}$. Because the resonance fields $R$ couples to the $%
O(p^{2})$ building blocks in $\mathcal{L}_{R}^{(4)}$ and the resonance
masses are counted as $O(p^{0})$, we are consistent with the chiral counting
(\ref{R_effective_counting}). Finally we get

\begin{equation*}
\mathcal{L}_{\chi PT}=\mathcal{L}_{\chi }^{(2)}+\mathcal{L}_{\chi }^{(4)}+%
\mathcal{L}_{\chi }^{(6)}+\dots
\end{equation*}
with explicit separate contribution from Goldstone bosons part of the R$\chi$%
T Lagrangian $\mathcal{L}_{GB}$ and the leading $N_{C}$ contribution of the
resonances
\begin{eqnarray}
\mathcal{L}_{\chi }^{(2)} &=&\mathcal{L}_{GB}^{(2)} \\
\mathcal{L}_{\chi }^{(n>2)} &=&\mathcal{L}_{GB}^{(n)}+\mathcal{L}_{\chi
,R}^{(n)}\,,
\end{eqnarray}%
where particularly%
\begin{eqnarray*}
\mathcal{L}_{\chi ,R}^{(4)} &=&\left(\mathcal{L}_{RR,kin}^{(4)}+\mathcal{L}%
_{R}^{(4)}\right)\big|_{R\rightarrow R^{(2)}} \\
\mathcal{L}_{\chi ,R}^{(6)} &=&\left(\mathcal{L}_{RR,kin}^{(6)}+\mathcal{L}%
_{R}^{(6)}+\mathcal{L}_{RR^{\prime }}^{(6)}+\mathcal{L}_{RR^{\prime
}R^{\prime \prime }}^{(6)}\right )\big|_{R\rightarrow R^{(2)}}.
\end{eqnarray*}%
The structure of Lagrangians $\mathcal{L}_{GB}^{(n)}$ and $\mathcal{L}_{\chi
,R}^{(n)}$ are identical to $\mathcal{L}_{\chi }^{(n)}$, just the couplings
are different. Then for generic chiral coupling constants $k_{\chi }$ of $%
\mathcal{L}_{\chi PT}$ we may write
\begin{equation}
k_{\chi }=k_{GB}+k_{\chi ,R}\,,
\end{equation}%
where $k_{\chi ,R}$ corresponds to the resonance contribution. The usual
hypothesis of resonance saturation assumes $k_{GB}$ to be very small and the
resonance contribution $k_{\chi ,R}$ is expected to be dominant.

The above resonance saturation strategy and the construction of all relevant
operators were studied already in the past. In \cite{Ecker:1988te} was found
the basis for all relevant resonance operators contributing to $\mathcal{O}%
(p^{4})$ and their contribution to LECs while in \cite{Cirigliano:2006hb}
the authors presented the extension to $\mathcal{O}(p^{6})$ in
even-intrinsic-parity sector.

In this paper, we complete this effort presenting the construction of basis
and resonance saturation at $\mathcal{O}(p^6)$ in the odd-intrinsic parity
sector.

\section{Lagrangian of R$\chi$T in odd-intrinsic parity sector\label{sec3}}

Before starting the construction of resonance monomials let us summarize the
structure of the pure Goldstone-boson part of the odd-intrinsic sector. The
leading order starts at $O(p^{4})$ and the parameters are set entirely by
the chiral anomaly. The Lagrangian is given by \cite{Witten:1983tw} (see
also \cite{Bijnens:2001bb}; note we have the same convention for the
Levi-Civita symbol, i.e. $\epsilon_{0123}=1$):
\begin{equation}
\mathcal{L}_{WZW}=\frac{N_{C}}{48\pi ^{2}}\epsilon ^{\mu \nu \alpha \beta }%
\Bigl\{\int_{0}^{1}d\xi \langle \sigma _{\mu }^{\xi }\sigma _{\nu }^{\xi
}\sigma _{\alpha }^{\xi }\sigma _{\beta }^{\xi }\,\frac{\phi }{F}\rangle -%
\mathrm{i}\langle W_{\mu \nu \alpha \beta }(U,l,r)-W_{\mu \nu \alpha \beta
}(1,l,r)\rangle \Bigr\}\,,
\end{equation}%
with
\begin{multline*}
W_{\mu \nu \alpha \beta }=L_{\mu }L_{\nu }L_{\alpha }R_{\beta }+\tfrac{1}{4}%
L_{\mu }R_{\nu }L_{\alpha }R_{\beta }+\mathrm{i}L_{\mu \nu }L_{\alpha
}R_{\beta }+\mathrm{i}R_{\mu \nu }L_{\alpha }R_{\beta }-\mathrm{i}\sigma
_{\mu }L_{\nu }R_{\alpha }L_{\beta }+\sigma _{\mu }R_{\nu \alpha }L_{\beta }
\\
-\sigma _{\mu }\sigma _{\nu }R_{\alpha }L_{\beta }+\sigma _{\mu }L_{\nu
}L_{\alpha \beta }+\sigma _{\mu }L_{\nu \alpha }L_{\beta }-i\sigma _{\mu
}L_{\nu }L_{\alpha }L_{\beta }+\tfrac{1}{2}\sigma _{\mu }L_{\nu }\sigma
_{\alpha }L_{\beta }-i\sigma _{\mu }\sigma _{\nu }\sigma _{\alpha }L_{\beta
}-(L\leftrightarrow R)\,,
\end{multline*}%
where we have defined
\begin{equation*}
L_{\mu }=u\,l_{\mu }u^{\dagger }\,,\quad L_{\mu \nu }=u\,\partial _{\mu
}l_{\nu }u^{\dagger }\,,\quad R_{\mu }=u^{\dagger }r_{\mu }u\,,\quad R_{\mu
\nu }=u\,\partial _{\mu }r_{\nu }u^{\dagger }\,,\quad \sigma _{\mu
}=\{u^{\dagger },\,\partial _{\mu }u\}
\end{equation*}%
and $(L\leftrightarrow R)$ stands also for $\sigma \leftrightarrow \sigma
^{\dagger }$ interchange. The power $\xi$ indicates a change of $u$ to $%
u^{\xi}=\exp (i\xi \phi /(F\sqrt{2}))$. Concerning the $O(p^6)$ part we will
stick to the form introduced in~\cite{Bijnens:2001bb}. Let us only note that
we will drop the index ${}^r$ and the explicit dependence on the
renormalization scale $\mu$ from the corresponding LECs $C_i^W$, but one
should have in mind that any $C_i^W$ studied in this text is a renormalized
LEC with the scale set to some reasonable value ($\sim M_\rho,
M_{\eta^{\prime }}$) to make good sense of the following study.

For the construction of the operator basis in the odd intrinsic parity
sector of R$\chi$T we use the same tools as in \cite{Cirigliano:2006hb},
where the reader can find further details. First we construct all possible
operators built from chiral building blocks and resonance fields that are
invariant under all symmetries. Then in order to find the independent basis
we use

\begin{enumerate}
\item Partial integration

\item Equation of motion
\begin{equation}
\nabla^\mu u_\mu = \frac{\mathrm{i}}{2}\left(\chi_- - \frac{1}{N_F}\langle
\chi_-\rangle\right)
\end{equation}

\item Bianchi identities
\begin{equation}
\nabla_\mu \Gamma_{\nu\rho} + \nabla_\nu \Gamma_{\rho \mu} + \nabla_\rho
\Gamma_{\mu \nu} = 0\qquad\qquad \mbox{for}\qquad
\Gamma_{\mu\nu}=\frac14[u_\mu,u_\nu]-\frac{i}{2}f_{+\mu\nu}
\end{equation}

\item Shouten identity \cite{schouten}
\begin{equation}
g_{\sigma\rho}\epsilon_{\alpha\beta\mu\nu}+g_{\sigma\alpha}\epsilon_{\beta%
\mu\nu\rho}
+g_{\sigma\beta}\epsilon_{\mu\nu\rho\alpha}+g_{\sigma\mu}\epsilon_{\nu\rho%
\alpha\beta} +g_{\sigma\nu}\epsilon_{\rho\alpha\beta\mu}=0
\end{equation}

\item Identity
\begin{equation}
\nabla^\mu h_{\mu\nu} = \nabla_\nu h^\mu_\mu - 2\left[u^\mu,
i\,\Gamma_{\mu\nu}\right] - \nabla^\mu f_{-\mu\nu}
\end{equation}
\end{enumerate}

All relevant operators in odd parity sector can be written in the form
\begin{equation}
\mathcal{O}_{i}^{X}=\varepsilon ^{\mu \nu \alpha \beta }\widehat{\mathcal{O}}%
_{i\,\mu \nu \alpha \beta }^{X}\,,
\end{equation}%
with the basis for individual monomials $\widehat{\mathcal{O}}_{i\,\mu \nu
\alpha \beta }^{X}$, with $X=V$, $A$, $P$, $S$, $VV$, $AA$, $SA$, $SV$, $VA$%
, $PA$, $PV$, $VVP$, $VAS$, $AAP$; so Lagrangian becomes:
\begin{equation}
\mathcal{L}_{R\chi T}^{(6,\text{ odd})}=\sum_{X}\sum_{i}\kappa _{i}^{X}%
\mathcal{O}_{i}^{X}\,.  \label{reslag}
\end{equation}%
The basis of the operators $\widehat{\mathcal{O}}_{i\,\mu \nu \alpha \beta
}^{X}$ is summarized in Tables~\ref{tab:prvni}-\ref{tab:posledni}. We have
included there only the operators \ relevant in the leading order in the $%
1/N_{C}$ expansion \emph{\ i.e.} operators with one flavour trace and those
with two traces that are enhanced by $\eta ^{\prime }$ exchange (see
Appendix~\ref{apA} for details). This represents main result of our work.

As is shown in \cite{Cirigliano:2006hb, {Kampf:2006yf}}, we can
further modify the resonance Lagrangian (\ref{reslag}). The reason
is that the resonance fields play merely the role of the
integration variables in the path integral (\ref{PI}) and can be
therefore freely redefined without changing the physical content
of the theory. As a consequence we can choose
appropriate field redefinition in order to eliminate some subset $\{\mathcal{%
O}_{i}^{X}\}_{(X,i)\in M}$ of $O(p^{6})$ operators from
$\mathcal{L}_{R\chi T}^{(6,\text{ odd})}$ and shift their
influence effectively to the $O(p^{6})$
terms including the remaining operators $\{\mathcal{O}_{i}^{X}\}_{(X,i)%
\notin M}$ and also to the higher chiral order terms
$\mathcal{L}_{R\chi
T}^{(>6,\text{ odd})}$, symbolically%
\begin{equation}
\mathcal{L}_{R\chi T}^{(6,\text{ odd})}=\sum_{(X,i)}\kappa _{i}^{X}\mathcal{O%
}_{i}^{X}\,\rightarrow \sum_{(X,i)\notin M}\overline{\kappa _{i}^{X}}%
\mathcal{O}_{i}^{X}+\mathcal{L}_{R\chi T}^{(>6,\text{ odd})}
\end{equation}
The possible new terms $\mathcal{L}_{R\chi T}^{(>6,\text{ odd})}$
of the \ order $O(p^{8})$ and higher generated by such a
redefinition can be neglected because they do not contribute to
the $O(p^{6})$ LECs when the resonance fields are integrated out.
Note however, that after such a
truncation we get new Lagrangian%
\begin{equation}
\overline{\mathcal{L}_{R\chi T}^{(6,\text{ odd})}}=\sum_{(X,i)\notin M}%
\overline{\kappa _{i}^{X}}\mathcal{O}_{i}^{X}
\end{equation}
which is not equivalent with the previous one on the resonance
level. On the
other hand, the LECs obtained form $\overline{\mathcal{L}_{R\chi T}^{(6,%
\text{ odd})}}$ coincide with those derived form $\mathcal{L}_{R\chi T}^{(6,%
\text{ odd})}$.

The stars in the Tables~\ref{tab:prvni}-\ref{tab:posledni}
indicate those operators which can be eliminated by the resonance
fields redefinitions discussed above and means therefore a
redundance of a given monomial as far as its contribution to the
resonance saturation is concerned. The details are shown in
Appendix~\ref{apB}. Note, however, that this redundance concerns
only the saturation and not the direct calculation of the
correlators and amplitudes with resonances in the initial and
final states. We will return to this point later on.

In the following section we will demonstrate the use of the resonance basis
for two classes of examples. The resonance saturation will be studied in
Section~\ref{sec:reslec}.

\begin{table}[tbp]
\begin{center}
\begin{tabular}{|c|c||c|c|}
\hline
$i$ & $\widehat{\mathcal{O}}^V_{i\,\mu\nu\alpha\beta}$ & $i$ & $\widehat{%
\mathcal{O}}^V_{i\,\mu\nu\alpha\beta}$ \\ \hline
1 & $\mathrm{i} {\langle} V^{\mu\nu}(h^{\alpha\sigma}u_\sigma u^\beta -
u^\beta u_\sigma h^{\alpha\sigma}){\rangle}$ & 11 & ${\langle}
V^{\mu\nu}\{f_+^{\alpha\rho},f_-^{\beta\sigma}\}{\rangle} g_{\rho\sigma}$ \\
\hline
2 & $\mathrm{i}{\langle} V^{\mu\nu}(u_\sigma h^{\alpha\sigma} u^\beta -
u^\beta h^{\alpha\sigma} u_\sigma)$ & 12 & ${\langle} V^{\mu\nu}
\{f_+^{\alpha\rho},h^{\beta\sigma}\}{\rangle} g_{\rho\sigma}$ \\ \hline
3 & $\mathrm{i}{\langle} V^{\mu\nu}(u_\sigma u^\beta h^{\alpha\sigma} -
h^{\alpha\sigma}u^\beta u_\sigma){\rangle}$ & 13 & $\mathrm{i}{\langle}
V^{\mu\nu}f_+^{\alpha\beta}{\rangle}{\langle} \chi_-{\rangle}$ \\ \hline
4 & $\mathrm{i}{\langle} [V^{\mu\nu},\nabla^\alpha\chi_+]u^\beta{\rangle} $
& 14 & $\mathrm{i} {\langle} V^{\mu\nu}\{f_+^{\alpha\beta},\chi_-\}{\rangle}$
\\ \hline
5 & $\mathrm{i}{\langle} V^{\mu\nu}[f_-^{\alpha\beta},u_\sigma u^\sigma]{%
\rangle}$ & 15 & $\mathrm{i} {\langle} V^{\mu\nu}[f_-^{\alpha\beta},\chi_+]{%
\rangle}$ \\ \hline
6 & $\mathrm{i}{\langle} V^{\mu\nu}(f_-^{\alpha\sigma}u^\beta u_\sigma -
u_\sigma u^\beta f_-^{\alpha\sigma}){\rangle}$ & 16 & ${\langle}
V^{\mu\nu}\{\nabla^\alpha f_+^{\beta\sigma},u_\sigma\}{\rangle}$ \\ \hline
7 & $\mathrm{i}{\langle} V^{\mu\nu}(u_\sigma f_-^{\alpha\sigma}u^\beta -
u^\beta f_-^{\alpha\sigma}u_\sigma){\rangle}$ & 17 & ${\langle}
V^{\mu\nu}\{\nabla_\sigma f_+^{\alpha\sigma},u^\beta\}{\rangle}$ \\ \hline
8 & $\mathrm{i} {\langle} V^{\mu\nu}(f_-^{\alpha\sigma} u_\sigma u^\beta -
u^\beta u_\sigma f_-^{\alpha\sigma}){\rangle} $ & 18 & ${\langle} V^{\mu\nu}
u^\alpha u^\beta {\rangle} {\langle} \chi_-{\rangle}$ \\ \hline
9 & ${\langle} V^{\mu\nu}\{\chi_-,u^\alpha u^\beta\}{\rangle} $ &  &  \\
\hline
10 & ${\langle} V^{\mu\nu} u^\alpha \chi_- u^\beta {\rangle}$ &  &  \\ \hline
\end{tabular}%
\end{center}
\caption{Monomials with one vector resonance field.}
\label{tab:prvni}
\end{table}

\begin{table}[tbp]
\begin{center}
\begin{tabular}{|c|c||c|c|}
\hline
$i$ & $\widehat{\mathcal{O}}^A_{i\,\mu\nu\alpha\beta}$ & $i $ & $\widehat{%
\mathcal{O}}^A_{i\,\mu\nu\alpha\beta}$ \\ \hline
1 & ${\langle} A^{\mu\nu}[u^\alpha u^\beta, u_\sigma u^\sigma]{\rangle}$ & 10
& $\mathrm{i}{\langle} A^{\mu\nu} u^\alpha{\rangle} {\langle} \nabla^\beta
\chi_- {\rangle}$ \\ \hline
2 & ${\langle} A^{\mu\nu}[u^\alpha u^\sigma u^\beta, u^\sigma]{\rangle}$ & 11
& $\mathrm{i}{\langle} A^{\mu\nu}\{f_-^{\alpha\beta},\chi_-\}{\rangle}$ \\
\hline
3 & ${\langle} A^{\mu\nu}\{\nabla^\alpha h^{\beta\sigma},u_\sigma\}{\rangle}$
& 12 & $\mathrm{i}{\langle} A^{\mu\nu}\{\nabla^\alpha \chi_-,u^\beta\}{%
\rangle}$ \\ \hline
4 & $\mathrm{i}{\langle} A^{\mu\nu}[f_+^{\alpha\beta},u^\sigma u_\sigma]{%
\rangle}$ & 13 & ${\langle} A^{\mu\nu}[\chi_+,u^\alpha u^\beta]{\rangle}$ \\
\hline
5 & $\mathrm{i} {\langle} A^{\mu\nu}(f_+^{\alpha\sigma}u_\sigma u^\beta -
u^\beta u_\sigma f_+^{\alpha\sigma}{\rangle}$ & 14 & $\mathrm{i}{\langle}
A^{\mu\nu}[f_+^{\alpha\beta},\chi_+]{\rangle}$ \\ \hline
6 & $\mathrm{i}{\langle} A^{\mu\nu}(f_+^{\alpha\sigma}u^\beta u_\sigma -
u_\sigma u^\beta f_+^{\alpha\sigma}{\rangle}$ & 15 & ${\langle}
A^{\mu\nu}\{\nabla^\alpha f_-^{\beta\sigma},u_\sigma\}{\rangle}$ \\ \hline
7 & $\mathrm{i}{\langle} A^{\mu\nu}(u_\sigma f_+^{\alpha\sigma}u^\beta -
u^\beta f_+^{\alpha\sigma}u_\sigma{\rangle}$ & 16 & ${\langle}
A^{\mu\nu}\{\nabla_\sigma f_-^{\alpha\sigma},u^\beta\}{\rangle}$ \\ \hline
8 & ${\langle} A^{\mu\nu}\{f_-^{\alpha\sigma},h^{\beta\sigma}\}{\rangle}$ &
&  \\ \hline
9 & $\mathrm{i}{\langle} A^{\mu\nu}f_-^{\alpha\beta}{\rangle} {\langle}
\chi_-{\rangle}$ &  &  \\ \hline
\end{tabular}%
\end{center}
\caption{Monomials with one axial-vector resonance field.}
\end{table}

\begin{table}[tbp]
\begin{center}
\begin{tabular}{|c|c||c|c||c|c|}
\hline
$i$ & $\widehat{\mathcal{O}}^{P}_{i\,\mu\nu\alpha\beta}$ & $i$ & $\widehat{%
\mathcal{O}}^{P}_{i\,\mu\nu\alpha\beta}$ & $i$ & $\widehat{\mathcal{O}}%
^{S}_{i\,\mu\nu\alpha\beta}$ \\ \hline\hline
1 & ${\langle} P \{f_-^{\mu\nu},f_-^{\alpha\beta}\}{\rangle}$ & 4 & ${\langle%
} P u^\mu u^\nu u^\alpha u^\beta {\rangle}$ & 1 & ${\langle} S
[f_-^{\alpha\beta},u^\mu u^\nu] {\rangle}$ \\ \hline
2 & $\mathrm{i}{\langle} P u^\alpha f_+^{\mu\nu} u^\beta{\rangle}$ & 5 & ${%
\langle} P \{f_+^{\mu\nu},f_+^{\alpha\beta}\}{\rangle}$ & 2 & $\mathrm{i}{%
\langle} S [f_+^{\mu\nu},f_-^{\alpha\beta}]{\rangle}$ \\ \hline
3 & $\mathrm{i}{\langle} P\{f_+^{\mu\nu},u^\alpha u^\beta\}{\rangle} $ & 6 &
&  &  \\ \hline
\end{tabular}%
\end{center}
\caption{Monomials with scalar or pseudo-scalar resonance field.}
\end{table}

\begin{table}[tbp]
\begin{center}
\begin{tabular}{|c|c||c|}
\hline
$i$ & Operator $\widehat{\mathcal{O}}^{RR}_{i\,\mu\nu\alpha\beta}$, $R=V,A$
& Operator $\widehat{\mathcal{O}}^{RR}_{i\,\mu\nu\alpha\beta}$, $R=P,S$ \\
\hline\hline
\phantom{$^{*}\!\!\!$}1$^{*}\!\!\!$ & $\mathrm{i}\langle
R^{\mu\nu}R^{\alpha\beta}\rangle \langle \chi_- \rangle $ &  \\ \hline
\phantom{$^{*}\!\!\!$}2$^{*}\!\!\!$ & $\mathrm{i}\langle
\{R^{\mu\nu},R^{\alpha\beta}\}\chi_-\rangle$ &  \\ \hline
3 & $\langle \{\nabla_\sigma R^{\mu\nu},R^{\alpha\sigma}\} u^\beta \rangle$
&  \\ \hline
4 & $\langle \{\nabla^\beta R^{\mu\nu},R^{\alpha\sigma}\}u_\sigma \rangle$ &
\\ \hline
\end{tabular}%
\end{center}
\caption{Monomials with two resonance fields of the same kind.}
\end{table}

\begin{table}[tbp]
\begin{center}
\begin{tabular}{|c|c||c|c||c|}
\hline
$i$ & Operator $\widehat{\mathcal{O}}^{SA}_{i\,\mu\nu\alpha\beta}$ & $i$ &
Operator $\widehat{\mathcal{O}}^{SV}_{i\,\mu\nu\alpha\beta}$ & Operator $%
\widehat{\mathcal{O}}^{SP}_{i\,\mu\nu\alpha\beta}$ \\ \hline\hline
\phantom{$^{*}\!\!\!$}1$^{*}\!\!\!$ & $\mathrm{i} {\langle}
[A^{\mu\nu},S]f_+^{\alpha\beta}{\rangle}$ & \phantom{$^{*}\!\!\!$}1$%
^{*}\!\!\!$ & $\mathrm{i}{\langle} [V^{\mu\nu},S]f_-^{\alpha\beta}{\rangle}$
&  \\ \hline
\phantom{$^{*}\!\!\!$}2$^{*}\!\!\!$ & ${\langle} A^{\mu\nu}[S,u^\alpha
u^\beta]{\rangle}$ & 2 & $\mathrm{i} {\langle} [V^{\mu\nu},\nabla^\alpha
S]u^\beta{\rangle}$ &  \\ \hline
\end{tabular}%
\end{center}
\caption{Monomials with two resonance fields of different kinds.}
\end{table}

\begin{table}[tbp]
\begin{center}
\begin{tabular}{|c|c||c|c||c|c|}
\hline
$i$ & Operator $\widehat{\mathcal{O}}^{VA}_{i\,\mu\nu\alpha\beta}$ & $i$ &
Operator $\widehat{\mathcal{O}}^{PA}_{i\,\mu\nu\alpha\beta}$ & $i$ &
Operator $\widehat{\mathcal{O}}^{PV}_{i\,\mu\nu\alpha\beta}$ \\ \hline\hline
\phantom{$^{*}\!\!\!$}1$^{*}\!\!\!$ & $\mathrm{i}{\langle}
V^{\mu\nu}[A^{\alpha\beta},u^\sigma u_\sigma]{\rangle}$ & %
\phantom{$^{*}\!\!\!$}1$^{*}\!\!\!$ & ${\langle} \{A^{\mu\nu},P\}f_-^{\alpha%
\beta}{\rangle}$ & \phantom{$^{*}\!\!\!$}1$^{*}\!\!\!$ & $\mathrm{i} {\langle%
} \{V^{\mu\nu},P\}u^\alpha u^\beta{\rangle} $ \\ \hline
\phantom{$^{*}\!\!\!$}2$^{*}\!\!\!$ & $\mathrm{i}{\langle}
V^{\mu\nu}(A^{\alpha\sigma}u_\sigma u^\beta - u^\beta u_\sigma
A^{\alpha\sigma}){\rangle}$ & 2 & ${\langle} \{A^{\mu\nu},\nabla^\alpha
P\}u^\beta{\rangle}$ & \phantom{$^{*}\!\!\!$}2$^{*}\!\!\!$ & $\mathrm{i}{%
\langle} V^{\mu\nu} u^\alpha P u^\beta{\rangle}$ \\ \hline
\phantom{$^{*}\!\!\!$}3$^{*}\!\!\!$ & $\mathrm{i}{\langle}
V^{\mu\nu}(A^{\alpha\sigma}u^\beta u_\sigma - u_\sigma u^\beta
A^{\alpha\sigma})$ &  &  & \phantom{$^{*}\!\!\!$}3$^{*}\!\!\!$ & ${\langle}
\{V^{\mu\nu},P\} f_+^{\alpha\beta}{\rangle}$ \\ \hline
\phantom{$^{*}\!\!\!$}4$^{*}\!\!\!$ & $\mathrm{i}{\langle}
V^{\mu\nu}(u_\sigma A^{\alpha\sigma}u^\beta - u^\beta
A^{\alpha\sigma}u_\sigma){\rangle}$ &  &  &  &  \\ \hline
\phantom{$^{*}\!\!\!$}5$^{*}\!\!\!$ & ${\langle} \{V^{\mu\nu},A^{\alpha\rho}%
\}f_+^{\beta\sigma}{\rangle} g_{\rho\sigma}$ &  &  &  &  \\ \hline
\phantom{$^{*}\!\!\!$}6$^{*}\!\!\!$ & $\mathrm{i}{\langle}
[V^{\mu\nu},A^{\alpha\beta}]\chi_+{\rangle}$ &  &  &  &  \\ \hline
\end{tabular}%
\end{center}
\caption{Monomials with two resonance fields of different kinds.}
\end{table}

\begin{table}[tbp]
\begin{center}
\begin{tabular}{|c||c|}
\hline
$i$ & Operator $\widehat{\mathcal{O}}^{RRR}_{i\,\mu\nu\alpha\beta}$ \\
\hline\hline
\phantom{$^{*}\!\!\!$}1$^{*}\!\!\!$ & $\langle V^{\mu\nu} V^{\alpha\beta} P
\rangle$ \\ \hline
\phantom{$^{*}\!\!\!$}2$^{*}\!\!\!$ & $\mathrm{i}\langle
[V^{\mu\nu},A^{\alpha\beta}]S\rangle$ \\ \hline
\phantom{$^{*}\!\!\!$}3$^{*}\!\!\!$ & $\langle A^{\mu\nu}A^{\alpha\beta} P
\rangle$ \\ \hline
\end{tabular}%
\end{center}
\caption{Monomials with three resonance fields.}
\label{tab:posledni}
\end{table}

\section{Applications}

In this section we illustrate applications of the Lagrangian $\mathcal{L}%
_{R\chi T}^{(6,\text{ odd})}$ using two examples. We study two three-point
correlators, namely ${\langle }VVP{\rangle }$ and ${\langle }VAS{\rangle }$,
and use both OPE constraints as well as phenomenological inputs to fix the
relevant coupling constants. In the first case we also discuss some
phenomenological applications in more detail.

\subsection{$VVP$ Green function revisited}

The standard definition of this correlator is
\begin{equation}
\Pi_{\mu\nu}^{abc}(p,q) = \int d^4x\, d^4y\, e^{ip\cdot x+iq\cdot y} \langle
0| T[V^a_\mu(x)V^b_\nu(y)P^c(0)|0\rangle
\end{equation}
with the vector current and the pseudoscalar density defined by
\begin{equation}
V_\mu^a(x) =\bar q(x) \gamma_\mu \frac{\lambda^a}{2}q(x)\,,\qquad P_\mu^a(x)
=\bar q(x) \mathrm{i}\gamma_5 \frac{\lambda^a}{2}q(x)\,,  \label{VPcur}
\end{equation}
(our convention is $\gamma_5 = i \gamma^0\gamma^1\gamma^2\gamma^3$). This
correlator was already studied in the past, see e.g. \cite{Moussallam:1994xp}%
, \cite{Knecht:2001xc}, \cite{RuizFemenia:2003hm}. Here we provide a
complete result based on our $\mathcal{L}_\text{R$\chi$T}$, i.e. also with
two- and three-resonance vertices that were not considered in \cite%
{RuizFemenia:2003hm}. Using Ward identities and Lorentz and parity
invariance one can define
\begin{equation}
\Pi (p)_{\mu \nu }^{abc}=d^{abc}\epsilon _{\mu \nu \alpha \beta }p^{\alpha
}q^{\beta }\Pi (p^{2},q^{2};r^{2})\,.
\end{equation}%
The OPE constraints dictate for high values of all independent momenta
\begin{equation}
\Pi ((\lambda p)^{2},(\lambda q)^{2};(\lambda r)^{2})=\frac{B_{0}F^{2}}{%
2\lambda ^{4}}\frac{p^{2}+q^{2}+r^{2}}{p^{2}q^{2}r^{2}}+\mathcal{O}\left(
\frac{1}{\lambda ^{6}}\right) \,,  \label{ope}
\end{equation}%
whereas in the case when only two operators are close to each other one gets
\begin{align}
& \Pi ((\lambda p)^{2},(q-\lambda p)^{2};q^{2})=\frac{B_{0}F^{2}}{\lambda
^{2}}\frac{1}{p^{2}q^{2}}+\mathcal{O}\left( \frac{1}{\lambda ^{3}}\right) \,,
\label{ope1} \\
& \Pi ((\lambda p)^{2},q^{2};(q+\lambda p)^{2})=\frac{1}{\lambda ^{2}}\frac{1%
}{p^{2}}\Pi _{VT}(q^{2})+\mathcal{O}\left( \frac{1}{\lambda ^{3}}\right) \,.
\label{ope2}
\end{align}%
In the following we will use only two-momentum OPE. The reason is that for
general correlator, not all the high energy constraints can be
simultaneously satisfied using only finite number of resonances in the
effective Lagrangian. This statement has been proved in \cite{Bijnens:2003rc}
for the case of the $\langle PPS\rangle $ three-point function. For the ${%
\langle }VVP{\rangle }$ this problem has been partially studied in \cite%
{Knecht:2001xc} (see also \cite{Nyffeler}).

\begin{figure}[tbh]
\begin{center}
\epsfig{width=0.8\textwidth,figure=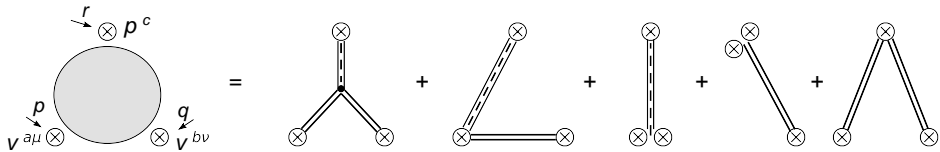}
\end{center}
\caption{Feynman graphs contributed to $VVP$ Green function. Double lines
stand for resonances and dash lines for GB (double lines together with dash
lines is the sum of both possible contributions). The crossing is implicitly
assumed.}
\label{fig:VVP}
\end{figure}

By means of an explicit calculation based on $\mathcal{L}_{R\chi T}^{(6,%
\text{ odd})}$ (the relevant Feynman graphs are depicted in Fig.\ref{fig:VVP}%
) we get
\begin{align}
&\frac{1}{B_{0}}\Pi ^{\text{R}\chi \text{T}}(p^{2},q^{2},r^{2})=  \notag \\
& =-\frac{N_{C}}{16\pi ^{2}r^{2}}+\frac{4F_{V}^{2}\kappa _{3}^{VV}p^{2}}{%
r^{2}(p^{2}-M_{V}^{2})(q^{2}-M_{V}^{2})}-\frac{16\sqrt{2}d_{m}F_{V}\kappa
_{3}^{PV}}{(p^{2}-M_{V}^{2})(r^{2}-M_{P}^{2})}-\frac{32d_{m}\kappa _{5}^{P}}{%
r^{2}-M_{P}^{2}}  \notag \\
& -\frac{8d_{m}F_{V}^{2}\kappa ^{VVP}}{%
(p^{2}-M_{V}^{2})(q^{2}-M_{V}^{2})(r^{2}-M_{P}^{2})}+\frac{2F_{V}^{2}}{%
(p^{2}-M_{V}^{2})(q^{2}-M_{V}^{2})}\left[ 8\kappa _{2}^{VV}-\kappa _{3}^{VV}%
\right]  \notag \\
& -\frac{2\sqrt{2}F_{V}}{r^{2}(p^{2}-M_{V}^{2})}\left[ p^{2}(\kappa
_{16}^{V}+2\kappa _{12}^{V})-q^{2}(\kappa _{16}^{V}-2\kappa
_{17}^{V}+2\kappa _{12}^{V})-r^{2}(8\kappa _{14}^{V}+\kappa
_{16}^{V}+2\kappa _{12}^{V})\right] \,  \notag \\
&+(p \leftrightarrow q).  \label{Pirchit}
\end{align}%
From OPE~(\ref{ope}) we get then the following constraints for the couplings
\begin{align}
& \kappa _{14}^{V}=\frac{N_{C}}{256\sqrt{2}\pi ^{2}F_{V}}\,,\qquad \kappa
_{16}^{V}+2\kappa _{12}^{V}=-\frac{N_{C}}{32\sqrt{2}\pi ^{2}F_{V}}\,,\qquad
\kappa _{17}^{V}=-\frac{N_{C}}{64\sqrt{2}\pi ^{2}F_{V}}\,,\qquad \kappa
_{5}^{P}=0\,,  \notag \\
& \kappa _{2}^{VV}=\frac{F^{2}+16\sqrt{2}d_{m}F_{V}\kappa _{3}^{PV}}{%
32F_{V}^{2}}-\frac{N_{C}M_{V}^{2}}{512\pi ^{2}F_{V}^{2}}\,,\qquad 8\kappa
_{2}^{VV}-\kappa _{3}^{VV}=\frac{F^{2}}{8F_{V}^{2}}\,.  \label{opecon}
\end{align}%
By employing these constraints one gets\footnote{%
Note that these constraints imply automatically also the fulfilment of (\ref%
{ope2}). However, the requirement (\ref{ope1}) cannot be satisfied until $%
\kappa_3^{PV}=0$, which is in contradiction with another high-energy
constraint for related pion transition form factor; see next subsection (cf.
also \cite{Knecht:2001xc}).}
\begin{align}
\frac{1}{B_{0}}\Pi ^{\text{R}\chi \text{T}}(p^{2},q^{2};r^{2})=& \phantom{+}%
\frac{F^{2}}{2}\frac{(p^{2}+q^{2}+r^{2})-\frac{N_{C}M_{V}^{4}}{4\pi ^{2}F^{2}%
}}{(p^{2}-M_{V}^{2})(q^{2}-M_{V}^{2})r^{2}}-\frac{16d_{m}F_{V}^{2}\kappa
^{VVP}}{(r^{2}-M_{P}^{2})(p^{2}-M_{V}^{2})(q^{2}-M_{V}^{2})}  \notag
\label{PcLMD} \\
& -\frac{16\sqrt{2}d_{m}F_{V}\kappa _{3}^{PV}\left[
(p^{2}+q^{2})M_{P}^{2}-2r^{2}M_{V}^{2}\right] }{%
r^{2}(r^{2}-M_{P}^{2})(q^{2}-M_{V}^{2})(p^{2}-M_{V}^{2})}\,.
\end{align}%
This should be equivalent with the LMD+P ansatz introduced in~\cite%
{Moussallam:1994xp} so that two independent constants $b$ and $c$ introduced
there are directly connected with phenomenological couplings $\kappa
_{3}^{PV}$ and $\kappa ^{VVP}$. Considering just vector resonance
interactions, $\kappa ^{VVP}=\kappa _{3}^{PV}=0$ (or equivalently taking the
limit $M_{P}\rightarrow \infty$ in~(\ref{Pirchit})), we can reconstruct the
LMD ansatz \cite{Knecht:2001xc}
\begin{equation}
\frac{1}{B_{0}}\Pi ^{\text{LMD}}(p^{2},q^{2};r^{2})=\frac{F^{2}}{2}\cdot
\frac{(p^{2}+q^{2}+r^{2})-\frac{N_{C}M_{V}^{4}}{4\pi ^{2}F^{2}}}{%
(p^{2}-M_{V}^{2})(q^{2}-M_{V}^{2})r^{2}}\,.
\end{equation}%
The result in ChPT up to $\mathcal{O}(p^{6})$ at the leading order in $%
1/N_{C}$ expansion includes two LECs from the $O(p^6)$ anomalous sector
\begin{equation}
\frac{1}{B_{0}}\Pi ^{\text{ChPT}}(p^{2},q^{2};r^{2})=-\frac{N_{C}}{8\pi
^{2}r^{2}}+32C_{7}^{W}-\frac{8C_{22}^{W}(p^{2}+q^{2})}{r^{2}}\,.
\label{Pichpt}
\end{equation}%
Comparing this with a low energy expansion of the $R\chi T$ result (\ref%
{Pirchit}) we give the following lowest-resonance contribution to $C_{7}^{W}$
and $C_{22}^{W}$ (cf. also Section~\ref{sec:reslec})
\begin{align}
C_{7}^{W} &=\frac{F_{V}^{2}(8\kappa _{2}^{VV}-\kappa _{3}^{VV})}{8M_{V}^{4}}+%
\frac{d_{m}F_{V}^{2}\kappa ^{VVP}}{2M_{P}^{2}M_{V}^{4}}-\frac{\sqrt{2}%
d_{m}F_{V}\kappa _{3}^{PV}}{M_{P}^{2}M_{V}^{2}}+\frac{2d_{m}\kappa _{5}^{P}}{%
M_{P}^{2}}-\frac{F_{V}(2\kappa _{12}^{V}+8\kappa _{14}^{V}+\kappa _{16}^{V})%
}{4\sqrt{2}M_{V}^{2}},  \notag \\
C_{22}^{W} &=-\frac{F_{V}\kappa _{17}^{V}}{\sqrt{2}M_{V}^{2}}-\frac{%
F_{V}^{2}\kappa _{3}^{VV}}{2M_{V}^{4}}\,.  \label{C7C22}
\end{align}%
Using the OPE constraints (\ref{opecon}) we obtain
\begin{align}
C_{7}^{W} &=\frac{F^{2}}{64M_{V}^{4}}+\frac{d_{m}F_{V}(-2\sqrt{2}\kappa
_{3}^{PV}M_{V}^{2}+F_{V}\kappa ^{VVP})}{2M_{P}^{2}M_{V}^{4}}\,,  \notag \\
C_{22}^{W} &=-\frac{F^{2}}{16M_{V}^{4}}+\frac{N_{C}}{64\pi ^{2}M_{V}^{2}}-%
\frac{2\sqrt{2}d_{m}F_{V}\kappa _{3}^{PV}}{M_{V}^{4}}\,.  \label{C7C22ope}
\end{align}

\subsubsection{Formfactors}

\begin{figure}[t]
\begin{center}
\epsfig{width=0.8\textwidth,figure=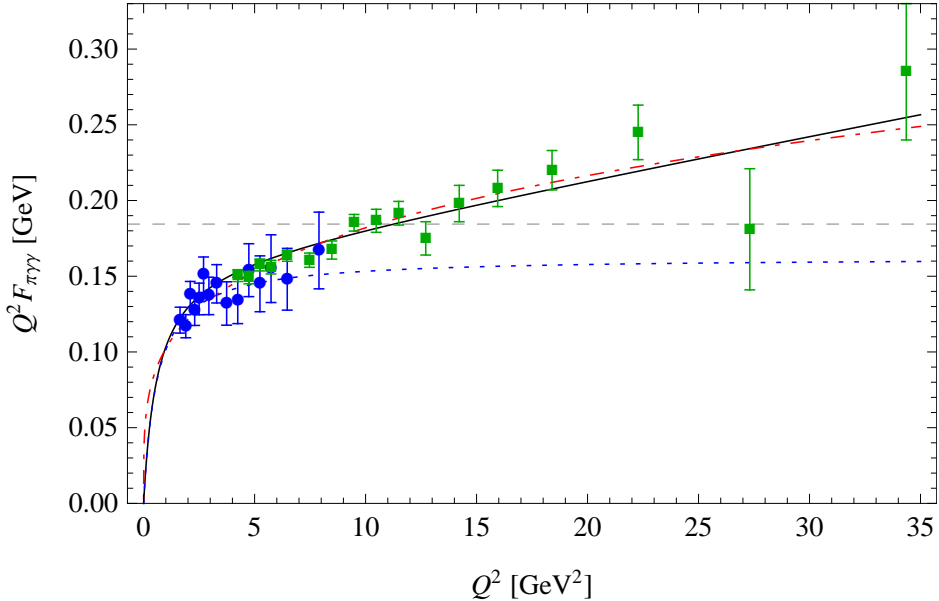}
\end{center}
\caption{CLEO (blue points) and BABAR (green squares) data with fitted function
$\mathcal{F}_{\pi^{0}\gamma \gamma }^{\text{R$\chi $T}}(0,-Q^{2};0)$
defined in (\ref{FcLMD}) using the modified Brodsky-Lepage condition
in~(\ref{mBL}). The full line is for $\delta _{\text{BL}}\
=-0.055$ and (blue) dotted line stands for standard B-L (i.e. $\delta_{\text{BL}}\ =0$). Dot-dashed (red) line shows fitted function $A(Q^{2}/(10\,%
\text{GeV}^{2}))^{\beta }$ with $A=0.182\pm 0.002\,\text{GeV}$ and $%
\beta =0.25\pm 0.02$ as obtained by BABAR collaboration~\cite%
{Aubert:2009mc}. The asymptotic $2F$ is represented by the horizontal dash
line.}
\label{fig:babar}
\end{figure}

Let us define fully off-shell formfactors for $\mathcal{P}^{\ast }\gamma
^{\ast }\gamma ^{\ast }$ vertex, where $\mathcal{P}$ can represents either
pion (or any other Goldstone boson) or pseudoscalar resonance via
\begin{equation}
\mathcal{F}_{\mathcal{P}\gamma \gamma }(p^{2},q^{2};r^{2})=\frac{1}{\mathcal{%
Z}_{\mathcal{P}}}(r^{2}-m_{\mathcal{P}}^{2})\Pi (p^{2},q^{2};r^{2})\,,
\label{FPgg}
\end{equation}%
where $\mathcal{Z}$ factor interpolates between pseudoscalar source and $%
\mathcal{P}$. Let us discuss in detail the $\pi ^{0}\gamma \gamma $
formfactor. We have $\mathcal{Z}_{\pi ^{0}}=3/2BF$ and using the OPE
constraints (\ref{opecon}) we can define (note we are working in the chiral
limit)
\begin{equation}
\mathcal{F}_{\pi ^{0}\gamma \gamma }^{\text{R}\chi \text{T}%
}(p^{2},q^{2};r^{2})=\frac{2}{3}\frac{1}{BF}r^{2}\Pi ^{\text{R}\chi \text{T}%
}(p^{2},q^{2};r^{2})\,,  \label{Fpigg}
\end{equation}%
where $\Pi ^{\text{R$\chi $T}}(p^{2},q^{2};r^{2})$ was introduced in~(\ref%
{PcLMD}). For on-shell pion the $\kappa ^{VVP}$ drops out (note that this is
not connected with the chiral limit simplification) and we get a simple
result
\begin{equation}
\mathcal{F}_{\pi ^{0}\gamma \gamma }^{\text{R}\chi \text{T}}(p^{2},q^{2};0)=%
\frac{F}{3}\frac{(p^{2}+q^{2})(1+32\sqrt{2}\frac{d_{m}F_{V}}{F^{2}}\kappa
_{3}^{PV})-\frac{N_{C}}{4\pi ^{2}}\frac{M_{V}^{4}}{F^{2}}}{%
(p^{2}-M_{V}^{2})(q^{2}-M_{V}^{2})}\,.  \label{FcLMD}
\end{equation}%
Dropping $\kappa _{3}^{PV}$ we can again reconstruct the LMD ansatz
\begin{equation}
\mathcal{F}_{\pi ^{0}\gamma \gamma }^{\text{R}\chi \text{T}}(p^{2},q^{2};0)%
\Bigr|_{\kappa _{3}^{PV}=0}=\mathcal{F}_{\pi ^{0}\gamma \gamma }^{\text{LMD}%
}(p^{2},q^{2};0)={\frac{F_{\pi }}{3}}\,{\frac{p^{2}+q^{2}-{\frac{N_{C}}{4\pi
^{2}}}{\frac{M_{V}^{4}}{F_{\pi }^{2}}}}{(p^{2}-M_{V}^{2})(q^{2}-M_{V}^{2})}}%
\,.  \label{zmena1}
\end{equation}%
Using Brodsky-Lepage (B-L) behaviour for large momentum \cite{Lepage:1980fj}
\cite{Brodsky:1981rp}
\begin{equation}
\text{B-L cond.:}\qquad \lim_{Q^{2}\rightarrow \infty }\mathcal{F}_{\pi
^{0}\gamma \gamma }(0,-Q^{2};m_{\pi }^{2})\sim -\frac{1}{Q^{2}}\,,
\end{equation}%
one can arrive to the following constraint
\begin{equation}
\text{B-L cond.:}\qquad \kappa _{3}^{PV}=-\frac{F^{2}}{32\sqrt{2}d_{m}F_{V}}%
\,.  \label{BLc}
\end{equation}%
Before discussing the possible violation of the Brodsky-Lepage
condition  let us study the influence of the constraint
(\ref{BLc})on the original VVP Green function. The $\Pi
^{\text{R}\chi \text{T}}$ correlator in~(\ref{PcLMD}) will now
depend only on one constant $\kappa ^{VVP}$ and we get
\begin{align}
\frac{1}{B_{0}}\Pi ^{\text{R}\chi \text{T}}(p^{2},q^{2};r^{2})=& \frac{F^{2}%
}{2}\frac{(p^{2}+q^{2}+r^{2})-\frac{N_{C}M_{V}^{4}}{4\pi ^{2}F^{2}}}{%
(p^{2}-M_{V}^{2})(q^{2}-M_{V}^{2})r^{2}}+\frac{F^{2}}{2}\frac{\left[
(p^{2}+q^{2})M_{P}^{2}-2r^{2}M_{V}^{2}\right] }{%
r^{2}(r^{2}-M_{P}^{2})(q^{2}-M_{V}^{2})(p^{2}-M_{V}^{2})}  \notag \\
& \,-\frac{16d_{m}F_{V}^{2}\kappa ^{VVP}}{%
(r^{2}-M_{P}^{2})(p^{2}-M_{V}^{2})(q^{2}-M_{V}^{2})}.
\end{align}%
The violation of the OPE (\ref{ope1}) is here manifest. The remaining
constant $\kappa ^{VVP}$ will drop out for an on-shell pion in the
formfactor and one gets:
\begin{equation}
\text{B-L cond.:}\qquad \mathcal{F}_{\pi ^{0}\gamma \gamma }^{\text{R}\chi
\text{T}}(p^{2},q^{2};0)=-\frac{N_{C}}{12\pi ^{2}F}\frac{M_{V}^{4}}{%
(p^{2}-M_{V}^{2})(q^{2}-M_{V}^{2})}\,.
\end{equation}%
For completeness let us also provide the ChPT result. From~(\ref{Pichpt})
using~(\ref{C7C22ope}) we have
\begin{equation}
\mathcal{F}_{\pi ^{0}\gamma \gamma }^{\text{ChPT}}(p^{2},q^{2};0)=-\frac{%
N_{C}}{12\pi ^{2}F}\biggl[1+\frac{p^{2}+q^{2}}{M_{V}^{2}}\Bigl(1-\frac{4\pi
^{2}F^{2}}{N_{C}M_{V}^{2}}(1+32\sqrt{2}\frac{d_{m}F_{V}}{F^{2}}\kappa
_{3}^{PV})\Bigr)\biggr]
\end{equation}%
and
\begin{equation}
\text{B-L cond.:}\qquad \mathcal{F}_{\pi ^{0}\gamma \gamma }^{\text{ChPT}%
}(p^{2},q^{2};0)=-\frac{N_{C}}{12\pi ^{2}F}\biggl(1+\frac{p^{2}+q^{2}}{%
M_{V}^{2}}\biggr)\,.
\end{equation}%
For a reader's convenience let us also summarize previous results
based on the  VMD and LMD+V ans\"{a}tze\footnote{The LMD+V ansatz
adds one extra vector multiplet in comparison with the LMD one.
This corresponds to MHA for which all the OPE and B-L constraints
can be satisfied simultaneously.} \cite{Knecht:2001xc}:
\begin{align}
& \mathcal{F}_{\pi ^{0}\gamma \gamma }^{\text{VMD}}(p^{2},q^{2};0)=-{\frac{%
N_{C}}{12\pi ^{2}F_{\pi }}}{\frac{M_{V}^{2}}{(p^{2}-M_{V}^{2})}}{\frac{%
M_{V}^{2}}{(q^{2}-M_{V}^{2})}}\,,  \label{VMD} \\
& \mathcal{F}_{\pi ^{0}\gamma \gamma }^{\text{LMD+V}}(p^{2},q^{2};0)={\frac{%
F_{\pi }}{3}}\,{\frac{%
p^{2}q^{2}(p^{2}+q^{2})+h_{1}(p^{2}+q^{2})^{2}+h_{2}p^{2}q^{2}+h_{5}(p^{2}+q^{2})+h_{7}%
}{%
(p^{2}-M_{V_{1}}^{2})(p^{2}-M_{V_{2}}^{2})(q^{2}-M_{V_{1}}^{2})(q^{2}-M_{V_{2}}^{2})%
}}\,,  \label{modelsMD}
\end{align}%
with (valid in the chiral limit)
\begin{equation*}
h_{7}\,=\,-{\frac{N_{C}}{4\pi ^{2}}}{\frac{M_{V_{1}}^{4}M_{V_{2}}^{4}}{%
F_{\pi }^{2}}}\,.
\end{equation*}%
We have therefore the following relation (compare with~(\ref{zmena1}))%
\begin{equation}
\mathcal{F}_{\pi ^{0}\gamma \gamma }^{\text{R}\chi \text{T}}(p^{2},q^{2};0)%
\Bigr|_{\text{B-L}}\ =\mathcal{F}_{\pi ^{0}\gamma \gamma }^{\text{VMD}%
}(p^{2},q^{2};0)\,.  \label{zmena2}
\end{equation}%
Now let us turn back to Brodsky-Lepage condition. We have seen it has
important consequences on the actual form of the $\pi ^{0}-\gamma -\gamma $
formfactor within R$\chi $T. However, recent BABAR measurement \cite{Aubert:2009mc} showed phenomenological disagreement with this condition.
There are also theoretical arguments \cite{Bijnens:2003rc} which showed that
high-energy constraints cannot be all satisfied for a given formfactor
within the ansatz with only finite number of resonance poles. (For a recent
study on Brodsky-Lepage revision see~\cite{Agaev:2010aq}; see also \cite{Dorokhov:2010bz} and references therein.) We will thus
relax the Brodsky-Lepage condition by allowing a small deviation from (\ref%
{BLc}) parameterized with $\delta _{\text{BL}}$
\begin{equation}
\kappa _{3}^{PV}=-\frac{F^{2}}{32\sqrt{2}d_{m}F_{V}}(1+\delta _{\text{BL}%
})\,.  \label{mBL}
\end{equation}%
Its actual value can be set by fitting the BABAR and CLEO data. In this fit
and also in the following phenomenological applications we set
\begin{equation}
M_{V}=m_{\rho }\approx 0.775\,\text{GeV},\quad M_{P}=m_{\pi ^{\prime
}}\approx 1.3\,\text{GeV},\quad F=F_{\pi }\approx 92.22\,\text{MeV}
\label{phen1}
\end{equation}%
and also (for details see \cite{Kampf:2006bn})
\begin{equation}
F_{V}=F_{\rho }=146.3\pm 1.2\,\text{MeV},\quad d_{m}\approx 26\,\text{MeV}\,.
\label{phen2}
\end{equation}%
The new BABAR data indicates important negative shift in $\delta _{\text{BL}}
$ with the result
\begin{equation}
\delta _{\text{BL}}\ =-0.055\pm 0.025\,.  \label{BL55}
\end{equation}%
Our fit together with CLEO and BABAR data is depicted in Fig.~\ref{fig:babar}%
.
%%%%

\subsubsection{Decay $\rho\to\pi\gamma$}

In this subsection we illustrate a particular phenomenological application
of the above results, namely a prediction for $\rho \rightarrow \pi \gamma $
decay. For this process we can use a connection with the off-shell $\pi
\gamma \gamma $ formfactor introduced in the previous subsection. First, let
us define the amplitude $\mathcal{A}$ for the process $\rho
^{+}(p)\rightarrow \pi ^{+}(p)\gamma (k)$ (we will use only the charged
decay process to avoid the discussion on $\omega -\rho $ mixing for the
neutral one):
\begin{equation}
\Gamma _{\rho \rightarrow \pi \gamma }=\frac{1}{8\pi }\frac{1}{3}\sum_{\text{%
pol.}}\ |\mathcal{A}_{\rho \rightarrow \pi \gamma }\varepsilon ^{\mu \nu
\alpha \beta }p_{\alpha }q_{\beta }\epsilon _{\mu }(p)\epsilon _{\nu }^{\ast
}(k)|^{2}\frac{m_{\rho }^{2}-m_{\pi }^{2}}{2m_{\rho }^{3}}\,,
\label{Gamma_rho_pi_gamma}
\end{equation}%
from which we have already factorized out the Levi-Civita and momentum
dependence. Similarly one can define the amplitude for $\pi
^{0}(p)\rightarrow \gamma (k)\gamma (l)$
\begin{equation}
\Gamma _{\pi ^{0}\rightarrow \gamma \gamma }=\frac{1}{32\pi }\sum_{\text{pol}%
}\ |\mathcal{A}_{\pi ^{0}\rightarrow \gamma \gamma }\varepsilon ^{\mu \nu
\alpha \beta }k_{\alpha }l_{\beta }\epsilon _{\mu }^{\ast }(k)\epsilon _{\nu
}^{\ast }(l)|^{2}\frac{1}{m_{\pi ^{0}}}\,.
\end{equation}%
The connection with $\pi \gamma \gamma $ formfactor is obtained via
\begin{equation}
\mathcal{A}_{\rho \rightarrow \pi \gamma }=\frac{e}{2}\frac{1}{F_{V}M_{V}}%
\lim_{q^{2}\rightarrow M_{V}^{2}}(q^{2}-M_{V}^{2})\mathcal{F}_{\pi \gamma
\gamma }(0,q^{2};0)  \label{Arhopig}
\end{equation}%
and quite simply
\begin{equation}
\mathcal{A}_{\pi \rightarrow \gamma \gamma }=e^{2}\mathcal{F}_{\pi \gamma
\gamma }(0,0;0)\,.  \label{Apigg}
\end{equation}%
Putting these two definitions together we can extract the ratio and
corresponding parameter $x$ \cite{Moussallam:1994xp}:
\begin{equation}
\frac{2eF_{V}}{M_{V}}\biggl|\frac{\mathcal{A}_{\rho \rightarrow \pi \gamma }%
}{\mathcal{A}_{\pi \rightarrow \gamma \gamma }}\biggr|\equiv 1+x\,.
\label{rAA}
\end{equation}%
Using the experimental value $\Gamma _{\rho \rightarrow \pi \gamma }=68\pm 7%
\text{keV}$ this parameter was obtained to be equal to $x=0.022\pm 0.051$ in~%
\cite{Moussallam:1994xp} based on the 1992 edition of the particle data book
(same number was also used later e.g. in~\cite{Knecht:2001xc}). Updating
this prediction with a new experimental input we can get flip in the sign
\begin{equation}
\text{exp:}\qquad x=-0.003\pm 0.054\,.  \label{newx}
\end{equation}%
The change is mainly due to a new value of $F_{V}$ (study e.g. in~\cite%
{Kampf:2006bn}) and a new precise measurement of $\pi ^{0}$ lifetime by
PrimEx group \cite{Larin:2010kq} (see also \cite{Kampf:2009tk}).

Within our formalism, the parameter $x$ defined above is
proportional to the deviation from the simple VMD ansatz
(\ref{VMD}), or in other words from the exact
Brodsky-Lepage condition (cf. (\ref{zmena2})). Using (\ref{FcLMD}) and (\ref%
{mBL}) we get in terms of $\delta_{BL}$
\begin{equation}
x = \frac{4\pi^2 F^2}{M_V^2 N_C} \delta_\text{BL}\,.
\end{equation}
The results of the previous subsection allows us to make rather precise
determination of this value
\begin{equation}
\text{R$\chi$T:}\qquad x = -0.010 \pm 0.005\,,
\end{equation}
which using (\ref{rAA}) and experimental input for $\Gamma_{\pi^0\to\gamma%
\gamma}$ leads to the following prediction:
\begin{equation}
\text{R$\chi$T:}\qquad \Gamma_{\rho\to\pi\gamma} = 67.0\pm 2.3 \text{ keV}\,.
\end{equation}
%%%%

\subsubsection{Decays of $\pi(1300)$}

In the previous section we have obtained a prediction for the $%
\rho\to\pi\gamma$ decay width. However, it was based on the ratio of two
decay widths (cf.~(\ref{rAA})) and experimental input of one of them. We
could predict also the absolute value for $\rho\to\pi\gamma$ directly from (%
\ref{Gamma_rho_pi_gamma}) and (\ref{Arhopig}) without the necessity to use
the experimental value of $\pi^0\to\gamma\gamma$ (in fact we will discuss a
little the latter process in the very next subsection) but one should
remember that we have been making several simplifications, namely: we are
working in large $N_C$, using only lowest-lying resonances and we are in the
chiral limit. All together within this approximation we cannot expect the
accuracy of the result being better than 30\%-40\%. On the other hand one
can expect that some of the systematic uncertainties will cancel out in the
ratios similar to one studied in the previous part.

\begin{figure}[tbh]
\begin{center}
\epsfig{width=0.6\textwidth,figure=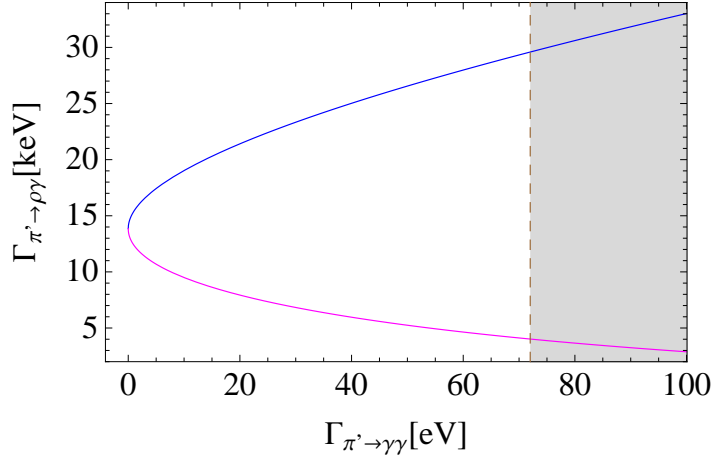}
\end{center}
\caption{The connection of decay width for $\pi (1300)\rightarrow
\gamma \gamma $ and $\pi (1300)\rightarrow %
\rho \gamma $ (note that we have two possible solutions). The dashed
line represents the Belle's limit~(\ref{c72}) on $\Gamma _{%
\pi^{\prime }\rightarrow \rho \pi }$ (grey area is thus
excluded by this experiment).}
\label{fig:GGP}
\end{figure}

The same strategy can be repeated for $\pi (1300)$ decays. In fact we can
work in exact correspondence; the two decays would be now: $\pi
(1300)\rightarrow \rho \gamma $ and $\pi (1300)\rightarrow \gamma \gamma $.
The only problem now is that none of these two processes have been seen so
far. The most recent limit on $\pi (1300)\rightarrow \gamma \gamma $ by
Belle collaboration \cite{Abe:2006by}
\begin{equation}
\Gamma _{\pi ^{\prime }\rightarrow \gamma \gamma }<72\,\text{eV}  \label{c72}
\end{equation}%
sets at least rough limit in our studies. Using the definition~(\ref{FPgg})
the main object here is
\begin{equation}
\mathcal{F}_{P\gamma \gamma }^{\text{R$\chi $T}}(p^{2},q^{2};m_{P}^{2})=%
\frac{8\sqrt{2}}{3}F_{V}\frac{\sqrt{2}\kappa
_{3}^{PV}(2M_{V}^{2}-p^{2}-q^{2})-F_{V}\kappa ^{VVP}}{%
(p^{2}-M_{V}^{2})(q^{2}-M_{V}^{2})}\,.  \label{cLMDP}
\end{equation}%
The amplitude for $\pi (1300)\rightarrow \gamma \gamma $ is given by
\begin{equation}
\mathcal{A}_{\pi ^{\prime }\rightarrow \gamma \gamma }=e^{2}\mathcal{F}%
_{P\gamma \gamma }(0,0;m_{\pi ^{\prime }}^{2})
\end{equation}%
and similarly for $\pi (1300)\rightarrow \rho \gamma $ (see also (\ref%
{Arhopig})). Then%
\begin{eqnarray}
\mathcal{A}_{\pi ^{\prime }\rightarrow \gamma \gamma }^{\text{R$\chi $T}}
&=&e^{2}\frac{8\sqrt{2}}{3}F_{V}\frac{2\sqrt{2}\kappa
_{3}^{PV}M_{V}^{2}-F_{V}\kappa ^{VVP}}{M_{V}^{4}}\,, \\
\mathcal{A}_{\pi ^{\prime }\rightarrow \rho \gamma }^{\text{R$\chi $T}} &=&-e%
\frac{4\sqrt{2}}{3M_{V}}\frac{\sqrt{2}\kappa _{3}^{PV}M_{V}^{2}-F_{V}\kappa
^{VVP}}{M_{V}^{2}}.
\end{eqnarray}%
Both these amplitudes depend on one so far undetermined constant $\kappa
^{VVP}$. Provided we have experimental values of both branching ratios we
could verify the consistency of our model. In the present situation we can
visualize how one decay mode depends on the second one, and this was done in
Fig.~\ref{fig:GGP}.
One can see that we have two solutions for $\kappa ^{VVP}$ as we
have quadratic equation for decay width as a function of $\kappa
^{VVP}$ and none of these two solutions can be ruled out. Note
that the full width for $\pi
(1300)$ is assumed to be between 200 and 600 MeV (see \cite{Nakamura:2010zzi}%
), so both processes are extremely suppressed for any of these two solutions.

The experimental bound on $\Gamma _{\pi ^{\prime }\rightarrow \gamma \gamma
} $ \ can be used to get estimate of $\kappa ^{VVP}$. In order to fulfill
the limit~(\ref{c72}) we expect the numerator in~(\ref{cLMDP}) to be
suppressed. This expected suppression leads in analogy with (\ref{mBL}) to
the following ansatz
\begin{equation}
\kappa _{VVP}=-\frac{F^{2}M_{V}^{2}}{16d_{m}F_{V}^{2}}(1+\delta _{\text{A}%
})\,,
\end{equation}%
with parameter $\delta _{\text{A}}$ which should be reasonably small. \ In
terms of this parameter we get the decay width in a compact form
\begin{equation}
\Gamma _{\pi ^{\prime }\rightarrow \gamma \gamma }=\Bigl(\frac{\alpha F^{2}}{%
6\sqrt{2}d_{m}M_{V}^{2}}\Bigr)^{2}\pi m_{\pi ^{\prime }}^{3}(\delta _{\text{%
BL}}-\delta _{\text{A}})^{2}\approx (1514.0\,\text{eV})\times (\delta _{%
\text{BL}}-\delta _{\text{A}})^{2}
\end{equation}%
and thus the extreme phenomenological suppression of $\pi (1300)\rightarrow
\gamma \gamma $ can be understood within our formalism to be due to the
small factor $(\delta _{\text{BL}}-\delta _{\text{A}})^{2}$. The
experimental limit together with~(\ref{BL55}) set the allowed range for the
parameter $\ \delta _{\text{A}}$
\begin{equation}
-0.27\lesssim \delta _{A}\lesssim 0.16\,,
\end{equation}%
which is good enough to set the value of $\kappa _{VVP}$ to
\begin{equation}
\kappa _{VVP}\approx (-0.57\pm 0.13)\,\text{GeV}\,.
\end{equation}

\subsubsection{Decay $\pi^0\rightarrow \gamma\gamma$
and $\eta\rightarrow \gamma\gamma$}

As we have stated, the absolute decay widths are accessible via our approach
only with the limited precision. For instance for the $\pi ^{0}\rightarrow
\gamma \gamma $ amplitude we have obtained only very simple prediction~(\ref%
{Apigg}). It turns out, however, that it agrees very well with the
experimental determination. On the other hand, similar determination for $%
\eta \rightarrow \gamma \gamma $ would be a phenomenological disaster.

In order to go beyond the leading order we can use the chiral corrections
calculated using ChPT. The most recent study of $\pi ^{0}\rightarrow \gamma
\gamma $ amplitude went up to NNLO \cite{Kampf:2009tk}. The motivation for
going beyond NLO lays in the fact that there are no chiral logarithms at NLO
\cite{Donoghue:1986wv,Bijnens:1988kx}. At NNLO these logarithms though
non-zero are relatively small so the $C_{i}^{W}$ play very important role.
We can therefore use existing calculations within ChPT with our estimate (%
\ref{C7C22ope}) of $C_{i}^{W}$. Here our approximation, namely the chiral
limit, does not make any difference as by construction LECs ($C_{i}^{W}$ in
our case) do not depend on light quark masses. With previous
phenomenological determination of the couplings $\kappa _{3}^{PV}$ and $%
\kappa _{VVP}$ we obtain
\begin{equation}
C_{7}^{W}=\frac{F^{2}}{64M_{V}^{4}}\Bigl(1+2\frac{M_{V}^{2}}{M_{P}^{2}}%
(\delta _{BL}-\delta _{A})\Bigr)\approx (0.35\pm 0.07)\times 10^{-3}\,\text{%
GeV}^{-2}\,.  \label{CW7}
\end{equation}%
The second and last unknown LEC at NLO for $\pi ^{0}\rightarrow \gamma
\gamma $ and $\eta \rightarrow \gamma \gamma $ is $C_{8}^{W}$. Anticipating
the result of the next section and using the OPE constraints (\ref{opecon})
we get%
\begin{equation}
C_{8}^{W}=\frac{N_C}{768M_{0}^{2}\pi ^{2}}-\frac{N_{C}}{512\pi ^{2}M_{V}^{2}}%
-\frac{F_{V}\kappa _{13}^{V}}{\sqrt{2}M_{V}^{2}}+\frac{F_{V}^{2}\kappa
_{1}^{VV}}{2M_{V}^{4}} + \frac{d_{m0} F^2}{96 d_m M_P^2 M_V^2} + \frac{%
d_{m0} F_V^2 \kappa_{VVP}}{6 M_P^2 M_V^4} \,,
\end{equation}%
where we have also dropped the term proportional to $\delta_{BL}$ because it
is not numerically relevant. Unfortunately at this moment similarly as for
already mentioned $d_{m0}$ we cannot make an estimate for $\kappa _{13}^{V}$
and $\kappa _{1}^{VV}$ (all these couplings are dominated by the $\eta
^{\prime }$ exchange, cf. Appendix~\ref{apA}). We may however again connect
two-gamma decay widths of $\pi^{0}$ and $\eta$. We may for example set the
unknown $C_8^W$ from the experimental value of $\Gamma(\eta\to\gamma\gamma)$%
. This was done for NLO $\eta\to\gamma\gamma$ expression in \cite%
{Kampf:2009tk}. There is ongoing project which should enlarge this
calculation to the NNLO within ChPT (for preliminary results in the chiral
limit calculation see~\cite{Bijnens:2010pa}). We thus rather postpone as a
future project the final determination of $\pi ^{0}\rightarrow \gamma \gamma
$ based on the experimental value $\Gamma _{\eta \rightarrow \gamma \gamma }$%
. Let us only mention, that if we assume that the NNLO corrections for $\eta$
are indeed small as for $\pi^0\to\gamma\gamma$, the value in~(\ref{CW7}) has
roughly the influence at 0.5\% level for $\Gamma _{\pi^0} \rightarrow \gamma
\gamma$ (with the opposite sign). A new study of isospin breaking effects in~%
\cite{Kampf:2011wr} indicates another shift of the similar size (however now
with a positive sign) and thus at this moment we do not expect quantitative
change of the prediction made in~\cite{Kampf:2009tk}.

\subsubsection{$g-2$}

Probably the main motivation for studying the $VVP$ correlator is hidden in
the determination of the muon $g-2$ factor. It is beyond the scope of this
paper to discuss this problem in great detail (for details see e.g.~\cite%
{Jegerlehner:2009ry}). Let us only mention that the main source of
theoretical error for its standard model prediction comes from
hadronic contributions, more precisely from the hadronic
light-by-light scattering which cannot be related to any available
data. The hadronic four-point correlator $VVVV$ is further
simplified into three classes of contributions: a) $\pi^\pm$ and
$K^\pm$ loops b) $\pi^0,\eta,\eta^{\prime }$ exchanges and finally
c) the rest, which is modelled usually via constituent quark
loops. It is clear that this separation is not without ambiguity
and different approaches can differently calculate contribution
especially between a) and c) or b) and c). Our contribution based
on the $VVP$ correlator study belongs to the group b). To avoid
inconsistency we will work in close relation with similar work
done for LMD or VMD ans\"{a}tze \cite{BijnensKnecht}. Using the
fully off-shell (i.e. including also the $\pi^0$ off-shellness) $%
\pi^0-\gamma-\gamma$ formfactor~(\ref{Fpigg}) we arrive to
\begin{equation}
a_\mu^{\text{LbyL;}\pi^0} = (65.8 \pm 1.2)\times 10^{-11}\,.
\end{equation}
In the error only the uncertainties of our model were included.
The systematic is mainly influenced by the above mentioned
ambiguity of how one defines and splits the pion-pole and regular
part from the $\langle VVVV \rangle$. We have put the cutoff
energy at $10$~GeV. For a better comparison let us present  in
Table~\ref{tab:g2} predictions for the studied
contribution to the muon $g-2$ for the different models summarized in~(\ref%
{modelsMD}).
\begin{table}[h]
\begin{center}
\begin{tabular}{cc}
\hline
model & $a_\mu^{\text{LbyL;}\pi^0} \times 10^{11}$ \\ \hline
VMD & $57.2$ \\
LMD & $73.7$ \\
LMD+V ``on-shell'' & 58.2 \\
LMD+V ``off-shell'' & $72\pm12$ \\ \hline
this work & $65.8\pm1.2$ \\ \hline
\end{tabular}%
\end{center}
\caption{Contribution of $\pi^0$ exchange to the muon $g-2$ factor
for different models.}
\label{tab:g2}
\end{table}
We have recalculated there the light-by-light contributions based
on VMD and LMD ans\"atze. We have also reevaluated the case of
LMD+V ansatz or more precisely its on-shell simplification as
defined in~(\ref{modelsMD}). Three unknown constants are set
similarly as
done in~\cite{Knecht:2001xc}, i.e. $h_{1,2}=0$ and $h_5$ is based on the $%
\rho\to\pi\gamma$ phenomenology $h_5=6.99$ (obtained for the updated value
in~(\ref{newx})). The full LMD+V ``off-shell'' ansatz has 7 parameters (for
details see~\cite{Nyffeler}). One relation can be obtained from the chiral
anomaly and others can be: Brodsky-Lepage behaviour, higher-twist
corrections in the OPE and one-large momentum OPE, together with data (CLEO
for this turn) we are still left with two undetermined parameters. Their
variations in reasonable range set the final error for the corresponding
LMD+V value in Tab.~\ref{tab:g2}. Let us note that also the possibility of
B-L violation together with new fit of two parameters ($h_1$ and $h_5$) was
studied in \cite{Nyffeler} with no influence on the central value of $g-2$
contribution. Too many parameters is not the only problem connected with the
LMD+V ansatz. Status of $\rho(1450)$ as a first radial excitation of $%
\rho(770)$ is doubted by the possible existence of lighter $\rho(1250)$ \cite%
{Bertin:1997vf}. Its presence is also supported by the study within AdS/QCD
approaches \cite{Gherghetta:2009ac}. On top of that the inclusion of the
complete set of all excitations in all channels (i.e. inclusion of $%
\pi^{\prime \prime }$) can change again the studied ansatz similarly as we
have encountered for the first excitation (see (\ref{zmena1}) and (\ref%
{zmena2})).

Let us also note quite astonishing coincidence of our result with the most
recent study based on AdS/QCD conjecture \cite{Cappiello:2010uy} $%
a_\mu^{\pi^0} = 65.4(2.5)\times 10^{-11}$.

\subsection{$VAS$ Green function}

The $\langle VVP \rangle$ Green's function studied in the previous section
represents without any doubts the most important example of the odd
intrinsic sector of QCD. However, it is not the only quantity one can
analyze using our complete lowest-lying resonance model. As the second
example we have chosen $\langle VAS \rangle$, which has not yet been studied
(to our knowledge) in the literature. It also enables to demonstrate the use
of the ``second half'' our the odd intrinsic resonance Lagrangian, i.e.
those with $1^{++}$ and $0^{++}$ states.

Defining (beware of the same symbol as for $\langle VVP\rangle$)
\begin{equation}
\Pi_{\mu\nu}^{abc}(p,q) = \int d^4x\, d^4y\, e^{ip\cdot x+iq\cdot y} \langle
0| T[V^a_\mu(x)A^b_\nu(y)S^c(0)|0\rangle\,,
\end{equation}
with (cf. also~(\ref{VPcur}))
\begin{equation*}
A_\mu^a(x) =\bar q(x) \gamma_\mu \gamma_5 \frac{\lambda^a}{2}q(x),\quad
S_\mu^a(x) =\bar q(x) \frac{\lambda^a}{2}q(x)\,.
\end{equation*}
Similarly as for $VVP$ one can write
\begin{equation}
\Pi(p,q)^{abc}_{\mu\nu} = f^{abc}\epsilon_{\mu\nu\alpha\beta}p^\alpha
q^\beta \Pi(p^2,q^2;r^2)\,,
\end{equation}
where $r=-(p+q)$. In resonance region, we have
\begin{multline}
\frac{1}{B_0}\Pi(p^2,q^2;r^2) = \frac{8\sqrt2F_V(\kappa^V_4-2\kappa^V_{15})}{%
p^2-M_V^2} + \frac{16\sqrt2F_A\kappa^A_{14}}{q^2-M_A^2} + \frac{%
32c_m\kappa^S_2}{r^2-M_S^2} + \frac{16\sqrt2F_Ac_m\kappa^{SA}_1}{%
(q^2-M_A^2)(r^2-M_S^2)} \\
-\frac{8\sqrt2F_Vc_m(2\kappa^{SV}_1+\kappa^{SV}_2)}{ (p^2-M_V^2)(r^2-M_S^2)}
- \frac{16F_AF_V\kappa^{VA}_6}{(q^2-M_A^2)(p^2-M_V^2)} + \frac{%
16F_AF_Vc_m\kappa^{VAS}}{(q^2-M_A^2)(p^2-M_V^2)(r^2-M_S^2)}\,.
\label{VASrcht}
\end{multline}
At high energies one can obtain the following OPE relation
\begin{equation}
\Pi((\lambda p)^2,(\lambda q)^2; (\lambda r)^2) = \frac{B_0F^2}{2\lambda^4}%
\frac{p^2-q^2-r^2}{p^2q^2r^2} + \mathcal{O}\left(\frac{1}{\lambda^6}\right)
\end{equation}
and again we will not consider here one-momentum OPE limits. The high-energy
constraint leads to
\begin{align}
&\kappa^{S}_{2} = \kappa^A_{14} = 0\,,\qquad
\kappa^V_4=2\kappa^V_{15}\,,\qquad \kappa^{VA}_6 = \frac{F^2}{32F_AF_V}\,,
\notag \\
&F_V(2\kappa^{SV}_1+\kappa^{SV}_2) = 2 F_A\kappa^{SA}_1 = \frac{F^2}{%
16\sqrt2c_m}\,.  \label{OPEvas}
\end{align}
If we use these relations, we have finally only one free parameter: $%
\kappa^{VAS}$; the result is
\begin{equation}
\frac{1}{B_0}\Pi^\text{R$\chi$T}(p^2,q^2;r^2) = \frac{%
F^2(p^2-q^2-r^2-M_V^2+M_A^2+M_S^2)+32F_AF_Vc_m\kappa^{VAS}}{
2(q^2-M_A^2)(p^2-M_V^2)(r^2-M_S^2)}\,.
\end{equation}
From the theoretical point of view we are thus in a better position than we
were for $\langle VVP \rangle$. After imposing OPE we are left with one free
parameter whereas in the case of $\langle VVP \rangle$ we had two (cf.~(\ref%
{PcLMD})). We can thus simply connect all processes schematically
represented as
\begin{equation}
(V: \rho, \omega, K^*, \gamma,\ldots)\, \sim \,(A: a_1,f_1,K_1, GB,
W\ldots)\,\sim \, (S: \sigma,\kappa,a_0, f_0, H\ldots)
\end{equation}
via a single parameter. The problem is that they are very rare and have not
yet been measured, on top of that the status of some of the particle content
is controversial by itself (especially if talking about a scalar sector).
The parameter $\kappa^{VAS}$ can be, however, set using other (not that
rare) processes it enters. One way is to check in next section to which of
23 parameters it contributes and use directly the system of LECs. This can
be done already here within the calculation of VAS. At low energies, up to $%
\mathcal{O}(p^6)$ one has
\begin{equation}
\frac{1}{B_0}\Pi(p^2,q^2,r^2) = -32C_{11}^W\,.
\end{equation}
Comparing with the low energy expansion of the full R$\chi$T result (\ref%
{VASrcht}) we get
\begin{eqnarray}
C_{11}^W &=& \frac{F_A\kappa^A_{14}}{\sqrt2M_A^2} +\frac{F_V(\kappa^V_4-2%
\kappa^V_{15})}{2\sqrt2M_V^2} +\frac{c_m\kappa^S_2}{M_S^2} +\frac{%
F_AF_V\kappa^{VA}_6}{2M_A^2M_V^2} + \frac{c_mF_V(2\kappa^{SV}_1+%
\kappa^{SV}_2)}{2\sqrt2M_S^2M_V^2}  \notag \\
&& -\frac{F_Ac_m\kappa^{SA}_1}{\sqrt2M_A^2M_S^2} +\frac{F_AF_Vc_m\kappa^{VAS}%
}{2M_A^2M_S^2M_V^2}\,.  \label{C11}
\end{eqnarray}
Using the OPE (\ref{OPEvas}) we obtain
\begin{eqnarray*}
C_{11}^W &=& \frac{F^2}{64}\left[\frac{1}{M_S^2M_V^2} +\frac{1}{M_A^2M_V^2}-%
\frac{1}{M_A^2M_S^2}\right] +\frac{F_AF_Vc_m\kappa^{VAS}}{2M_A^2M_S^2M_V^2}%
\,.
\end{eqnarray*}
The knowledge of $C_{11}^W$ leads directly to the value of $\kappa^{VAS}$
and thus to the rare processes schematically specified above. The current
attempts for a $C_{11}^W$ estimation were summarized in Table~1 of~\cite%
{Jiang:2010wa} with rather inconsistent values obtained both from the
phenomenology (\cite{Unterdorfer:2008zz}, \cite{Strandberg:2003zf}) and by a
model-dependent determination \cite{Jiang:2010wa}. The most precise value
seems to be obtained from $K^+\to l^+\nu\gamma$ data \cite{Poblaguev:2002ug}%
: $C_{11}^W = (0.68\pm 0.21)\times 10^{-3}\,\text{GeV}^{-2}$ \cite%
{Unterdorfer:2008zz}. Using the values set in~(\ref{phen1}) and~(\ref{phen2}%
), together with
\begin{equation}
M_S = m_{a_0} \approx 984.7\,\text{MeV},\qquad c_m \approx 42\,\text{MeV}
\end{equation}
and the Weinberg sum rules (to get values of $M_A$ and $F_A$) we arrive at
\begin{equation}
\kappa^{VAS} = 0.61 \pm 0.40\,\text{GeV}\,.
\end{equation}

\subsection{Short note on the field redefinition}

The previous two examples were calculated using the full resonance
Lagrangian ${\mathcal{L}_{R\chi T}^{(6,\text{ odd})}}$. Here we
would like to address  a question what would happened if one would
repeat the same calculation but instead use the reduced resonance
Lagrangian $\overline{\mathcal{L}_{R\chi T}^{(6,\text{ odd})}}$.
This Lagrangian is established in Appendix~\ref{apB} and can be
obtained from the full Lagrangian~(\ref{reslag}) by means dropping
the operators marked with a star in the
Tables~\ref{tab:prvni}-\ref{tab:posledni} i.e. by means of
omitting 20 parameters: $\kappa^{RR}_{1,2}, \kappa^{SA}_i,
\kappa^{SV}_1, \kappa^{VA}_i, \kappa^{PA}_1, \kappa^{PV}_i,
\kappa^{RRR}_i$ and using the bar over the rest of $\kappa^X_i$
(see Section~\ref{aprel}). This can be motivated by its equivalent
contribution to the saturation of LECs. This exercise was already
performed in \cite{Cirigliano:2006hb} for $\langle VAP \rangle$
with an interesting finding, that after imposing the OPE condition
the both results are the same. In our case the conclusion is,
however, different. Using the reduced resonance Lagrangian we
would not be able to simply fulfill the OPE constraints by
imposing some conditions on $\overline{\kappa^X_i}$. In the first
case,
the OPE for $\langle VVP \rangle$ requires an additional relation, namely $%
M_V = \frac{4\pi F}{\sqrt{N_C}}$. In the second case, $\langle
VAS\rangle$, the OPE cannot be satisfied at all.

Thus we have to conclude that the equivalence of both calculation in the
even sector for $\langle VAP \rangle$ was just a coincidence and it is not a
general feature.

\section{Resonance contributions to the LECs of the anomalous sector}

\label{sec:reslec}

We have seen in the previous two applications the explicit examples of the
calculation with the resonance fields. A match between this result in a
region of small momenta (i.e. $p\ll M_R$) at one side and the ChPT result at
other side enables to extract the dependence of LECs on resonances. In this
way we have obtained within $VVP$ calculation $C^W_7$ and $C^W_{22}$ (see~(%
\ref{C7C22})) and from $VAS$ it was possible to extract $C^W_{11}$ (\ref{C11}%
). The dependence of all others $C^W_i$ on the parameters of the resonance
model can be obtained by systematic integration-out of all resonances. So
obtained Lagrangian can be expand over the canonical basis of NLO
odd-intrinsic Lagrangian established for example in~\cite{Bijnens:2001bb}.
In this way we have saturated 21 of 23 constants and only $C^W_3$ and $%
C^W_{18}$ stayed intact as they are subleading in large $N_C$. The $%
\eta^{\prime }$ was explicitly considered (see Appendix~\ref{apA} and \cite%
{Kaiser:2000gs}) and it contributes in $C^W_6$, $C^W_8$ and $C^W_{10}$. It
is always the first term in these LECs and we put it in the boldface font to
stress its large $N_C$ dominance over the rest. Generally we have the
following expansion in large $N_C$ for all $C_i^W$, schematically
\begin{equation}
C^W_i = a_i N_C^2 + b_i N_C + O(N_C^0)\,,
\end{equation}
where $a_i \neq 0$ for $i=6,8,10$ and $b_i=0$ for $i=3,18$.

The field redefinition similarly as done in \cite{Cirigliano:2006hb} was
performed and details are summarized in Appendix~\ref{apB}. All 20
parameters denoted by stars in Tab.\ref{tab:prvni}--\ref{tab:posledni} can
be dropped in the following and for all others a bar should be added (bar
parameters $\overline{\kappa _{i}^{X}}$ are defined in the last section of
Appendix~\ref{apB}). We prefer, however, to use the original parametrization
as it represents direct connection with the resonance phenomenology and is
thus simpler to use.

The explicit form of the resonance saturation generated by the resonance
Lagrangian~(\ref{reslag}) is:
\begin{align}
&C^W_1 = \frac{d_m \kappa _ 4^P}{M_P^2}+\frac{2 \sqrt{2} d_m G_V \kappa _ 1^{%
\text{PV}}}{M_P^2 M_V^2}-\frac{\sqrt{2} d_m G_V \kappa _ 2^{\text{PV}}}{%
M_P^2 M_V^2}+\frac{\sqrt{2} G_V \kappa _ 3^V}{M_V^2}-\frac{2 \sqrt{2} G_V
\kappa _ 9^V}{M_V^2}+\frac{\sqrt{2} G_V \kappa _ {10}^V}{M_V^2}  \notag \\
&\phantom{C^W_1} -\frac{4 G_V^2 \kappa _ 2^{\text{VV}}}{M_V^4}-\frac{3 G_V^2
\kappa _ 3^{\text{VV}}}{2 M_V^4}-\frac{2 \kappa ^{\text{VVP}} d_m G_V^2}{%
M_P^2 M_V^4} \,,  \notag \\
&C^W_ 2 = \frac{F_A \kappa _ {13}^A}{\sqrt{2} M_A^2}+\frac{c_m \kappa _ 1^S}{%
M_S^2}+\frac{c_m F_A \kappa _ 2^{\text{SA}}}{\sqrt{2} M_A^2 M_S^2}+\frac{%
\sqrt{2} c_m G_V \kappa _ 1^{\text{SV}}}{M_S^2 M_V^2}+\frac{c_m F_V \kappa _
2^{\text{SV}}}{2 \sqrt{2} M_S^2 M_V^2}+\frac{F_V \kappa _ 4^V}{2 \sqrt{2}
M_V^2}-\frac{\sqrt{2} G_V \kappa _ {15}^V}{M_V^2}  \notag \\
&\phantom{C^W_1} +\frac{F_A G_V \kappa _ 6^{\text{VA}}}{M_A^2 M_V^2}+\frac{%
\kappa ^{\text{VAS}} c_m F_A G_V}{M_A^2 M_S^2 M_V^2} \,,  \notag \\
&C^W_ 3=0 \,,  \notag \\
&C^W_ 4=-\frac{F_A \kappa _ 3^A}{2 \sqrt{2} M_A^2}-\frac{d_m \kappa _ 3^P}{%
M_P^2}+\frac{\sqrt{2} d_m G_V \kappa _ 3^{\text{PV}}}{M_P^2 M_V^2}+\frac{F_A
F_V \kappa _ 3^{\text{VA}}}{2 M_A^2 M_V^2}+\frac{\sqrt{2} F_A \kappa _ 4^A}{%
M_A^2}-\frac{F_A \kappa _ {12}^A}{2 \sqrt{2} M_A^2}+\frac{F_A \kappa _ {15}^A%
}{2 \sqrt{2} M_A^2}  \notag \\
&\phantom{C^W_1} -\frac{F_A^2 \kappa _ 4^{\text{AA}}}{4 M_A^4}-\frac{d_m F_A
\kappa _ 2^{\text{PA}}}{2 \sqrt{2} M_A^2 M_P^2}+\frac{d_m F_V \kappa _ 1^{%
\text{PV}}}{\sqrt{2} M_P^2 M_V^2}+\frac{2 c_d \kappa _ 2^S}{M_S^2}-\frac{%
\sqrt{2} c_d F_A \kappa _ 1^{\text{SA}}}{M_A^2 M_S^2}+\frac{\sqrt{2} c_d F_V
\kappa _ 1^{\text{SV}}}{M_S^2 M_V^2}  \notag \\
&\phantom{C^W_1} +\frac{c_d F_V \kappa _ 2^{\text{SV}}}{\sqrt{2} M_S^2 M_V^2}
+\frac{F_V \kappa _ 1^V}{\sqrt{2} M_V^2}-\frac{\sqrt{2} F_V \kappa _ 5^V}{%
M_V^2}-\frac{F_V \kappa _ 6^V}{\sqrt{2} M_V^2}-\frac{F_V \kappa _ 8^V}{\sqrt{%
2} M_V^2}-\frac{F_V \kappa _ 9^V}{\sqrt{2} M_V^2} +\frac{\sqrt{2} G_V \kappa
_ {14}^V}{M_V^2}  \notag \\
&\phantom{C^W_1} +\frac{F_A F_V \kappa _ 1^{\text{VA}}}{M_A^2 M_V^2}+\frac{%
F_A F_V \kappa _ 2^{\text{VA}}}{2 M_A^2 M_V^2} -\frac{2 F_V G_V \kappa _ 2^{%
\text{VV}}}{M_V^4}+\frac{\kappa ^{\text{VAS}} c_d F_A F_V}{M_A^2 M_S^2 M_V^2}%
-\frac{\kappa ^{\text{VVP}} d_m F_V G_V}{M_P^2 M_V^4} \,,  \notag \\
&C^W_ 5 = \frac{F_A \kappa _ {12}^A}{\sqrt{2} M_A^2}-\frac{d_m \kappa _ 2^P}{%
M_P^2}+\frac{d_m F_A \kappa _ 2^{\text{PA}}}{\sqrt{2} M_A^2 M_P^2}-\frac{d_m
F_V \kappa _ 2^{\text{PV}}}{\sqrt{2} M_P^2 M_V^2}+\frac{F_V \kappa _ {10}^V}{%
\sqrt{2} M_V^2}-\frac{G_V \kappa _ {17}^V}{\sqrt{2} M_V^2}-\frac{F_V G_V
\kappa _ 3^{\text{VV}}}{M_V^4} \,,  \notag \\
&C^W_ 6=\mathbf{-\frac{N_C}{576 M_0^2 \pi ^2}} + \frac{F_A \kappa _ 3^A}{3
\sqrt{2} M_A^2}-\frac{F_A F_V \kappa _ 3^{\text{VA}}}{3 M_A^2 M_V^2}-\frac{%
F_V G_V \kappa _ 3^{\text{VV}}}{3 M_V^4}-\frac{2 \sqrt{2} F_A \kappa _ 4^A}{%
3 M_A^2}+\frac{F_A \kappa _ {10}^A}{\sqrt{2} M_A^2}-\frac{F_A \kappa _ {15}^A%
}{3 \sqrt{2} M_A^2}  \notag \\
&\phantom{C^W_1} +\frac{F_A^2 \kappa _ 4^{\text{AA}}}{6 M_A^4} -\frac{4 c_d
\kappa _ 2^S}{3 M_S^2}+\frac{2 \sqrt{2} c_d F_A \kappa _ 1^{\text{SA}}}{3
M_A^2 M_S^2}-\frac{2 \sqrt{2} c_d F_V \kappa _ 1^{\text{SV}}}{3 M_S^2 M_V^2}-%
\frac{\sqrt{2} c_d F_V \kappa _ 2^{\text{SV}}}{3 M_S^2 M_V^2}-\frac{\sqrt{2}
F_V \kappa _ 1^V}{3 M_V^2}  \notag \\
&\phantom{C^W_1} +\frac{2 \sqrt{2} F_V \kappa _ 5^V}{3 M_V^2}+\frac{\sqrt{2}
F_V \kappa _ 6^V}{3 M_V^2}+\frac{\sqrt{2} F_V \kappa _ 8^V}{3 M_V^2}+\frac{%
\sqrt{2} G_V \kappa _ {13}^V}{M_V^2}-\frac{G_V \kappa _ {17}^V}{3 \sqrt{2}
M_V^2}-\frac{F_V \kappa _ {18}^V}{\sqrt{2} M_V^2}  \notag \\
&\phantom{C^W_1} -\frac{2 F_A F_V \kappa _ 1^{\text{VA}}}{3 M_A^2 M_V^2}-%
\frac{F_A F_V \kappa _ 2^{\text{VA}}}{3 M_A^2 M_V^2}-\frac{2 F_V G_V \kappa
_ 1^{\text{VV}}}{M_V^4}-\frac{2 \kappa ^{\text{VAS}} c_d F_A F_V}{3 M_A^2
M_S^2 M_V^2} -\frac{\sqrt{2} F_A d_{\text{m0}} \kappa _2^{\text{PA}}}{3
M_A^2 M_p^2} -\frac{2 d_{\text{m0}} \kappa _3^P}{3 M_p^2}  \notag \\
&\phantom{C^W_1} +\frac{d_{\text{m0}} \kappa _2^P}{3 M_p^2} +\frac{d_{\text{%
m0}} F_V \kappa _2^{\text{PV}}}{3 \sqrt{2} M_p^2 M_V^2}+\frac{\sqrt{2} d_{%
\text{m0}} F_V \kappa _1^{\text{PV}}}{3 M_p^2 M_V^2}+\frac{2 \sqrt{2} d_{%
\text{m0}} G_V \kappa _3^{\text{PV}}}{3 M_p^2 M_V^2}-\frac{2 d_{\text{m0}}
F_V G_V \kappa ^{\text{VVP}}}{3 M_p^2 M_V^4} \,,  \notag \\
&C^W_ 7= \frac{2 d_m \kappa _ 5^P}{M_P^2}-\frac{\sqrt{2} d_m F_V \kappa _ 3^{%
\text{PV}}}{M_P^2 M_V^2}-\frac{F_V \kappa _ {12}^V}{2 \sqrt{2} M_V^2}-\frac{%
\sqrt{2} F_V \kappa _ {14}^V}{M_V^2}-\frac{F_V \kappa _ {16}^V}{4 \sqrt{2}
M_V^2}  \notag \\
&\phantom{C^W_1} +\frac{F_V^2 \kappa _ 2^{\text{VV}}}{M_V^4}-\frac{F_V^2
\kappa _ 3^{\text{VV}}}{8 M_V^4}+\frac{\kappa ^{\text{VVP}} d_m F_V^2}{2
M_P^2 M_V^4} \,,  \notag \\
&C^W_ 8=\mathbf{\frac{N_C}{768 M_0^2 \pi ^2}} + \frac{F_V \kappa _ {12}^V}{6
\sqrt{2} M_V^2}-\frac{F_V \kappa _ {13}^V}{\sqrt{2} M_V^2}+\frac{F_V \kappa
_ {16}^V}{12 \sqrt{2} M_V^2}+\frac{F_V^2 \kappa _ 1^{\text{VV}}}{2 M_V^4}+%
\frac{F_V^2 \kappa _ 3^{\text{VV}}}{24 M_V^4}  \notag \\
&\phantom{C^W_1} -\frac{\sqrt{2} d_{\text{m0}} F_V \kappa _3^{\text{PV}}}{3
M_p^2 M_V^2}+\frac{d_{\text{m0}} F_V^2 \kappa ^{\text{VVP}}}{6 M_p^2 M_V^4}+%
\frac{2 d_{\text{m0}} \kappa _5^P}{3 M_p^2} \,,  \notag \\
&C^W_ 9=-\frac{F_A \kappa _ 3^A}{4 \sqrt{2} M_A^2}-\frac{F_A^2 \kappa _ 3^{%
\text{AA}}}{8 M_A^4}-\frac{F_A \kappa _ 8^A}{2 \sqrt{2} M_A^2}-\frac{\sqrt{2}
F_A \kappa _ {11}^A}{M_A^2}-\frac{F_A \kappa _ {12}^A}{\sqrt{2} M_A^2}-\frac{%
F_A \kappa _ {15}^A}{4 \sqrt{2} M_A^2}  \notag \\
&\phantom{C^W_1} +\frac{F_A^2 \kappa _ 2^{\text{AA}}}{M_A^4}+\frac{2 d_m
\kappa _ 1^P}{M_P^2}-\frac{\sqrt{2} d_m F_A \kappa _ 1^{\text{PA}}}{M_A^2
M_P^2}-\frac{d_m F_A \kappa _ 2^{\text{PA}}}{\sqrt{2} M_A^2 M_P^2}+\frac{%
\kappa ^{\text{AAP}} d_m F_A^2}{2 M_A^4 M_P^2} \,,  \notag \\
&C^W_ {10}=\mathbf{\frac{N_C}{768 M_0^2 \pi ^2}} + \frac{F_A \kappa _ 3^A}{%
12 \sqrt{2} M_A^2}+\frac{F_A^2 \kappa _ 3^{\text{AA}}}{24 M_A^4}+\frac{F_A
\kappa _ 8^A}{6 \sqrt{2} M_A^2}-\frac{F_A \kappa _ 9^A}{\sqrt{2} M_A^2}+%
\frac{F_A \kappa _ {10}^A}{2 \sqrt{2} M_A^2}+\frac{F_A \kappa _ {15}^A}{12
\sqrt{2} M_A^2}  \notag \\
&\phantom{C^W_1} +\frac{F_A^2 \kappa _ 1^{\text{AA}}}{2 M_A^4} +\frac{F_A^2
\kappa ^{\text{AAP}} d_{\text{m0}}}{6 M_A^4 M_p^2}-\frac{\sqrt{2} F_A d_{%
\text{m0}} \kappa _1^{\text{PA}}}{3 M_A^2 M_p^2}-\frac{F_A d_{\text{m0}}
\kappa _2^{\text{PA}}}{3 \sqrt{2} M_A^2 M_p^2}+\frac{2 d_{\text{m0}} \kappa
_1^P}{3 M_p^2} \,,  \notag \\
&C^W_ {11}= \frac{F_A \kappa _ {14}^A}{\sqrt{2} M_A^2}+\frac{c_m \kappa _ 2^S%
}{M_S^2}-\frac{c_m F_A \kappa _ 1^{\text{SA}}}{\sqrt{2} M_A^2 M_S^2}+\frac{%
c_m F_V \kappa _ 1^{\text{SV}}}{\sqrt{2} M_S^2 M_V^2}+\frac{c_m F_V \kappa _
2^{\text{SV}}}{2 \sqrt{2} M_S^2 M_V^2}+\frac{F_V \kappa _ 4^V}{2 \sqrt{2}
M_V^2}-\frac{F_V \kappa _ {15}^V}{\sqrt{2} M_V^2}  \notag \\
&\phantom{C^W_1} +\frac{F_A F_V \kappa _ 6^{\text{VA}}}{2 M_A^2 M_V^2}+\frac{%
\kappa ^{\text{VAS}} c_m F_A F_V}{2 M_A^2 M_S^2 M_V^2} \,,  \notag \\
&C^W_ {12}= \frac{\sqrt{2} G_V \kappa _ 1^V}{M_V^2}-\frac{\sqrt{2} G_V
\kappa _ 2^V}{M_V^2}-\frac{\sqrt{2} G_V \kappa _ 3^V}{M_V^2}+\frac{G_V^2
\kappa _ 3^{\text{VV}}}{M_V^4} \,,  \notag \\
&C^W_ {13}= -\frac{F_A \kappa _ 3^A}{\sqrt{2} M_A^2}+\frac{F_A F_V \kappa _
3^{\text{VA}}}{M_A^2 M_V^2}+\frac{F_V G_V \kappa _ 3^{\text{VV}}}{M_V^4}+%
\frac{2 \sqrt{2} F_A \kappa _ 4^A}{M_A^2}+\frac{F_A \kappa _ {15}^A}{\sqrt{2}
M_A^2}-\frac{F_A^2 \kappa _ 4^{\text{AA}}}{2 M_A^4}+\frac{4 c_d \kappa _ 2^S%
}{M_S^2}  \notag \\
&\phantom{C^W_1} -\frac{2 \sqrt{2} c_d F_A \kappa _ 1^{\text{SA}}}{M_A^2
M_S^2}+\frac{2 \sqrt{2} c_d F_V \kappa _ 1^{\text{SV}}}{M_S^2 M_V^2}+\frac{%
\sqrt{2} c_d F_V \kappa _ 2^{\text{SV}}}{M_S^2 M_V^2}+\frac{\sqrt{2} F_V
\kappa _ 1^V}{M_V^2}-\frac{2 \sqrt{2} F_V \kappa _ 5^V}{M_V^2}-\frac{\sqrt{2}
F_V \kappa _ 6^V}{M_V^2}  \notag \\
&\phantom{C^W_1} -\frac{\sqrt{2} F_V \kappa _ 8^V}{M_V^2}-\frac{\sqrt{2} G_V
\kappa _ {12}^V}{M_V^2}-\frac{G_V \kappa _ {16}^V}{\sqrt{2} M_V^2}+\frac{%
\sqrt{2} G_V \kappa _ {17}^V}{M_V^2}+\frac{2 F_A F_V \kappa _ 1^{\text{VA}}}{%
M_A^2 M_V^2}+\frac{F_A F_V \kappa _ 2^{\text{VA}}}{M_A^2 M_V^2}  \notag \\
&\phantom{C^W_1} +\frac{2 \kappa ^{\text{VAS}} c_d F_A F_V}{M_A^2 M_S^2 M_V^2%
} \,,  \notag \\
&C^W_ {14}= -\frac{F_A \kappa _ 3^A}{2 \sqrt{2} M_A^2}-\frac{F_V \kappa _ 3^V%
}{\sqrt{2} M_V^2}+\frac{F_A F_V \kappa _ 3^{\text{VA}}}{2 M_A^2 M_V^2}+\frac{%
\sqrt{2} F_A \kappa _ 4^A}{M_A^2}+\frac{F_A \kappa _ {15}^A}{2 \sqrt{2} M_A^2%
}-\frac{F_A^2 \kappa _ 4^{\text{AA}}}{4 M_A^4}+\frac{2 c_d \kappa _ 2^S}{%
M_S^2}  \notag \\
&\phantom{C^W_1} -\frac{\sqrt{2} c_d F_A \kappa _ 1^{\text{SA}}}{M_A^2 M_S^2}%
+\frac{\sqrt{2} c_d F_V \kappa _ 1^{\text{SV}}}{M_S^2 M_V^2}+\frac{c_d F_V
\kappa _ 2^{\text{SV}}}{\sqrt{2} M_S^2 M_V^2}+\frac{F_V \kappa _ 1^V}{\sqrt{2%
} M_V^2}-\frac{\sqrt{2} F_V \kappa _ 5^V}{M_V^2}-\frac{F_V \kappa _ 6^V}{%
\sqrt{2} M_V^2}-\frac{F_V \kappa _ 8^V}{\sqrt{2} M_V^2}  \notag \\
&\phantom{C^W_1} +\frac{F_A F_V \kappa _ 1^{\text{VA}}}{M_A^2 M_V^2}+\frac{%
F_A F_V \kappa _ 2^{\text{VA}}}{2 M_A^2 M_V^2}+\frac{\kappa ^{\text{VAS}}
c_d F_A F_V}{M_A^2 M_S^2 M_V^2} \,,  \notag \\
&C^W_ {15}= -\frac{F_A \kappa _ 3^A}{2 \sqrt{2} M_A^2}+\frac{F_A F_V \kappa
_ 3^{\text{VA}}}{2 M_A^2 M_V^2}-\frac{F_V G_V \kappa _ 3^{\text{VV}}}{M_V^4}+%
\frac{\sqrt{2} F_A \kappa _ 4^A}{M_A^2}+\frac{F_A \kappa _ {15}^A}{2 \sqrt{2}
M_A^2}-\frac{F_A^2 \kappa _ 4^{\text{AA}}}{4 M_A^4}+\frac{2 c_d \kappa _ 2^S%
}{M_S^2}  \notag \\
&\phantom{C^W_1} -\frac{\sqrt{2} c_d F_A \kappa _ 1^{\text{SA}}}{M_A^2 M_S^2}%
+\frac{\sqrt{2} c_d F_V \kappa _ 1^{\text{SV}}}{M_S^2 M_V^2}+\frac{c_d F_V
\kappa _ 2^{\text{SV}}}{\sqrt{2} M_S^2 M_V^2}+\frac{F_V \kappa _ 2^V}{\sqrt{2%
} M_V^2}-\frac{\sqrt{2} F_V \kappa _ 5^V}{M_V^2}-\frac{F_V \kappa _ 6^V}{%
\sqrt{2} M_V^2}-\frac{F_V \kappa _ 8^V}{\sqrt{2} M_V^2}  \notag \\
&\phantom{C^W_1} -\frac{G_V \kappa _ {17}^V}{\sqrt{2} M_V^2}+\frac{F_A F_V
\kappa _ 1^{\text{VA}}}{M_A^2 M_V^2}+\frac{F_A F_V \kappa _ 2^{\text{VA}}}{2
M_A^2 M_V^2}+\frac{\kappa ^{\text{VAS}} c_d F_A F_V}{M_A^2 M_S^2 M_V^2} \,,
\notag \\
&C^W_ {16}= \frac{\sqrt{2} F_A \kappa _ 2^A}{M_A^2}+\frac{F_A G_V \kappa _
2^{\text{VA}}}{M_A^2 M_V^2}-\frac{\sqrt{2} G_V \kappa _ 6^V}{M_V^2}+\frac{%
\sqrt{2} G_V \kappa _ 7^V}{M_V^2}-\frac{\sqrt{2} G_V \kappa _ 8^V}{M_V^2}+%
\frac{F_A G_V \kappa _ 3^{\text{VA}}}{M_A^2 M_V^2}  \notag \\
&\phantom{C^W_1} -\frac{F_A G_V \kappa _ 4^{\text{VA}}}{M_A^2 M_V^2}-\frac{%
G_V^2 \kappa _ 3^{\text{VV}}}{M_V^4} \,,  \notag \\
&C^W_ {17}= \frac{F_A \kappa _ 1^A}{\sqrt{2} M_A^2}-\frac{c_d \kappa _ 1^S}{%
M_S^2}-\frac{\sqrt{2} c_d F_A \kappa _ 1^{\text{SA}}}{M_A^2 M_S^2}+\frac{%
\sqrt{2} c_d F_V \kappa _ 1^{\text{SV}}}{M_S^2 M_V^2}-\frac{\sqrt{2} c_d G_V
\kappa _ 1^{\text{SV}}}{M_S^2 M_V^2}+\frac{F_A F_V \kappa _ 1^{\text{VA}}}{%
M_A^2 M_V^2}  \notag \\
&\phantom{C^W_1} -\frac{F_A G_V \kappa _ 1^{\text{VA}}}{M_A^2 M_V^2}+\frac{%
F_A \kappa _ 2^A}{\sqrt{2} M_A^2}+\frac{\sqrt{2} F_A \kappa _ 4^A}{M_A^2}+%
\frac{F_A \kappa _ {15}^A}{2 \sqrt{2} M_A^2}-\frac{F_A^2 \kappa _ 4^{\text{AA%
}}}{4 M_A^4}+\frac{2 c_d \kappa _ 2^S}{M_S^2}-\frac{c_d F_A \kappa _ 2^{%
\text{SA}}}{\sqrt{2} M_A^2 M_S^2}  \notag \\
&\phantom{C^W_1} +\frac{c_d F_V \kappa _ 2^{\text{SV}}}{2 \sqrt{2} M_S^2
M_V^2}-\frac{\sqrt{2} F_V \kappa _ 5^V}{M_V^2}+\frac{\sqrt{2} G_V \kappa _
5^V}{M_V^2}-\frac{F_V \kappa _ 6^V}{\sqrt{2} M_V^2}-\frac{F_V \kappa _ 8^V}{%
\sqrt{2} M_V^2}+\frac{F_A F_V \kappa _ 2^{\text{VA}}}{2 M_A^2 M_V^2}+\frac{%
F_A F_V \kappa _ 3^{\text{VA}}}{2 M_A^2 M_V^2}  \notag \\
&\phantom{C^W_1} -\frac{F_V G_V \kappa _ 3^{\text{VV}}}{4 M_V^4}+\frac{G_V^2
\kappa _ 4^{\text{VV}}}{2 M_V^4}-\frac{F_V G_V \kappa _ 4^{\text{VV}}}{4
M_V^4}+\frac{\kappa ^{\text{VAS}} c_d F_A F_V}{M_A^2 M_S^2 M_V^2}-\frac{%
\kappa ^{\text{VAS}} c_d F_A G_V}{M_A^2 M_S^2 M_V^2} \,,  \notag \\
&C^W_ {18}=0 \,,  \notag \\
&C^W_ {19}= -\frac{\sqrt{2} F_A \kappa _ 5^A}{M_A^2}-\frac{F_A G_V \kappa _
5^{\text{VA}}}{M_A^2 M_V^2}-\frac{\sqrt{2} F_V \kappa _ 6^V}{M_V^2}-\frac{%
\sqrt{2} G_V \kappa _ {11}^V}{M_V^2}+\frac{F_V \kappa _ {16}^V}{\sqrt{2}
M_V^2}-\frac{G_V \kappa _ {16}^V}{\sqrt{2} M_V^2}+\frac{\sqrt{2} G_V \kappa
_ {17}^V}{M_V^2}  \notag \\
&\phantom{C^W_1} +\frac{F_A F_V \kappa _ 3^{\text{VA}}}{M_A^2 M_V^2}+\frac{%
F_V G_V \kappa _ 3^{\text{VV}}}{M_V^4}-\frac{F_V^2 \kappa _ 4^{\text{VV}}}{2
M_V^4}+\frac{F_V G_V \kappa _ 4^{\text{VV}}}{M_V^4} \,,  \notag \\
&C^W_ {20}= \frac{\sqrt{2} F_A \kappa _ 4^A}{M_A^2}-\frac{F_A^2 \kappa _ 4^{%
\text{AA}}}{2 M_A^4}-\frac{F_V^2 \kappa _ 4^{\text{VV}}}{4 M_V^4}+\frac{F_V
G_V \kappa _ 4^{\text{VV}}}{2 M_V^4}-\frac{F_A \kappa _ 5^A}{\sqrt{2} M_A^2}-%
\frac{F_A \kappa _ 6^A}{\sqrt{2} M_A^2}+\frac{F_A \kappa _ {15}^A}{\sqrt{2}
M_A^2}+\frac{2 c_d \kappa _ 2^S}{M_S^2}  \notag \\
&\phantom{C^W_1} -\frac{\sqrt{2} c_d F_A \kappa _ 1^{\text{SA}}}{M_A^2 M_S^2}%
+\frac{\sqrt{2} c_d F_V \kappa _ 1^{\text{SV}}}{M_S^2 M_V^2}+\frac{c_d F_V
\kappa _ 2^{\text{SV}}}{\sqrt{2} M_S^2 M_V^2}-\frac{\sqrt{2} F_V \kappa _ 5^V%
}{M_V^2}-\frac{\sqrt{2} F_V \kappa _ 6^V}{M_V^2}-\frac{\sqrt{2} F_V \kappa _
8^V}{M_V^2}  \notag \\
&\phantom{C^W_1} +\frac{F_V \kappa _ {16}^V}{2 \sqrt{2} M_V^2}-\frac{G_V
\kappa _ {16}^V}{\sqrt{2} M_V^2}+\frac{F_A F_V \kappa _ 1^{\text{VA}}}{M_A^2
M_V^2}+\frac{F_A F_V \kappa _ 2^{\text{VA}}}{M_A^2 M_V^2}+\frac{F_A F_V
\kappa _ 3^{\text{VA}}}{M_A^2 M_V^2}-\frac{F_V G_V \kappa _ 3^{\text{VV}}}{%
M_V^4}  \notag \\
&\phantom{C^W_1} +\frac{\kappa ^{\text{VAS}} c_d F_A F_V}{M_A^2 M_S^2 M_V^2}
\,,  \notag \\
&C^W_ {21}= -\frac{\sqrt{2} F_A \kappa _ 4^A}{M_A^2}+\frac{F_A^2 \kappa _ 4^{%
\text{AA}}}{2 M_A^4}-\frac{F_A F_V \kappa _ 4^{\text{VA}}}{2 M_A^2 M_V^2}+%
\frac{F_V^2 \kappa _ 4^{\text{VV}}}{4 M_V^4}-\frac{F_V G_V \kappa _ 4^{\text{%
VV}}}{2 M_V^4}+\frac{F_A \kappa _ 7^A}{\sqrt{2} M_A^2}-\frac{F_A \kappa _
{15}^A}{\sqrt{2} M_A^2}  \notag \\
&\phantom{C^W_1} -\frac{2 c_d \kappa _ 2^S}{M_S^2}+\frac{\sqrt{2} c_d F_A
\kappa _ 1^{\text{SA}}}{M_A^2 M_S^2}-\frac{\sqrt{2} c_d F_V \kappa _ 1^{%
\text{SV}}}{M_S^2 M_V^2}-\frac{c_d F_V \kappa _ 2^{\text{SV}}}{\sqrt{2}
M_S^2 M_V^2}+\frac{\sqrt{2} F_V \kappa _ 5^V}{M_V^2}+\frac{F_V \kappa _ 6^V}{%
\sqrt{2} M_V^2}+\frac{F_V \kappa _ 7^V}{\sqrt{2} M_V^2}  \notag \\
&\phantom{C^W_1} +\frac{F_V \kappa _ 8^V}{\sqrt{2} M_V^2}-\frac{F_V \kappa _
{16}^V}{2 \sqrt{2} M_V^2}+\frac{G_V \kappa _ {16}^V}{\sqrt{2} M_V^2}-\frac{%
G_V \kappa _ {17}^V}{\sqrt{2} M_V^2}-\frac{F_A F_V \kappa _ 1^{\text{VA}}}{%
M_A^2 M_V^2}-\frac{F_A F_V \kappa _ 2^{\text{VA}}}{2 M_A^2 M_V^2}-\frac{F_A
F_V \kappa _ 3^{\text{VA}}}{2 M_A^2 M_V^2}  \notag \\
&\phantom{C^W_1} -\frac{F_V G_V \kappa _ 3^{\text{VV}}}{2 M_V^4}-\frac{%
\kappa ^{\text{VAS}} c_d F_A F_V}{M_A^2 M_S^2 M_V^2} \,,  \notag \\
&C^W_ {22}= -\frac{F_V \kappa _ {17}^V}{\sqrt{2} M_V^2}-\frac{F_V^2 \kappa _
3^{\text{VV}}}{2 M_V^4} \,,  \notag \\
&C^W_ {23}= -\frac{F_A \kappa _ {16}^A}{\sqrt{2} M_A^2}-\frac{F_A^2 \kappa _
3^{\text{AA}}}{2 M_A^4}\,.  \label{satCW}
\end{align}
Apart from already mentioned relations $C^W_3 =0$ and $C^W_{18}=0$ we have
found out one further relation free from $\kappa_i^X$
\begin{equation}
\frac{F_V^2}{2 G_V} C^W_{12} = F_V (C^W_{14} - C^W_{15}) + G_V C^W_{22}\,.
\end{equation}
The transformation established in Appendix~\ref{apB} was employed as an
independent check of the previous relations.

\section{Summary}

In this paper we have studied the odd-intrinsic sector of the
low-energy QCD. We have constructed the most general resonance
Lagrangian that describes the interactions of the Goldstone bosons
and the lowest-lying vector-, axial-,
scalar-, pseudoscalar-resonance multiplets. We were working in the large $%
N_C $ approximation and considered only those terms that contributes to $%
O(p^6)$ anomalous Lagrangian (i.e. to the first non-trivial order). This was
the main aim of our work. We then demonstrated the use of this Lagrangian
for three different applications. The first two represent calculations of
two three-point Green functions $\langle VVP \rangle$ and $\langle VAS
\rangle$. The third application was the complete integration out of the
resonance fields and establishing the so-called saturation of LECs by
resonance fields.

The first application $VVP$ is the most important example of the
odd-intrinsic sector, both from the theoretical and phenomenological point
of view. We have discussed different aspects of this Green functions. First,
after calculating this three-point correlator within our model and imposing
a certain high-energy constraint we ended up with the result which depends
only on two parameters. These were further set using new BABAR data on $%
\pi\gamma\gamma$ off-shell formfactor and Belle collaboration's limit on $%
\pi^{\prime }\to\gamma\gamma$ decay. After setting these two parameters we
can make further predictions. The outcome of our analysis is for example a
very precise determination of the decay width of a process $\rho\to\pi\gamma$:
$%
\Gamma_{\rho\to\pi\gamma} = 67(2.3)\text{ keV}$. We have also studied a
relative dependence of the rare decays $\pi^{\prime }\to\gamma\gamma$ and $%
\pi^{\prime }\to\rho\gamma$. Based on the experimental upper limit of the
former one can set the lowest limit of the latter. Prediction of our model
is $30\text{ keV} \gtrsim\Gamma_{\pi^{\prime }\to\rho\gamma} \gtrsim 4\text{
keV}$ (based on Belle's $\Gamma_{\pi^{\prime }\to\gamma\gamma} \lesssim 72%
\text{ eV}$). Next, we have also evaluate the value of $C_7^W$ LEC together
with short discussion on $\pi^0$ and $\eta$ two photon decays. Last but not
least a very precise determination of the off-shell $\pi^0$-pole
contribution to the muon $g-2$ factor was provided. Our final determination
of this factor is $a_\mu^{\pi^0} = 65.8(1.2) \times 10^{-11}$. The $R\chi T$
approach has thus reduced the error of the similar determination based on
lowest-meson saturation ansatz by factor of ten and is in exact agreement
with the most recent determination based on AdS/QCD assumptions. Let us note
that the present theoretical error for the complete anomalous magnetic
moment of the muon is around $50\times 10^{-11}$ and the experimental error
around $60\times 10^{-11}$ \cite{Bennett:2006fi} (with the well-know
discrepancy above $3\sigma$). A new proposed experiment at Fermilab E989
\cite{prop989} plans to go down with the precision to the preliminary value $%
16\times 10^{-11}$ and thus the reduction of the error in the theoretical
light-by-light calculation is more than desirable.

If $VVP$ represents very important and rich phenomenological
example, the  three-point correlator $\langle VAS \rangle$ is
connected with very rare processes  and represents so far never
studied example of the odd-sector. We have established its OPE
behaviour which enabled us to reduce the dependence of the $VAS$
Green function to one parameter. This opens the possibility of a
future study of these rare but interesting processes.

In the last section we have studied the resonance saturation at low
energies. We have integrated out the resonance fields to establish the
dependence of LECs of odd-sector $C_i^W$ on our parameters. As we are
limited by large $N_C$ we cannot make prediction for $C_3^W$ and $C_{18}^W$
but we have set all other 21 LECs. We have found one relation among $C_{12}^W
$, $C_{14}^W$, $C_{15}^W$ and $C_{22}^W$ free from our parameters.

\section*{Acknowledgement}

We would especially like to thank Jarda Trnka for initiating this project
and his contribution at the early stage. We thank also Hans Bijnens and Bachir
Moussallam for valuable discussions and comments.
This work is supported in part by the European Community-Research Infrastructure
Integrating Activity ``Study of
Strongly Interacting Matter'' (HadronPhysics2, Grant Agreement n.\ 227431)
and the Center for Particle Physics (project no.\ LC 527) of the Ministry of
Education of the Czech Republic.

\appendix

\section{The large $N_{C}$ counting}

\label{apA}

\subsection{General considerations}

Let us start with the $U_{L}(N_{F})\times U_{R}(N_{F})$ invariant Lagrangian
for the nonet of the GB and resonances without using the equations of motion
and the Cayley-Hamilton identities. Then the large $N_{C}$ behaviour of the
couplings accompanying individual operators in the effective Lagrangian with
octet GB (after $\eta^{\prime}$ has been integrated out) can be understood
as follows.

Let us write in the same way as in \cite{Cirigliano:2006hb}
\begin{equation}
\widetilde{u}=\mathrm{e}^{\mathrm{i}\phi ^{0}T^{0}/F\sqrt{2}}u\,,
\end{equation}%
where $T^{0}=\sqrt{1/N_{F}}\mathbf{1}$ and
\begin{equation*}
u=\mathrm{e}^{\mathrm{i}\phi ^{a}T^{a}/F\sqrt{2}}
\end{equation*}%
is the $SU_{L}(N_{F})\times SU_{R}(N_{F})$ basic building block, and
therefore%
\begin{equation}
\phi ^{0}=\frac{F}{\mathrm{i}}\sqrt{\frac{2}{N_{F}}}\ln (\det \widetilde{u}%
)\,.
\end{equation}%
Let us also remind \cite{Cirigliano:2006hb}, that $\phi ^{0}$ and $\phi^{a}$
do not mix under the nonlinearly realized $U_{L}(N_{F})\times U_{R}(N_{F})$
symmetry. For the construction of the $U_{L}(N_{F})\times U_{R}(N_{F})$
effective Lagrangian, we have the usual building blocks constructed from $%
\widetilde{u}$ and the usual external sources $l_{\mu },r_{\mu },\chi $ and $%
\chi ^{+}$ (now also with singlet components) \emph{e.g.}
\begin{align}
\widetilde{u}_{\mu } &=u_{\mu }-D_{\mu }\phi ^{0}\frac{\sqrt{2}T^{0}}{F}\,,
\notag \\
\widetilde{\chi }_{\pm } &=\mathrm{e}^{-\mathrm{i}\phi ^{0}\sqrt{2}%
T^{0}/F}u^{+}\chi u^{+}\pm \mathrm{e}^{\mathrm{i}\phi ^{0}\sqrt{2}%
T^{0}/F}u\chi ^{+}u  \notag \\
&=\chi _{\pm }-\frac{\mathrm{i}}{F}\sqrt{\frac{2}{N_{F}}}\phi ^{0}\chi _{\mp
}+\ldots\,,  \notag \\
\langle l_{\mu }\rangle &=l_{\mu }^{0}\sqrt{\frac{N_{F}}{2}}\,,
\end{align}%
\emph{etc.} at our disposal. In the above formulae, the covariant (in fact
invariant) derivative of $\phi ^{0}$ is defined as
\begin{equation}
D_{\mu }\phi ^{0}=\partial _{\mu }\phi ^{0}-2a_{\mu }^{0}F\,,
\end{equation}%
however, it does not represent an independent building block because of the
identity
\begin{equation}
\langle \widetilde{u}_{\mu }\rangle =\sqrt{\frac{N_{F}}{2}}\frac{D_{\mu
}\phi ^{0}}{F}\,.
\end{equation}%
The above set of building block have to be further enlarged including also
the external sources $\theta $ for the winding number density%
\begin{equation}
\omega =\frac{g^{2}}{16\pi ^{2}}tr_{c}G_{\mu \nu }\widetilde{G}^{\mu \nu }\,,
\end{equation}%
with covariant derivative
\begin{equation*}
D_{\mu }\theta =\partial _{\mu }\theta +2a_{\mu }^{0}\,.
\end{equation*}%
We have to include also the following invariant combination%
\begin{equation}
X=\theta +\frac{\phi ^{0}}{F}\,.
\end{equation}%
Let us remind the large $N_{C}$ counting for the generating functional of
the connected Green function of quark bilinears and winding number densities%
\begin{equation}
Z[l,r,\chi ,\chi ^{+},\theta ]=N_{C}^{2}Z_{0}[\theta
/N_{C}]+N_{C}Z_{1}[l,r,\chi ,\chi ^{+},\theta /N_{C}]+\ldots\,,
\end{equation}%
where the ellipses stay for the subleading terms in the $1/N_{C}$ expansion.
This implies the usual $N_{C}$ counting of the physical amplitudes with $g$
glueballs and $m$ mesons
\begin{equation}
{\mathcal{A}}_{g,m}=O(N_{C}^{1+\delta _{m0}-g-\frac{m}{2}})\,.
\label{Nc_couts}
\end{equation}%
This counting should be reflected within the construction of the effective
chiral Lagrangian of R$\chi$T.

According to the (\ref{Nc_couts}), the explicit resonance fields have to be
counted as $O(N_{C}^{-1/2})$. As far as the GB are concerned, within the
tilded building blocs, each member of the pseudoscalar nonet is
automatically accompanied with (minus)one power of the decay constant $%
F=O(N_{C}^{1/2})$, which ensures the right counting of the vertices with GB,
provided the corresponding fields are counted as $O(N_{C}^{0})$. The only
subtlety is connected with the field $\phi ^{0}$.

The origin of the field $\phi ^{0}$ in the individual terms of the
Lagrangian is twofold. It can either come from the tilded building blocks $Y=%
\widetilde{u}_{\mu }$, $\widetilde{h}_{\mu \nu }$, $\widetilde{\chi }_{\pm }$
(and from their covariant derivatives $\widetilde{D}_{\mu }Y$ ; note that it
completely decouples from $\Gamma _{\mu }$ and $f_{\pm }^{\mu \nu }$) or
from the $X-$dependence of the Lagrangian. Each operator $\widetilde{%
\mathcal{O}}$ constructed form the tilded building blocks only (and
therefore including at least one flavour trace, the only exception is $%
\widetilde{\mathcal{O}}=1$) is in general accompanied by a potential $V_{%
\widetilde{\mathcal{O}}}(X)$ which is a function of the variable $X$ only,
\begin{equation}
\widetilde{\mathcal{L}}=\sum_{\widetilde{\mathcal{O}}}V_{\widetilde{\mathcal{%
O}}}(X)\widetilde{\mathcal{O}}\,.
\end{equation}%
While $\phi ^{0}$ originating from the tilded operators is counted as $%
O(N_{C}^{0})$ as the other GB, however, the same field coming from the power
expansion of the potentials counts as $O(1/N_{C})$ within the large $N_{C}$
expansion. Therefore, expanding the general operator $\widetilde{\mathcal{O}}
$ and the corresponding potential $V_{\widetilde{\mathcal{O}}}(X)$ in powers
of $\phi ^{0}$ and its derivatives (and taking into account that $%
F=O(N_{C}^{1/2})$) we have the following natural rule for the order $%
O(N_{C}^{n})$ of the resulting coupling constant at a term of this expansion
with $T$ \ flavour traces, $R$ resonance fields and $n_{0}$ fields $\phi
^{0} $
\begin{equation}
2-T-\frac{1}{2}R-\frac{3}{2}n_{0}\leq n\leq 2-T-\frac{1}{2}R-\frac{1}{2}%
n_{0}\,.  \label{operator_order}
\end{equation}%
The lower or higher bounds are saturated in the case when all $\phi ^{0}$'s
come exclusively either from $V_{\widetilde{\mathcal{O}}}(X)$ or from $%
\widetilde{\mathcal{O}}$.

Suppose that we had used the LO GB\ equations of motion prior to the
expansion in powers of $\phi ^{0}$. This allows to eliminate the terms with
derivatives, namely \cite{Cirigliano:2006hb}
\begin{equation}
\nabla^{\mu }\widetilde{u}_{\mu }=\widetilde{\chi }_{-}+\frac{4}{\sqrt{2N_{F}%
}}M_{0}^{2}\frac{\phi ^{0}}{F}\,.
\end{equation}%
Such a transformation of the original tilded operator do not create any
extra trace in contrast to the octet case. Because the singlet mass $%
M_{0}^{2}=O(1/N_{C})$, the $\phi ^{0}$ dependence of the resulting operator
brings about a factor of the order $O(N_{C}^{-3/2})$ (the same, as if $\phi
^{0}$ came from the potential) and the above bounds on $n$ remain therefore
valid. On the other hand, the further simplification using the
Cayley-Hamilton identity can destroy them, provided we use it in order to
eliminate terms with less traces in favour of the terms with more traces.

The next step is to integrate out $\phi ^{0}$ treating the mass $M_{0}^{2}$
as $O(p^{0})$. This can be done using its equation of motion, derived from
the corresponding part of the LO Lagrangian expanded in powers of $\phi ^{0}$%
\begin{equation}
{\mathcal{L}}_{0}^{(2)}=\frac{1}{2}D\phi ^{0}\cdot D\phi ^{0}-\frac{1}{2}%
M_{0}^{2}\left( \phi ^{0}\right) ^{2}-\mathrm{i}\frac{F}{2\sqrt{2N_{F}}}%
\langle \chi _{-}\rangle \phi ^{0}+d_{0}\langle P\rangle \phi ^{0}+\ldots\,,
\label{L_etaprim}
\end{equation}%
where $d_{0}$ term comes from the expansion of the potential and is
therefore of the order $O(N_{C}^{-1})$. The solution for $\phi ^{0}$ reads
in the leading order of the $p$ expansion\footnote{%
Here we have took into account, that the resonance fields should be counted
as $O(p^{2})$.}%
\begin{align}
\phi ^{0(2)}& \,=\,\frac{1}{M_{0}^{2}}\left( \mathrm{i}\frac{F}{2\sqrt{2N_{F}%
}}\langle \chi _{-}\rangle +d_{0}\langle P\rangle \right)  \notag \\
& \,=\;O(N_{C}^{3/2})\,+\,O(N_{C}^{0})\,,  \label{subst_eta_prim}
\end{align}%
where we have depicted the orders of both terms. $\phi ^{0(2)}$ should then
be inserted into the original Lagrangian expanded in powers of $\phi ^{0}$.
As a result, taking (\ref{operator_order}) into account, the orders of the
multiple trace operators within the $SU_{L}(N_{F})\times SU_{R}(N_{F})$
operator basis are enhanced. Namely, we have the following bound for the
corresponding couplings%
\begin{equation}
2-T_{0}-\frac{1}{2}R_{0}-\frac{3}{2}n_{\langle P\rangle }\leq n\leq 2-T_{0}-%
\frac{1}{2}R_{0}+n_{\langle \chi _{-}\rangle }-\frac{1}{2}n_{\langle
P\rangle }\,,
\end{equation}%
where $T_{0}$ and $R_{0}$ are the numbers of the traces and resonance fields
before elimination of $\phi ^{0}$ and $n_{\langle \chi _{-}\rangle }$ and $%
n_{\langle P\rangle }$ are the numbers of the new factors $\langle \chi
_{-}\rangle $ and $\langle P\rangle $ (which appear after $\phi ^{0}$ is
integrated out) respectively. More conveniently this can be expressed in
terms of the actual number of traces $T=T_{0}+n_{\langle P\rangle
}+n_{\langle \chi _{-}\rangle }$ and resonances $R=R_{0}+n_{\langle P\rangle
}$ as%
\begin{equation}
2-T-\frac{1}{2}R+n_{\langle \chi _{-}\rangle }\leq n\leq 2-T-\frac{1}{2}%
R+n_{\langle P\rangle }+2n_{\langle \chi _{-}\rangle }\,.  \label{bound1}
\end{equation}%
The loophole of this formula is, that for its application one has to trace
back which of the factors $\langle P\rangle $ and $\langle \chi _{-}\rangle $
originate in the $\phi ^{0}$ dependence of the tilded Lagrangian. The
extreme cases are either none or all of them, which gives a much raw estimate%
\begin{equation}
2-T-\frac{1}{2}R\leq n\leq 2-T-\frac{1}{2}R+N_{\langle P\rangle
}+2N_{\langle \chi _{-}\rangle }\,,  \label{bound2}
\end{equation}%
where now $N_{\langle \chi _{-}\rangle }$ and $N_{\langle P\rangle }$ are
the total numbers of $\langle P\rangle $ and $\langle \chi _{-}\rangle $
traces in the operator, the lower bound corresponds now to the usual trace
and resonance counting.

\subsection{Explicit examples}

Let us illustrate the above statements by means of an explicit examples. For
instance, the coupling at the term $\langle S\chi _{-}\rangle \langle \chi
_{-}\rangle $, at first sight of the order $O(N_{C}^{-1/2})$ might be of the
order $O(N_{C}^{1/2})$ or even $O(N_{C}^{3/2})$, because it can originate
either from the term%
\begin{multline}
\mathrm{i}\langle S\widetilde{\chi }_{-}\rangle W_{\langle S\chi _{-}\rangle
}(X)=\mathrm{i}\langle S\left( \chi _{-}+\ldots \right) \rangle \left(
w_{\langle S\chi _{-}\rangle }^{1}X+\ldots \right) \\
\rightarrow -\langle S\chi _{-}\rangle \frac{1}{M_{0}^{2}}\left( w_{\langle
S\chi _{-}\rangle }^{1}\frac{1}{2\sqrt{2N_{F}}}\langle \chi _{-}\rangle
+\ldots \right)\,,
\end{multline}%
which has the constant $w_{\langle S\chi _{-}\rangle }^{1}=O(N_{C}^{-1/2})$,
(this corresponds to the lower bound (\ref{bound1})) or from the term%
\begin{multline}
\langle S\widetilde{\chi }_{+}\rangle W_{\langle S\chi _{+}\rangle }(X)
=\langle S\left( \chi _{+}-\frac{\mathrm{i}}{F}\sqrt{\frac{2}{N_{F}}}\phi
^{0}\chi _{-}+\ldots \right) \rangle \left( w_{\langle S\chi _{+}\rangle
}^{0}+\ldots \right) \\
=-w_{\langle S\chi _{+}\rangle }^{0}\frac{\mathrm{i}}{F}\sqrt{\frac{2}{N_{F}}%
}\phi ^{0}\langle S\chi _{-}\rangle +\ldots \\
\rightarrow \frac{1}{M_{0}^{2}}w_{\langle S\chi _{+}\rangle }^{0}\left(
\frac{1}{2N_{F}}\right) \langle S\chi _{-}\rangle \langle \chi _{-}\rangle
+\ldots\,,
\end{multline}%
where $w_{\langle S\chi _{+}\rangle }^{0}=O(N_{C}^{1/2})$; (this corresponds
to the upper bound (\ref{bound1})).

Similarly the coupling $d_{m0}$ at the operator $\mathrm{i}\langle P\rangle
\langle \chi _{-}\rangle $ (see (\ref{LR})), naively of the order $%
O(N_{C}^{-1/2})$ can be enhanced by the $\phi ^{0}$ exchange. Indeed,
inserting (\ref{subst_eta_prim}) to the term $d_{0}\langle P\rangle \phi
^{0} $ of the Lagrangian (\ref{L_etaprim}), we get the following
contribution to $d_{m0}$%
\begin{equation}
d_{m0}=\frac{d_{0}}{M_{0}^{2}}\frac{F\sqrt{N_{F}}}{2\sqrt{2}}%
=O(N_{C}^{1/2})\,,  \label{dm0con}
\end{equation}%
where we have taken into account that $d_{0}=O(N_{C}^{-1})$.

Let us give also some examples of the odd intrinsic parity terms with
resonances, which similarly to the previous example lead to $N_{C}$ enhanced
multiple trace terms when $\phi ^{0}$ is integrated out. Some terms with one
resonance are for example%
\begin{equation}
\widetilde{\mathcal{L}}_{R} =\varepsilon _{\mu \nu \alpha \beta }\langle
V^{\mu \nu }[\widetilde{u}^{\alpha },\widetilde{u}^{\beta }]\rangle
W_{R1}(X)+\varepsilon _{\mu \nu \alpha \beta }\langle V^{\mu \nu
}f_{+}^{\alpha \beta }\rangle W_{R2}(X)+\varepsilon _{\mu \nu \alpha \beta
}\langle A^{\mu \nu }f_{+}^{\alpha \beta }\rangle W_{R3}(X)\,,
\end{equation}
where
\begin{eqnarray}
W_{Ri}(X) &=&\sum_{k}w_{Ri}^{(k)}X^{k}
\end{eqnarray}
and where
\begin{equation}
w_{Ri}^{(0)} =0,\qquad w_{Ri}^{(1)}=O(N_{C}^{-1/2}),\qquad \text{for}\quad
i=1,2,3\,.
\end{equation}%
These generate the operators%
\begin{eqnarray}
\hat{{\mathcal{O}}}^{V}_{18}&=&\varepsilon _{\mu \nu \alpha \beta }\langle
V^{\mu \nu }[u^{\alpha },u^{\beta }]\rangle \langle \chi _{-}\rangle \,,
\notag \\
\hat{{\mathcal{O}}}^{V}_{13}&=&\mathrm{i}\varepsilon _{\mu \nu \alpha \beta
}\langle V^{\mu \nu }f_{+}^{\alpha \beta }\rangle \langle \chi _{-}\rangle\,,
\notag \\
\hat{{\mathcal{O}}}^{A}_{9}&=&\mathrm{i}\langle A^{\mu \nu }f_{+}^{\alpha
\beta }\rangle \langle \chi _{-}\rangle
\end{eqnarray}%
with the couplings of the order $O(N_{C}^{1/2})$ (\emph{i.e} of the same
order as analogous single trace operators and therefore included in our
basis) and
\begin{eqnarray}
&&\varepsilon _{\mu \nu \alpha \beta }\langle V^{\mu \nu }[u^{\alpha
},u^{\beta }]\rangle \langle P\rangle\,,  \notag \\
&&\varepsilon _{\mu \nu \alpha \beta }\langle V^{\mu \nu }f_{+}^{\alpha
\beta }\rangle \langle P\rangle\,,  \notag \\
&&\langle A^{\mu \nu }f_{+}^{\alpha \beta }\rangle \langle P\rangle
\end{eqnarray}%
with the couplings of the order $O(N_{C}^{-1})$ suppressed with respect to
the single trace operators.

The two-resonance example is%
\begin{equation}
\widetilde{\mathcal{L}}_{RR} =\varepsilon _{\mu \nu \alpha \beta }\langle
V^{\mu \nu }V^{\alpha \beta }\rangle W_{RR1}(X)+\varepsilon _{\mu \nu \alpha
\beta }\langle A^{\mu \nu }A^{\alpha \beta }\rangle W_{RR1}(X)\,,
\end{equation}
where
\begin{equation*}
W_{RRi}(X) =\sum_{k}w_{RRi}^{(k)}X^{k} \qquad \text{with}\quad w_{RRi}^{(0)}
=0,\; w_{RRi}^{(1)}=O(N_{C}^{-1}), \qquad \text{for}\quad i=1,2\,.
\end{equation*}%
It gives rise to the operators%
\begin{eqnarray}
\hat{{\mathcal{O}}}^{VV}_{1}&=&\mathrm{i} \varepsilon _{\mu \nu \alpha \beta
}\langle V^{\mu \nu }V^{\alpha \beta }\rangle \langle \chi _{-}\rangle\,,
\notag \\
\hat{{\mathcal{O}}}^{AA}_{1}&=&\mathrm{i}\varepsilon _{\mu \nu \alpha \beta
}\langle A^{\mu \nu }A^{\alpha \beta }\rangle \langle \chi _{-}\rangle
\end{eqnarray}%
with the couplings of the order $O(N_{C}^{0})$ (the same order as the
analogous single trace operators and therefore included in our basis) and $%
O(N_{C}^{-3/2})$ operators%
\begin{eqnarray}
&&\varepsilon _{\mu \nu \alpha \beta }\langle V^{\mu \nu }V^{\alpha \beta
}\rangle \langle P\rangle\,,  \notag \\
&&\varepsilon _{\mu \nu \alpha \beta }\langle A^{\mu \nu }A^{\alpha \beta
}\rangle \langle P\rangle\,,
\end{eqnarray}
which are suppressed with respect to the single trace ones.

As the last step, we integrate out the resonance fields in order to get the
resonance contribution to the odd parity sector LECs of the resulting $\chi
PT$ Lagrangian. It can be done using the $O(p^{2})$ EOM for the resonance
fields and inserting their solution $R^{(2)}$ back to the $R\chi T$
Lagrangian. The general form reads%
\begin{equation}
R^{(2)}=\frac{1}{M_{R}^{2}}J_{R}^{(2)}\,,
\end{equation}%
where $J_{R}^{(2)}=O(p^{2})$ comes from the LO resonance Lagrangian (\ref{LR}%
). Because $J_{R}^{(2)}=O(N_{C}^{1/2})$, the order of the contribution of
the individual terms of the $R\chi T$ Lagrangian (with $\phi ^{0}$
integrated out) can be obtained counting the resonance fields as $%
O(N_{C}^{1/2})$. This gives finally the following simple bound on the order
of the contribution of the operator with $T$ traces, total $N_{\langle
P\rangle }$ factors $\langle P\rangle $ and total $N_{\langle \chi
_{-}\rangle }$ factors $\langle \chi _{-}\rangle $ originating in the to the
LECs%
\begin{equation}
2-T\leq n\leq 2-T+N_{\langle P\rangle }+2N_{\langle \chi _{-}\rangle }\,.
\end{equation}%
The lower bound represents the usual trace counting. Note however, that the
upper bound have to be taken with some caution, because it can be saturated
only in the case when all $\langle P\rangle $ and $\langle \chi _{-}\rangle $%
~traces appear as a consequence of the $\phi ^{0}$ dependence and that this $%
\phi ^{0}$dependence comes solely from the tilded operators and not from the
potentials. For a given operator these two conditions need not to be
satisfied simultaneously.

The fact that the $N_{C}$ of order some operators can be enhanced could
further complicate the usual way of the saturation of the ChPT LECs. Namely,
in the process of integrating out the resonances, it is assumed, that loops
can give only NLO contribution suppressed by the factor $1/N_{C}$ for each
loop. This counting could be apparently complicated by the enhanced
operators. Let us illustrate this point assuming the contribution of the
following term of the odd R$\chi$T Lagrangian $\widetilde{\mathcal{L}}$%
\begin{eqnarray}
\widetilde{\mathcal{L}} &=&\ldots +W_{2}^{AP}(X)\varepsilon _{\mu \nu \alpha
\beta }\langle \{A^{\mu \nu },\nabla^{\alpha }P\}\widetilde{u}^{\beta
}\rangle +\ldots  \notag \\
&=&\ldots -2w_{2}^{AP}\varepsilon _{\mu \nu \alpha \beta }\langle A^{\mu \nu
}\nabla^{\alpha }P\rangle \sqrt{\frac{2}{N_{F}}}\frac{D^{\beta }\phi ^{0}}{F}%
+\ldots
\end{eqnarray}%
with $w_{2}^{AP}=O(N_{C}^{0})$. This gives rise to the following enhanced $%
N_{C}$ term
\begin{equation}
-\frac{\mathrm{i}}{N_{F}}w_{2}^{AP}\frac{1}{M_{0}^{2}}\varepsilon _{\mu \nu
\alpha \beta }\langle A^{\mu \nu }\nabla^{\alpha }P\rangle \partial ^{\beta
}\langle \chi _{-}\rangle =O(N_{C})\,.
\end{equation}%
Apparently, this term contributes to the $O(p^{8})$ LECs, when the
resonances are integrated out at the \emph{tree level}. However, the bubble
with two such vertices gives a contribution to the $O(p^{6})$ operator $%
\partial _{\alpha }\langle \chi _{-}\rangle \partial ^{\alpha }\langle \chi
_{-}\rangle $ of the enhanced order $O(N_{C}^{2})$. The same is true also
for analogous operators from the even sector, \emph{e.g.}%
\begin{equation}
V_{1}^{SP}(X)\langle \{D_{\mu }S,P\}\widetilde{u}^{\mu }\rangle
=2v_{1}^{SP}\langle PD_{\mu }S\rangle \sqrt{\frac{2}{N_{F}}}\frac{D^{\mu
}\phi ^{0}}{F}+\ldots
\end{equation}%
with $v_{1}^{SP}=O(N_{C}^{0})$ which leads to the enhanced operator
\begin{equation}
\frac{\mathrm{i}}{N_{F}}v_{1}^{SP}\frac{1}{M_{0}^{2}}\langle P\nabla_{\mu
}S\rangle \partial ^{\mu }\langle \chi _{-}\rangle =O(N_{C})
\end{equation}%
counted as $O(p^{8})$ in the tree level saturation process. The bubble with
two vertices%
\begin{equation}
\frac{\mathrm{i}}{N_{F}}v_{1}^{SP}\frac{1}{M_{0}^{2}}\langle P\partial _{\mu
}S\rangle \partial ^{\mu }\langle \chi _{-}\rangle
\end{equation}%
leads to the expression%
\begin{eqnarray}
&&\frac{N_{F}^{2}}{2}\left( \frac{\mathrm{i}}{N_{F}}v_{1}^{SP}\frac{1}{%
M_{0}^{2}}\right) ^{2}\int \mathrm{d}^{d}x\mathrm{d}^{d}y\partial ^{\mu
}\langle \chi _{-}(x)\rangle \partial ^{\nu }\langle \chi _{-}(y)\rangle
\notag \\
&&\times \int \frac{\mathrm{d}^{d}k}{(2\pi )^{d}}\mathrm{e}^{\mathrm{i}%
k\cdot (x-y)}\frac{\mathrm{d}^{d}p}{(2\pi )^{d}}\frac{p_{\mu }p_{\nu }}{%
\left( p^{2}-M_{S}^{2}+i0\right) \left( (p-k)^{2}-M_{P}^{2}+i0\right) } \\
&=&\frac{\mathrm{i}}{2}\left( \frac{v_{1}^{SP}}{M_{0}^{2}}\right) ^{2}\int
\mathrm{d}^{d}x\partial ^{\mu }\langle \chi _{-}(x)\rangle \partial _{\mu
}\langle \chi _{-}(x)\rangle \frac{\left( M_{P}^{2}\right) ^{2-\varepsilon
}-\left( M_{S}^{2}\right) ^{2-\varepsilon }}{M_{P}^{2}-M_{S}^{2}}\frac{1}{%
32\pi ^{2}}\Gamma (\varepsilon -2)\left( 4\pi \right) ^{\varepsilon }  \notag
\\
&&+O(p^{8})  \notag
\end{eqnarray}%
and (after addition of appropriate counterterm) results in the following $%
O(N_{C}^{2})$ contribution to the coupling $C_{\partial _{\alpha }\langle
\chi _{-}\rangle \partial ^{\alpha }\langle \chi _{-}\rangle }$ associated
with $O(p^{6})$ operator $\partial _{\alpha }\langle \chi _{-}\rangle
\partial ^{\alpha }\langle \chi _{-}\rangle $%
\begin{equation}
C_{\partial _{\alpha }\langle \chi _{-}\rangle \partial ^{\alpha }\langle
\chi _{-}\rangle }^{PS-loop}=-\frac{1}{64\pi ^{2}}\left( \frac{v_{1}^{SP}}{%
M_{0}^{2}}\right) ^{2}\frac{M_{P}^{4}\left( \ln \frac{M_{P}^{2}}{\mu ^{2}}%
+\gamma -\frac{1}{2}\right) -M_{S}^{4}\left( \ln \frac{M_{S}^{2}}{\mu ^{2}}%
+\gamma -\frac{1}{2}\right) }{M_{P}^{2}-M_{S}^{2}}\,.
\end{equation}

Though the above loop contribution are enhanced by the factor $N_{C}^{2}$
with respect to the naive trace counting, it does not mean, that loop
counting fails. The reason is that the LO contribution to $C_{\partial
_{\alpha }\langle \chi _{-}\rangle \partial ^{\alpha }\langle \chi
_{-}\rangle }$ that comes from the tree level and originates in the kinetic
term of the field $\phi ^{0}$%
\begin{equation}
\frac{1}{2}\partial _{\mu }\phi ^{0}\cdot \partial ^{\mu }\phi
^{0}\rightarrow -\frac{1}{2}\left( \frac{F}{2M_{0}^{2}\sqrt{2N_{F}}}\right)
^{2}\partial _{\alpha }\langle \chi _{-}\rangle \partial ^{\alpha }\langle
\chi _{-}\rangle =O(N_{C}^{3})
\end{equation}%
so that the loop contribution is suppressed by $1/N_{C}$ as usual.

Let us finally comment briefly on one point, which also might lead to
confusion. In \cite{Cirigliano:2006hb}, the following operators are abandon
using the large $N_{C}$ arguments, namely
\begin{align}
&\mathrm{i}\langle Pu_{\mu }u^{\mu }\rangle \langle \chi _{-}\rangle\,,
\notag \\
&\mathrm{i}\langle SP\rangle \langle \chi _{-}\rangle\,,  \notag \\
&\mathrm{i}\langle \nabla_{\mu }\nabla^{\mu }\chi _{-}\rangle \langle
P\rangle\,.
\end{align}%
These can be, however, derived from the operators (before doing any
transformations)%
\begin{align}
&\mathrm{i}\langle P\widetilde{u}_{\mu }\widetilde{u}^{\mu }\rangle V(X)\,,
\notag \\
&\mathrm{i}\langle SP\rangle V(X)\,,  \notag \\
&\mathrm{i}\langle \nabla_{\mu }\nabla^{\mu }\widetilde{\chi }_{+}\rangle
V(X)\,,
\end{align}%
by means of integrating out the field $\phi ^{0}$, which appears from the
potential for the first two operators (and saturates therefore the lower
bound of (\ref{bound1})) and from the building block $\widetilde{\chi }_{+}$
for the last one (and corresponds therefore to the upper bound of (\ref%
{bound1})). According to our rules the operators are of the order $%
O(N_{C}^{1/2})$, $O(N_{C}^{0})$ and $O(N_{C}^{1/2})$ respectively (as the
similar operators without additional trace) and all of them contribute
therefore at the $O(N_{C})$ order of the LECs of the effective chiral
Lagrangian staying at the operators%
\begin{align}
&\langle \chi _{-}u_{\mu }u^{\mu }\rangle \langle \chi _{-}\rangle\,,  \notag
\\
&\langle \chi _{+}\chi _{-}\rangle \langle \chi _{-}\rangle\,,  \notag \\
&\langle \nabla_{\mu }\nabla^{\mu }\chi _{-}\rangle \langle \chi
_{-}\rangle\,.
\end{align}%
However, these operators can be derived analogously as above from%
\begin{align}
&\langle \widetilde{\chi }_{+}\widetilde{u}_{\mu }\widetilde{u}^{\mu
}\rangle\,,  \notag \\
&\langle \widetilde{\chi }_{+}\widetilde{\chi }_{+}\rangle\,,  \notag \\
&\langle \nabla_{\mu }\nabla^{\mu }\widetilde{\chi }_{+}\rangle\,,
\end{align}%
by the process which saturates the upper bound of (\ref{bound1}) and results
in the order $O(N_{C}^{2})$. The abandoned operators lead therefore to the
NLO contribution to the corresponding LECs.

\section{Field redefinition}

\label{apB}

As we have discussed in detail in Section \ref{sec3}, by means of
appropriate  field redefinition we can effectively eliminate
subset of the $O(p^6)$ operators from the Lagrangian
$\mathcal{L}_{R\chi T}^{(6,\text{ odd})}$ and shift their
influence on the ChPT LECs into the effective coefficients
$\overline{\kappa _{i}^{X}}$ which stay at the remaining operators
of the chiral order $O(p^6)$ and higher. As a consequence, the
$O(p^6)$ LECs resulting from the process of integrating out the
resonance fields from the Lagrangian $\mathcal{L}_{R\chi T}$
depend only on these effective couplings $\overline{\kappa
_{i}^{X}}$ which are particular linear combinations of the
original resonance couplings $\kappa _{i}^{X}$. In order to
identify these relevant combinations and the redundant operators,
we can proceed in several steps.

\subsection{Elimination of $\mathcal{O}_{1,2}^{VV}$, $\mathcal{O}_{1,2}^{AA}
$, $\mathcal{O}^{VVP}$ and $\mathcal{O}^{AAP}$}

With the field redefinitions%
\begin{eqnarray*}
V_{\mu \nu } &\rightarrow &V_{\mu \nu }-\frac{2}{M_{V}^{2}}\varepsilon _{\mu
\nu \alpha \beta }\left( \mathrm{i}\kappa _{1}^{VV}\langle \chi _{-}\rangle
V^{\alpha \beta }+\mathrm{i}\kappa _{2}^{VV}\{\chi _{-},V^{\alpha \beta }\}+%
\frac{1}{2}\kappa ^{VVP}\{P,V^{\alpha \beta }\}\right)\,, \\
A_{\mu \nu } &\rightarrow &A_{\mu \nu }-\frac{2}{M_{A}^{2}}\varepsilon _{\mu
\nu \alpha \beta }\left( \mathrm{i}\kappa _{1}^{AA}\langle \chi _{-}\rangle
A^{\alpha \beta }+\mathrm{i}\kappa _{2}^{AA}\{\chi _{-},A^{\alpha \beta }\}+%
\frac{1}{2}\kappa ^{AAP}\{P,A^{\alpha \beta }\}\right)\,,
\end{eqnarray*}%
we get for the $O(p^{4})$ part of the Lagrangian%
\begin{eqnarray*}
&&\mathcal{L}_{RR,kin}^{(4)}+\mathcal{L}_{R}^{(4)}\rightarrow \mathcal{L}%
_{RR,kin}^{(4)}+\mathcal{L}_{R}^{(4)} \\
&=&\mathcal{L}_{RR,kin}^{(4)}+\mathcal{L}_{R}^{(4)} \\
&&-\kappa _{1}^{VV}\mathcal{O}_{1}^{VV}-\kappa _{2}^{VV}\mathcal{O}%
_{2}^{VV}-\kappa ^{VVP}\mathcal{O}^{VVP}-\kappa _{1}^{AA}\mathcal{O}%
_{1}^{AA}-\kappa _{2}^{AA}\mathcal{O}_{2}^{AA}-\kappa ^{AAP}\mathcal{O}^{AAP}
\\
&&-\frac{F_{V}}{\sqrt{2}M_{V}^{2}}\left( \kappa _{1}^{VV}\mathcal{O}%
_{13}^{V}+\kappa _{2}^{VV}\mathcal{O}_{14}^{V}+\frac{1}{2}\kappa ^{VVP}%
\mathcal{O}_{3}^{PV}\right) \\
&&-\frac{\mathrm{i}G_{V}}{\sqrt{2}M_{V}^{2}}\left( 2\mathrm{i}\kappa
_{1}^{VV}\mathcal{O}_{18}^{V}+2\mathrm{i}\kappa _{2}^{VV}\mathcal{O}_{9}^{V}-%
\mathrm{i}\kappa ^{VVP}\mathcal{O}_{1}^{PV}\right) \\
&&-\frac{F_{A}}{\sqrt{2}M_{A}^{2}}\left( \kappa _{1}^{AA}\mathcal{O}%
_{9}^{A}+\kappa _{2}^{AA}\mathcal{O}_{11}^{A}+\frac{1}{2}\kappa ^{AAP}%
\mathcal{O}_{1}^{PA}\right) +O(p^{8})\,.
\end{eqnarray*}%
At the same time, the same redefinition applied to $\mathcal{L}_{R\chi
T}^{(6,\text{ odd})}$ generates only the additional terms of the order $%
O(p^{8})$ and higher, which can be neglected as described above. We can thus
eliminate the operators $\mathcal{O}_{1,2}^{VV}$, , $\mathcal{O}^{VVP}$, $%
\mathcal{O}_{1,2}^{AA}$ and $\mathcal{O}^{AAP}$ and include their influence
on the $O(p^{6})$ LECs effectively into the constants $\overline{\kappa
_{13}^{V}}$, $\overline{\kappa _{14}^{V}}$, $\overline{\kappa _{3}^{PV}}$, $%
\overline{\kappa _{18}^{V}}$, $\overline{\kappa _{9}^{V}}$, $\overline{%
\kappa _{1}^{PV}}$ $\ $and $\overline{\kappa _{9}^{A}}$, $\overline{\kappa
_{11}^{A}}$, $\overline{\kappa _{1}^{PA}}$.

\subsection{Elimination of $\mathcal{O}_{i}^{VA}$ and $\mathcal{O}^{VAS}$}

In the same way we can eliminate also the mixed bilinear terms using the
field redefinition%
\begin{eqnarray*}
V_{\mu \nu } &\rightarrow &V_{\mu \nu }-\frac{1}{M_{V}^{2}}\varepsilon _{\mu
\nu \alpha \sigma }\left( \mathrm{i}\kappa _{1}^{VA}g_{\beta }^{\sigma
}[A^{\alpha \beta },u^{\rho }u_{\rho }]+\mathrm{i}\kappa _{2}^{VA}(A^{\alpha
\beta }u_{\beta }u^{\sigma }-u^{\sigma }u_{\beta }A^{\alpha \beta })\right.
\\
&&\left. +\mathrm{i}\kappa _{3}^{VA}(A^{\alpha \beta }u^{\sigma }u_{\beta
}-u_{\beta }u^{\sigma }A^{\alpha \beta })+\mathrm{i}\kappa
_{4}^{VA}(u_{\beta }A^{\alpha \beta }u^{\sigma }-u^{\sigma }A^{\alpha \beta
}u_{\beta })\right. \\
&&\left. +\kappa _{5}^{VA}\{A^{\alpha \beta },f_{+}^{\sigma \rho }\}g_{\beta
\rho }+\mathrm{i}\kappa _{6}^{VA}[A^{\alpha \beta },\chi _{+}]g_{\beta
}^{\sigma }+\mathrm{i}\kappa ^{VAS}[A^{\alpha \beta },S]g_{\beta }^{\sigma
}\right)\,, \\
A_{\mu \nu } &\rightarrow &A_{\mu \nu }-\frac{1}{M_{A}^{2}}\varepsilon
_{\alpha \beta \mu \sigma }\left( \mathrm{i}\kappa _{1}^{VA}g_{\nu }^{\sigma
}[u^{\rho }u_{\rho },V^{\alpha \beta }]+\mathrm{i}\kappa _{2}^{VA}(u^{\nu
}u^{\sigma }V^{\alpha \beta }-V^{\alpha \beta }u^{\sigma }u^{\nu })\right. \\
&&\left. +\mathrm{i}\kappa _{3}^{VA}(u^{\sigma }u^{\nu }V^{\alpha \beta
}-V^{\alpha \beta }u^{\nu }u^{\sigma })+\mathrm{i}\kappa _{4}^{VA}(u^{\sigma
}V^{\alpha \beta }u^{\nu }-u^{\nu }V^{\alpha \beta }u^{\sigma })\right. \\
&&\left. +\kappa _{5}^{VA}\{V^{\alpha \beta },f_{+}^{\sigma \rho }\}g_{\rho
\nu }+\mathrm{i}\kappa _{6}^{VA}[\chi _{+},V^{\alpha \beta }]g_{\nu
}^{\sigma }+\mathrm{i}\kappa ^{VAS}[S,V^{\alpha \beta }]g_{\nu }^{\sigma
}\right)\,.
\end{eqnarray*}%
We get then%
\begin{eqnarray*}
&&\frac{1}{4}M_{V}^{2}\langle V^{\mu \nu }V_{\mu \nu }\rangle +\frac{1}{4}%
M_{A}^{2}\langle A^{\mu \nu }A_{\mu \nu }\rangle \rightarrow \frac{1}{4}%
M_{V}^{2}\langle V^{\mu \nu }V_{\mu \nu }\rangle +\frac{1}{4}%
M_{A}^{2}\langle A^{\mu \nu }A_{\mu \nu }\rangle \\
&&-\kappa _{1}^{VA}\mathcal{O}_{1}^{VA}-\kappa _{2}^{VA}\mathcal{O}%
_{2}^{VA}-\kappa _{3}^{VA}\mathcal{O}_{3}^{VA}-\kappa _{4}^{VA}\mathcal{O}%
_{4}^{VA}-\kappa _{5}^{VA}\mathcal{O}_{5}^{VA}-\kappa _{6}^{VA}\mathcal{O}%
_{6}^{VA}-\kappa ^{VAS}\mathcal{O}^{VAS}
\end{eqnarray*}%
and the operators $\mathcal{O}_{i}^{VA}$ and $\mathcal{O}^{VAS}$ are thus
eliminated. The only relevant additional effect of the redefinition comes
from transformation of $\mathcal{L}_{R}^{(4)}$
\begin{align*}
&\frac{F_{V}}{2\sqrt{2}}\langle V_{\mu \nu }f_{+}^{\mu \nu }\rangle
\;\rightarrow\;
\frac{F_{V}}{2\sqrt{2}}\langle V_{\mu \nu }f_{+}^{\mu \nu
}\rangle -\frac{F_{V}}{2\sqrt{2}M_{V}^{2}}\Bigl[ -\kappa _{1}^{VA}\mathcal{O}%
_{4}^{A}+\kappa _{2}^{VA} (\mathcal{O}_{6}^{A}-\frac{1}{2}\mathcal{O}_{4}^{A})\\
&\qquad\qquad\qquad\qquad
+\kappa_{3}^{VA}( \mathcal{O}_{5}^{A}-\frac{1}{2}%
\mathcal{O}_{4}^{A})
+\kappa _{4}^{VA}\mathcal{O}_{7}^{A}-\kappa _{6}^{VA}\mathcal{O}%
_{14}^{A}+\kappa ^{VAS}\mathcal{O}_{1}^{SA}\Bigr]\,, \\
&\frac{F_{A}}{2\sqrt{2}}\langle A_{\mu \nu }f_{-}^{\mu \nu }\rangle
\;\rightarrow\; \frac{F_{A}}{2\sqrt{2}}\langle A_{\mu \nu }f_{-}^{\mu \nu}\rangle
-\frac{F_{A}}{2\sqrt{2}M_{A}^{2}}\Bigl[ \kappa _{1}^{VA}\mathcal{O}_{5}^{V}+\kappa _{2}^{VA}\mathcal{O}_{8}^{V}+\kappa _{3}^{VA}\mathcal{O}%
_{6}^{V}+\kappa _{4}^{VA}\mathcal{O}_{7}^{V} \\
&\qquad\qquad\qquad\qquad
-\kappa _{5}^{VA}\mathcal{O}_{11}^{V}+\kappa_{6}^{VA}\mathcal{O}%
_{15}^{V}-\kappa ^{VAS}\mathcal{O}_{1}^{SV}\Bigr]\,, \\
&\frac{\mathrm{i}G_{V}}{2\sqrt{2}}\langle V_{\mu \nu }[u^{\mu },u^{\nu
}]\rangle\; \rightarrow\; \frac{\mathrm{i}G_{V}}{2\sqrt{2}}\langle V_{\mu \nu
}[u^{\mu },u^{\nu }]\rangle
+\frac{G_{V}}{2\sqrt{2}M_{V}^{2}}\Bigl[ -2\kappa
_{1}^{VA}\mathcal{O}_{1}^{A}+\kappa _{2}^{VA}( \mathcal{O}_{2}^{A}-\mathcal{O%
}_{1}^{A})\\ & \qquad\qquad\qquad\qquad
+\kappa _{3}^{VA}( \mathcal{O}_{2}^{A}-\mathcal{O}_{1}^{A})
+\kappa _{4}^{VA}( \mathcal{O}_{1}^{A}-\mathcal{O}_{2}^{A})
\\&\qquad\qquad\qquad\qquad
+\kappa
_{5}^{VA}( \mathcal{O}_{5}^{A}-\mathcal{O}_{6}^{A}) +2\kappa _{6}^{VA}%
\mathcal{O}_{13}^{A}+2\kappa ^{VAS}\mathcal{O}_{2}^{SA}\Bigr]\, .
\end{align*}%
Here we have used that%
\begin{equation*}
\langle \{A^{\alpha \beta },f_{+}^{\sigma \rho }\}f_{+}^{\mu \nu }\rangle
g_{\beta \rho }\varepsilon _{\mu \nu \alpha \sigma }=0
\end{equation*}%
and other similar consequences of the Shouten identity.

\subsection{Elimination of $\mathcal{O}_{1}^{PA}$, $\mathcal{O}_{1}^{SV}$, $%
\mathcal{O}_{i}^{PV}$and $\mathcal{O}_{i}^{SA}$}

Finally we can further eliminate another terms by the redefinitions%
\begin{eqnarray*}
S &\rightarrow &S+\frac{1}{2M_{S}^{2}}\varepsilon _{\mu \nu \alpha \beta
}\left( \mathrm{i}\kappa _{1}^{SA}[f_{+}^{\alpha \beta },A^{\mu \nu
}]+\kappa _{2}^{SA}[u^{\alpha }u^{\beta },A^{\mu \nu }]\right)\,, \\
P &\rightarrow &P+\frac{1}{2M_{P}^{2}}\varepsilon _{\mu \nu \alpha \beta
}\left( \kappa _{1}^{PA}\{A^{\mu \nu },f_{-}^{\alpha \beta }\}+\mathrm{i}%
\kappa _{1}^{PV}\{V^{\mu \nu },u^{\alpha }u^{\beta }\}\right.\\
&&\qquad\qquad\qquad\quad\left.+\mathrm{i}\kappa_{2}^{PV}u^{\beta }V^{\mu \nu }u^{\alpha }+\kappa _{3}^{PV}\{V^{\mu \nu
},f_{+}^{\alpha \beta }\}\right)\,, \\
A_{\mu \nu } &\rightarrow &A_{\mu \nu }-\frac{1}{M_{A}^{2}}\varepsilon _{\mu
\nu \alpha \beta }\left( \mathrm{i}\kappa _{1}^{SA}[S,f_{+}^{\alpha \beta
}]+\kappa _{2}^{SA}[S,u^{\alpha }u^{\beta }]+\kappa
_{1}^{PA}\{P,f_{-}^{\alpha \beta }\}\right) \\
V_{\mu \nu } &\rightarrow &V_{\mu \nu }-\frac{1}{M_{V}^{2}}\varepsilon _{\mu
\nu \alpha \beta }\left( \mathrm{i}\kappa _{1}^{PV}\{P,u^{\alpha }u^{\beta
}\}+\mathrm{i}\kappa _{2}^{PV}u^{\alpha }Pu^{\beta }+\kappa
_{3}^{PV}\{P,f_{+}^{\alpha \beta }\}\right)\,.
\end{eqnarray*}%
We get then%
\begin{eqnarray*}
&&-\frac{1}{2}M_{P}^{2}\langle PP\rangle -\frac{1}{2}M_{S}^{2}\langle
SS\rangle +\frac{1}{4}M_{A}^{2}\langle A^{\mu \nu }A_{\mu \nu }\rangle +%
\frac{1}{4}M_{V}^{2}\langle V^{\mu \nu }V_{\mu \nu }\rangle \\
&\rightarrow &-\frac{1}{2}M_{P}^{2}\langle PP\rangle -\frac{1}{2}%
M_{S}^{2}\langle SS\rangle +\frac{1}{4}M_{A}^{2}\langle A^{\mu \nu }A_{\mu
\nu }\rangle +\frac{1}{4}M_{V}^{2}\langle V^{\mu \nu }V_{\mu \nu }\rangle \\
&&-\kappa _{1}^{PA}\mathcal{O}_{1}^{PA}-\kappa _{1}^{PV}\mathcal{O}%
_{1}^{PV}-\kappa _{2}^{PV}\mathcal{O}_{2}^{PV}-\kappa _{3}^{PV}\mathcal{O}%
_{3}^{PV}-\kappa _{1}^{SA}\mathcal{O}_{1}^{SA}-\kappa _{2}^{SA}\mathcal{O}%
_{2}^{SA}-\kappa _{1}^{SV}\mathcal{O}_{1}^{SV}\,,
\end{eqnarray*}%
therefore the operators $\mathcal{O}_{i}^{SA}$, $\mathcal{O}_{i}^{PV}$ and $%
\mathcal{O}_{1}^{PA}$ are eliminated. We get additional contributions

\begin{eqnarray*}
c_{d}\langle Su^{\mu }u_{\mu }\rangle &\rightarrow &c_{d}\langle Su^{\mu
}u_{\mu }\rangle +\frac{c_{d}}{2M_{S}^{2}}\left( -\kappa _{1}^{SA}\mathcal{O}%
_{4}^{A}-\kappa _{2}^{SA}\mathcal{O}_{1}^{A}-\kappa _{1}^{SV}\mathcal{O}%
_{5}^{V}\right) \\
c_{m}\langle S\chi _{+}\rangle &\rightarrow &c_{m}\langle S\chi _{+}\rangle +%
\frac{c_{m}}{2M_{S}^{2}}\left( -\kappa _{1}^{SA}\mathcal{O}_{14}^{A}+\kappa
_{2}^{SA}\mathcal{O}_{13}^{A}-\kappa _{1}^{SV}\mathcal{O}_{15}^{V}\right) \\
\mathrm{i}d_{m}\langle P\chi _{-}\rangle &\rightarrow &\mathrm{i}%
d_{m}\langle P\chi _{-}\rangle +\frac{d_{m}}{2M_{P}^{2}}(\kappa _{1}^{PA}%
\mathcal{O}_{11}^{A}-\kappa _{1}^{PV}\mathcal{O}_{9}^{V}-\kappa _{2}^{PV}%
\mathcal{O}_{10}^{V}+\kappa _{3}^{PV}\mathcal{O}_{14}^{V}) \\
\mathrm{i}\frac{d_{m0}}{N_{F}}\langle P\rangle \langle \chi _{-}\rangle
&\rightarrow &\mathrm{i}\frac{d_{m0}}{N_{F}}\langle P\rangle \langle \chi
_{-}\rangle +\frac{d_{m0}}{2N_{F}M_{P}^{2}}\left( 2\kappa _{1}^{PA}\mathcal{O%
}_{9}^{A}-2\kappa _{1}^{PV}\mathcal{O}_{18}^{V}-\kappa _{2}^{PV}\mathcal{O}%
_{18}^{V}+2\kappa _{3}^{PV}\mathcal{O}_{13}^{V}\right) \\
\frac{F_{A}}{2\sqrt{2}}\langle A_{\mu \nu }f_{-}^{\mu \nu }\rangle
&\rightarrow &\frac{F_{A}}{2\sqrt{2}}\langle A_{\mu \nu }f_{-}^{\mu \nu
}\rangle -\frac{F_{A}}{2\sqrt{2}M_{A}^{2}}\left( \kappa _{1}^{SA}\mathcal{O}%
_{2}^{S}-\kappa _{2}^{SA}\mathcal{O}_{1}^{S}+\kappa _{1}^{PA}\mathcal{O}%
_{1}^{P}\right) \\
\frac{F_{V}}{2\sqrt{2}}\langle V_{\mu \nu }f_{+}^{\mu \nu }\rangle
&\rightarrow &\frac{F_{V}}{2\sqrt{2}}\langle V_{\mu \nu }f_{+}^{\mu \nu
}\rangle -\frac{F_{V}}{2\sqrt{2}M_{V}^{2}}\left( \kappa _{1}^{PV}\mathcal{O}%
_{3}^{P}-\kappa _{2}^{PV}\mathcal{O}_{2}^{P}+\kappa _{3}^{PV}\mathcal{O}%
_{5}^{P}-\kappa _{1}^{SV}\mathcal{O}_{2}^{S}\right) \\
\frac{\mathrm{i}G_{V}}{2\sqrt{2}}\langle V_{\mu \nu }[u^{\mu },u^{\nu
}]\rangle &\rightarrow &\frac{\mathrm{i}G_{V}}{2\sqrt{2}}\langle V_{\mu \nu
}[u^{\mu },u^{\nu }]\rangle +\frac{G_{V}}{2\sqrt{2}M_{V}^{2}}\left( 4\kappa
_{1}^{PV}\mathcal{O}_{4}^{P}-2\kappa _{2}^{PV}\mathcal{O}_{4}^{P}-2\kappa
_{3}^{PV}\mathcal{O}_{3}^{P}+2\kappa _{1}^{SV}\mathcal{O}_{1}^{S}\right)
\end{eqnarray*}

\subsection{The effective couplings $\overline{\kappa _{i}^{X}}$}

\label{aprel}

Putting the result of previous subsections together we get the parameters $%
\overline{\kappa _{i}^{X}}$ of the reparameterized and truncated Lagrangian $%
\overline{\mathcal{L}_{R\chi T}^{(6,\text{ odd})}}$, which is relevant for
the saturation of ChPT LECs, as a functions of the parameters $\kappa
_{i}^{X}$. As we have discussed above, the LECs have to depend on the
couplings $\kappa _{i}^{X}$ of the original Lagrangian $\mathcal{L}_{R\chi
T}^{(6,\text{ odd})}$ only through their particular combinations $\overline{%
\kappa _{i}^{X}}$. We have proved this by means of direct calculation as a
nontrivial check of the formulae (\ref{satCW}).
\begin{eqnarray*}
\overline{\kappa _{1}^{V}} &=&\kappa _{1}^{V} \\
\overline{\kappa _{2}^{V}} &=&\kappa _{2}^{V} \\
\overline{\kappa _{3}^{V}} &=&\kappa _{3}^{V} \\
\overline{\kappa _{4}^{V}} &=&\kappa _{4}^{V} \\
\overline{\kappa _{5}^{V}} &=&\kappa _{5}^{V}-\frac{c_{d}}{2M_{S}^{2}}\left(
\kappa _{1}^{SV}+\frac{F_{A}}{2\sqrt{2}M_{A}^{2}}\kappa ^{VAS}\right) -\frac{%
F_{A}}{2\sqrt{2}M_{A}^{2}}\kappa _{1}^{VA} \\
\overline{\kappa _{6}^{V}} &=&\kappa _{6}^{V}-\frac{F_{A}}{2\sqrt{2}M_{A}^{2}%
}\kappa _{3}^{VA} \\
\overline{\kappa _{7}^{V}} &=&\kappa _{7}^{V}-\frac{F_{A}}{2\sqrt{2}M_{A}^{2}%
}\kappa _{4}^{VA} \\
\overline{\kappa _{8}^{V}} &=&\kappa _{8}^{V}-\frac{F_{A}}{2\sqrt{2}M_{A}^{2}%
}\kappa _{2}^{VA} \\
\overline{\kappa _{9}^{V}} &=&\kappa _{9}^{V}+\frac{2G_{V}\kappa _{2}^{VV}}{%
\sqrt{2}M_{V}^{2}}-\frac{d_{m}}{2M_{P}^{2}}\left( \kappa _{1}^{PV}-\frac{%
2G_{V}\kappa ^{VVP}}{2\sqrt{2}M_{V}^{2}}\right) \\
\overline{\kappa _{10}^{V}} &=&\kappa _{10}^{V}-\frac{d_{m}}{2M_{P}^{2}}%
\kappa _{2}^{PV} \\
\overline{\kappa _{11}^{V}} &=&\kappa _{11}^{V}+\frac{F_{A}}{2\sqrt{2}%
M_{A}^{2}}\kappa _{5}^{VA} \\
\overline{\kappa _{12}^{V}} &=&\kappa _{12}^{V} \\
\overline{\kappa _{13}^{V}} &=&\kappa _{13}^{V}+\frac{d_{m0}}{N_{F}M_{P}^{2}}%
\left( \kappa _{3}^{PV}-\frac{F_{V}\kappa ^{VVP}}{2\sqrt{2}M_{V}^{2}}\right)
-\frac{F_{V}\kappa _{1}^{VV}}{\sqrt{2}M_{V}^{2}} \\
\overline{\kappa _{14}^{V}} &=&\kappa _{14}^{V}-\frac{F_{V}\kappa _{2}^{VV}}{%
\sqrt{2}M_{V}^{2}}+\frac{d_{m}}{2M_{P}^{2}}\left( \kappa _{3}^{PV}-\frac{%
F_{V}\kappa ^{VVP}}{2\sqrt{2}M_{V}^{2}}\right) \\
\overline{\kappa _{15}^{V}} &=&\kappa _{15}^{V}-\frac{c_{m}}{2M_{S}^{2}}%
\left( \kappa _{1}^{SV}+\frac{F_{A}}{2\sqrt{2}M_{A}^{2}}\kappa ^{VAS}\right)
-\frac{F_{A}}{2\sqrt{2}M_{A}^{2}}\kappa _{6}^{VA} \\
\overline{\kappa _{16}^{V}} &=&\kappa _{16}^{V} \\
\overline{\kappa _{17}^{V}} &=&\kappa _{17}^{V} \\
\overline{\kappa _{18}^{V}} &=&\kappa _{18}^{V}-\frac{d_{m0}}{2N_{F}M_{P}^{2}%
}\left( 2\left( \kappa _{1}^{PV}-\frac{2G_{V}\kappa ^{VVP}}{2\sqrt{2}%
M_{V}^{2}}\right) +\kappa _{2}^{PV}\right) +\frac{2G_{V}\kappa _{1}^{VV}}{%
\sqrt{2}M_{V}^{2}} \\
\overline{\kappa _{1}^{A}} &=&\kappa _{1}^{A}-\frac{c_{d}}{2M_{S}^{2}}\left(
\kappa _{2}^{SA}+\frac{G_{V}}{\sqrt{2}M_{V}^{2}}\kappa ^{VAS}\right) -\frac{%
G_{V}}{2\sqrt{2}M_{V}^{2}}\left( 2\kappa _{1}^{VA}+\kappa _{2}^{VA}+\kappa
_{3}^{VA}-\kappa _{4}^{VA}\right) \\
\overline{\kappa _{2}^{A}} &=&\kappa _{2}^{A}+\frac{G_{V}}{2\sqrt{2}M_{V}^{2}%
}\left( \kappa _{2}^{VA}+\kappa _{3}^{VA}-\kappa _{4}^{VA}\right) \\
\overline{\kappa _{3}^{A}} &=&\kappa _{3}^{A} \\
\overline{\kappa _{4}^{A}} &=&\kappa _{4}^{A}-\frac{c_{d}}{2M_{S}^{2}}\left(
\kappa _{1}^{SA}-\frac{F_{V}}{2\sqrt{2}M_{V}^{2}}\kappa ^{VAS}\right) +\frac{%
F_{V}}{2\sqrt{2}M_{V}^{2}}\left( \kappa _{1}^{VA}+\frac{1}{2}\kappa
_{2}^{VA}+\frac{1}{2}\kappa _{3}^{VA}\right) \\
\overline{\kappa _{5}^{A}} &=&\kappa _{5}^{A}-\frac{F_{V}}{2\sqrt{2}M_{V}^{2}%
}\kappa _{3}^{VA}+\frac{G_{V}}{2\sqrt{2}M_{V}^{2}}\kappa _{5}^{VA} \\
\overline{\kappa _{6}^{A}} &=&\kappa _{6}^{A}-\frac{F_{V}}{2\sqrt{2}M_{V}^{2}%
}\kappa _{2}^{VA}-\frac{G_{V}}{2\sqrt{2}M_{V}^{2}}\kappa _{5}^{VA} \\
\overline{\kappa _{7}^{A}} &=&\kappa _{7}^{A}-\frac{F_{V}}{2\sqrt{2}M_{V}^{2}%
}\kappa _{4}^{VA} \\
\overline{\kappa _{8}^{A}} &=&\kappa _{8}^{A} \\
\overline{\kappa _{9}^{A}} &=&\kappa _{9}^{A}+\frac{d_{m0}}{N_{F}M_{P}^{2}}%
\left( \kappa _{1}^{PA}-\frac{F_{A}\kappa ^{AAP}}{2\sqrt{2}M_{A}^{2}}\right)
-\frac{F_{A}\kappa _{1}^{AA}}{\sqrt{2}M_{A}^{2}} \\
\overline{\kappa _{10}^{A}} &=&\kappa _{10}^{A} \\
\overline{\kappa _{11}^{A}} &=&\kappa _{11}^{A}-\frac{F_{A}\kappa _{2}^{AA}}{%
\sqrt{2}M_{A}^{2}}+\frac{d_{m}}{2M_{P}^{2}}\left( \kappa _{1}^{PA}-\frac{%
F_{A}\kappa ^{AAP}}{2\sqrt{2}M_{A}^{2}}\right) \\
\overline{\kappa _{12}^{A}} &=&\kappa _{12}^{A} \\
\overline{\kappa _{13}^{A}} &=&\kappa _{13}^{A}+\frac{G_{V}}{\sqrt{2}%
M_{V}^{2}}\kappa _{6}^{VA}+\frac{c_{m}}{2M_{S}^{2}}\left( \kappa _{2}^{SA}+%
\frac{G_{V}}{\sqrt{2}M_{V}^{2}}\kappa ^{VAS}\right) \\
\overline{\kappa _{14}^{A}} &=&\kappa _{14}^{A}+\frac{F_{V}}{2\sqrt{2}%
M_{V}^{2}}\kappa _{6}^{VA}-\frac{c_{m}}{2M_{S}^{2}}\left( \kappa _{1}^{SA}-%
\frac{F_{V}}{2\sqrt{2}M_{V}^{2}}\kappa ^{VAS}\right) \\
\overline{\kappa _{15}^{A}} &=&\kappa _{15}^{A} \\
\overline{\kappa _{16}^{A}} &=&\kappa _{16}^{A} \\
\overline{\kappa _{1}^{S}} &=&\kappa _{1}^{S}+\frac{F_{A}}{2\sqrt{2}M_{A}^{2}%
}\left( \kappa _{2}^{SA}+\frac{G_{V}}{\sqrt{2}M_{V}^{2}}\kappa ^{VAS}\right)
+\frac{G_{V}}{\sqrt{2}M_{V}^{2}}\left( \kappa _{1}^{SV}+\frac{F_{A}}{2\sqrt{2%
}M_{A}^{2}}\kappa ^{VAS}\right) \\
\overline{\kappa _{2}^{S}} &=&\kappa _{2}^{S}-\frac{F_{A}}{2\sqrt{2}M_{A}^{2}%
}\left( \kappa _{1}^{SA}-\frac{F_{V}}{2\sqrt{2}M_{V}^{2}}\kappa
^{VAS}\right) +\frac{F_{V}}{2\sqrt{2}M_{V}^{2}}\left( \kappa _{1}^{SV}+\frac{%
F_{A}}{2\sqrt{2}M_{A}^{2}}\kappa ^{VAS}\right) \\
\overline{\kappa _{1}^{P}} &=&\kappa _{1}^{P}-\frac{F_{A}}{2\sqrt{2}M_{A}^{2}%
}\left( \kappa _{1}^{PA}-\frac{F_{A}\kappa ^{AAP}}{2\sqrt{2}M_{A}^{2}}\right)
\\
\overline{\kappa _{2}^{P}} &=&\kappa _{2}^{P}+\frac{F_{V}}{2\sqrt{2}M_{V}^{2}%
}\kappa _{2}^{PV} \\
\overline{\kappa _{3}^{P}} &=&\kappa _{3}^{P}-\frac{F_{V}}{2\sqrt{2}M_{V}^{2}%
}\left( \kappa _{1}^{PV}-\frac{2G_{V}\kappa ^{VVP}}{2\sqrt{2}M_{V}^{2}}%
\right) -\frac{G_{V}}{\sqrt{2}M_{V}^{2}}\left( \kappa _{3}^{PV}-\frac{%
F_{V}\kappa ^{VVP}}{2\sqrt{2}M_{V}^{2}}\right) \\
\overline{\kappa _{4}^{P}} &=&\kappa _{4}^{P}+\frac{G_{V}}{\sqrt{2}M_{V}^{2}}%
\left( 2\left( \kappa _{1}^{PV}-\frac{2G_{V}\kappa ^{VVP}}{2\sqrt{2}M_{V}^{2}%
}\right) -\kappa _{2}^{PV}\right) \\
\overline{\kappa _{5}^{P}} &=&\kappa _{5}^{P}-\frac{F_{V}}{2\sqrt{2}M_{V}^{2}%
}\left( \kappa _{3}^{PV}-\frac{F_{V}\kappa ^{VVP}}{2\sqrt{2}M_{V}^{2}}\right)
\\
\overline{\kappa _{3}^{VV}} &=&\kappa _{3}^{VV} \\
\overline{\kappa _{4}^{VV}} &=&\kappa _{4}^{VV} \\
\overline{\kappa _{2}^{SV}} &=&\kappa _{2}^{SV} \\
\overline{\kappa _{2}^{PA}} &=&\kappa _{2}^{PA}
\end{eqnarray*}

\end{document}